\documentclass[prd,showpacs,altaffilletter,twocolumn,
superscriptaddress,floatfix,nofootinbib,preprintnumbers]{revtex4-1} 

\pdfoutput=1
\usepackage{amsfonts,amsmath,graphicx,bm} %,siunitx}
\usepackage{dcolumn}
\usepackage[pdftex,
       colorlinks=true,
       urlcolor=blue,       % \href{...}{...} external (URL)
       linkcolor=red,       % \ref{...} and \pageref{...}
       pdfpagemode=None,
       bookmarksopen=true]{hyperref}
\usepackage[capitalise]{cleveref}
\usepackage{ifpdf}
\usepackage{multirow}
\usepackage[colorlinks=true]{hyperref}
\usepackage{url}

\newcommand{\glasgow}{SUPA, School of Physics and Astronomy,
  University of Glasgow, Glasgow, G12 8QQ, UK}

\newcommand{\INFN}{INFN,  Sezione  di  Roma  Tor  Vergata,  Via  della  Ricerca  Scientifica  1,  00133  Roma  RM,  Italy}

\def\today{\number\day\space\ifcase\month\or
January\or February\or March\or April\or May\or June\or
July\or August\or September\or October\or November\or December\fi
\space\number\year}
\newcount\mins \newcount\hours
\def\now{\hours=\time \mins=\time
	\divide\hours by60 \multiply\hours by60 \advance\mins by-\hours
	\divide\hours by60 
	\number\hours:\ifnum\mins<10 0\fi\number\mins }
\newcommand{\RATIOEMU}{1.004018(95)}
\newcommand{\GAMMATOTAL}{1.73(12)\times 10^{13} ~\mathrm{s}^{-1}}
\newcommand{\GAMMAGEV}{11.36(81)\times 10^{-12} ~\mathrm{GeV}}
\newcommand{\GAMMAeTOTAL}{1.73(12)\times 10^{13} ~\mathrm{s}^{-1}}
\newcommand{\GAMMAeGEV}{11.40(82)\times 10^{-12} ~\mathrm{GeV}}
\newcommand{\GAMMATOTALincVCBETAEW}{2.94(21)(20)\times 10^{10} ~\mathrm{s}^{-1}}
\newcommand{\BRANCHINGFRAC}{0.0150(11)(10)(3)} 
\newcommand{\BbBc}{0.00581(58)(67)}
\begin{document}

\title{$B_c \rightarrow J/\psi$ Form Factors for the full $q^2$ range from Lattice QCD}
%The dispersion relation of lattice non-relativistic QCD to $\mathcal{O}(p^6)$ with kinetic couplings corrected through $\mathcal{O}(\alpha_s p^4)$} %\bm

\author{Judd \surname{Harrison}}
\email[]{judd.harrison@glasgow.ac.uk}
\affiliation{\glasgow}

\author{Christine~T.~H.~\surname{Davies}} 
\email[]{christine.davies@glasgow.ac.uk}
\affiliation{\glasgow}

\author{Andrew \surname{Lytle}} 
\affiliation{\INFN}

\collaboration{HPQCD Collaboration}
\email[]{http://www.physics.gla.ac.uk/HPQCD}

%\pacs{12.38.Gc, 13.20.Gd, 13.40.Hq, 14.40.Pq}
%\preprint{DAMTP-2015-xx}

\begin{abstract}
We present the first lattice QCD determination of the $B_c \rightarrow J/\psi$ vector and axial-vector form factors. These will enable experimental information
on the rate for $B_c$ semileptonic decays to $J/\psi$ to be converted into a 
value for $V_{cb}$. Our calculation covers the full physical $q^2$ range of the decay 
and uses non-perturbatively renormalised lattice currents. We use the 
Highly Improved Staggered Quark (HISQ) action for all valence quarks 
on the second generation MILC ensembles of gluon field configurations 
including $u$, $d$, $s$ and $c$ HISQ sea quarks. 
Our HISQ heavy quarks have masses ranging upwards from that of $c$; we 
are able to reach that of the $b$ on our finest lattices. 
This enables us to map out the dependence on heavy quark mass and determine 
results in the continuum limit at the $b$. 
We use our form factors to construct the differential 
rates for $B_c^- \rightarrow J/\psi \mu^- \overline{\nu}_\mu$ and 
obtain a total rate with 7\% uncertainty: $\Gamma(B_c^-\rightarrow J/\psi \mu^-\overline{\nu}_{\mu})/|\eta_{\mathrm{EW}}V_{cb}|^2 = \GAMMATOTAL$. 
Including values for $V_{cb}$, $\eta_{\mathrm{EW}}$ and $\tau_{B_c}$ yields a 
branching fraction for this decay mode of \BRANCHINGFRAC ~with uncertainties 
from lattice QCD, $\eta_\mathrm{EW}V_{cb}$ and $\tau_{B_c}$ respectively. 
\end{abstract}

\maketitle

%Version compiled on \today~at \now.

%===========================================================================8

\section{Introduction}
\label{sec:intro}

Accurate calculations of hadronic parameters are needed for the 
determination of Cabibbo-Kobayashi-Maskawa (CKM) matrix elements from 
the comparison of Standard Model theory and experimental results 
for exclusive flavour-changing decay rates. 
Leptonic decay rates require decay constants 
to be determined and semileptonic decay rates require form factors. 
Lattice QCD is now established as the method of choice 
for the calculation of these hadronic parameters and efforts are ongoing 
both to improve the accuracy of the results and to expand the range of 
processes for which calculational results are available. 
Here we report on the first lattice QCD calculation of the form 
factors for $B_c$ to $J/\psi$ semileptonic decay, a process under active 
study by the LHCb experiment~\cite{PhysRevLett.120.121801}. Because the valence quarks involved in 
this process are all heavy, the calculation is under very good 
control in lattice QCD as we will show. This opens the prospect 
of a new exclusive process that can be used for the determination of $|V_{cb}|$
that has a reduced theoretical uncertainty. 

Currently exclusive determinations of $|V_{cb}|$ are focussed on 
$B \rightarrow D^*$ and $B \rightarrow D$ semileptonic decays. The
emphasis is on the former pseudoscalar to vector meson transition 
because of more favourable kinematic factors 
for the differential decay rate towards the zero recoil region. 
Lattice QCD calculations are generally more accurate in this 
region because the daughter meson has small spatial momentum in 
the lattice frame (where the parent meson is usually taken to be 
at rest). The emphasis on this region led to the early lattice 
QCD $B \rightarrow D^*$ form factor calculations being purely 
done at zero recoil, where there is a single form 
factor~\cite{Bernard:2008dn, Bailey:2014tva, Harrison:2017fmw}. 
Comparison was then made to results derived from experimental 
differential rates in this same limit. More recently 
(see, for example,~\cite{BIGI2017441,Bordone:2019vic}) it has become clear that 
extrapolating the experimental results to the zero recoil 
point comes with significant systematic uncertainties, associated 
with the underlying model dependence of such extrapolations, that 
were previously being underestimated. The way forward      
requires a much more direct comparison $q^2$-bin 
by $q^2$-bin of the experimental decay rate and that 
from theory, determined from lattice 
QCD form factors. For this we need lattice QCD form
factors as a function of $q^2$ and preferably covering the 
full $q^2$ range of the decay. This allows the uncertainty 
in the extracted value of $|V_{cb}|$ to be optimised 
between the changing experimental and theoretical uncertainties 
as a function of $q^2$. We show here that it is possible 
to calculate the form factors for 
$B_c \rightarrow J/\psi$ semileptonic decay across the full 
$q^2$ range in lattice QCD.    

The $B_c \rightarrow J/\psi$ form factor calculation that 
we describe here acts as a prototype for upcoming calculations 
of form factors for $B \rightarrow D^*$ and $B_s \rightarrow D_s^*$. 
$B_c \rightarrow J/\psi$ is a more attractive starting point 
for lattice QCD, however, because the mesons are 
`gold-plated' (with tiny widths) and the valence quarks involved 
are all relatively heavy. This means that the valence quark 
propagators, from which appropriate correlation functions are 
constructed, are inexpensive to calculate. The correlation functions 
then have small statistical errors, even when the daughter has 
maximum spatial momentum. The absence of valence 
light quarks means that finite-volume effects are negligible 
and sensitivity to the $u/d$ quark mass in the sea should be small. 

The main obstacle for the calculation  of $B_c \rightarrow J/\psi$ 
form factors is that of the discretisation effects associated 
with the heavy quarks. The $c$ quark 
is handled very accurately in lattice QCD as long as improved 
discretisations of the Dirac equation are used. A particularly 
accurate approach is that of HPQCD's Highly Improved Staggered Quark (HISQ) 
action~\cite{PhysRevD.75.054502} and it is the one that we will use here for all quarks. 
For $c$ quarks a recent calculation using HISQ~\cite{Hatton:2020qhk} obtained a 0.4\% 
uncertainty in the $J/\psi$ decay constant and showed good control 
of lattice discretisation effects all the way to very coarse 
lattices with a spacing of 0.15 fm. Discretisation effects are larger 
for the $b$ quark. Indeed, since we expect discretisation effects  
to grow as a power of the heavy quark mass in lattice units, $am_h$, 
we need to work on fine lattices to approach the $b$ quark mass, 
and this is what we will do here. 
The dominant discretisation effects in the HISQ action behave 
as $a^4$, since tree-level $a^2$ errors are removed and those 
that include powers of $\alpha_s$ are heavily suppressed. 
This means that discretisation effects can be 
controlled for quarks with masses $m_h$ between that of the $c$ and the $b$ 
on lattices with lattice spacing below 0.1 fm. By working 
with a range of quark masses reaching up to that of the $b$ 
on a range of lattice spacings, we can map out both the dependence 
on $m_h$ and the dependence on $am_h$, both of which are 
smooth functions, and obtain a physical value in the 
continuum limit for the $b$ quark. This `heavy-HISQ' approach 
was developed by HPQCD for $B$ meson decay constants~\cite{McNeile:2011ng, McNeile:2012qf} and 
is now being extended to form factors~\cite{EuanBsDsstar}. Its efficacy has 
been demonstrated for $B_s \rightarrow D_s$ form factors for the
full $q^2$ range in~\cite{EuanBsDs} 
and here we apply it to $B_c \rightarrow J/\psi$.  
A big advantage of this approach is that the lattice current operators
that couple to the $W$ boson can be normalised fully 
nonperturbatively, e.g.\cite{Hatton:2019sfi,Hatton:2019gha,Cooper:2020wnj,EuanBsDs}, avoiding the systematic errors associated 
with the perturbative normalisation needed if a nonrelativistic 
approach is used for the $b$ quark in lattice QCD~\cite{Bailey:2014tva, Harrison:2017fmw}.

A further motivation for studying $B_c \rightarrow J/\psi$ 
semileptonic decay in detail is to calculate the ratio, $R(J/\psi)$, 
of the partial widths for the outgoing lepton to be a
$\tau$ compared to that for it to be an $e$ or $\mu$. The analogous results for $B$ decays, 
$R(D)$ and $R(D^*)$, 
e.g.\cite{PhysRevLett.109.101802,PhysRevLett.103.171801,PhysRevD.86.032012,PhysRevD.88.072012},
have been a source of tension 
between experiment and the Standard Model~\cite{Gambino:2019sif}, implying 
lepton-universality violation. 
Recent results from Belle, on the contrary, 
show good agreement with the Standard Model~\cite{Belle:2019rba}.   
This makes it very important to test lepton-universality violation 
in other processes and we can obtain $R(J/\psi)$ from our form 
factors for comparison with ongoing LHCb analyses. We will 
present those results and analyses of other lepton-universality 
violation tests separately~\cite{inprep}; here we focus on the form factors and 
differential rates for $W$ decay to $\mu$ or $e$.   

The subsequent sections are organised as follows:
\begin{itemize}
\item  In section \ref{sec:theory} we begin by outlining the relevant experimental observables and relate them to the invariant form factors coming from QCD matrix elements. 
\item  Section \ref{sec:calcdetails} gives an overview of methods generic to the extraction of matrix elements from HISQ three-point and two-point correlation functions in lattice QCD, and discusses our choices of operators, polarisations and momenta appropriate to extract the form factors specified in \ref{sec:theory}.
\item  Section \ref{sec:lattdeets} details the specifics of our lattice calculation including our non-perturbative current renormalisation.
\item  In section \ref{sec:results} we present the direct results of the lattice calculation and discuss the extraction of the physical continuum form factors as a function of $q^2$
\item  Finally in sections \ref{sec:discussion} and \ref{sec:conclusion} we use our form factors to compute the physical differential rates for $B_c^- \rightarrow J/\psi \mu^- \overline{\nu}_\mu$ and discuss the significance of these results and implications for future work.
\end{itemize}

\begin{figure}
\centering
\includegraphics[scale=0.35]{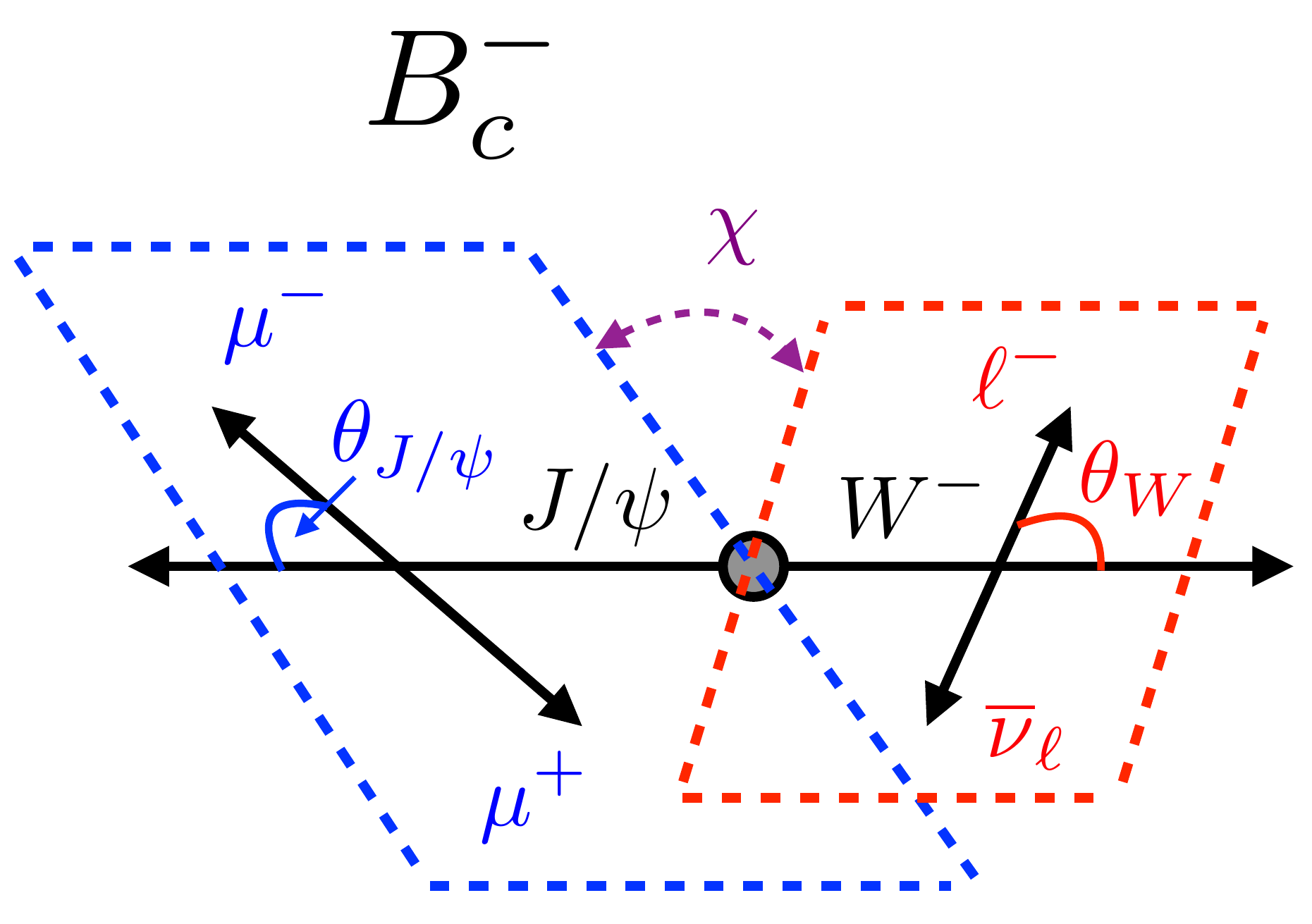}
\caption{\label{angles} Definitions of the angular variables $\chi$, $\theta_W$ and $\theta_{J/\psi}$ defined in the rest frame of the decaying $B_c^-$, $W^-$ and $J/\psi$ respectively. }
\end{figure}

\section{Theoretical Background}
\label{sec:theory}
Here we give the partial rates for $B_c^- \rightarrow J/\psi(\rightarrow \mu^+\mu^-) \ell^- \overline{\nu}_\ell$ where $\ell$ is the final state lepton as differentials with respect to 
$q^2$ and angular variables defined in the standard way in Fig.~\ref{angles}. In this work we consider only the cases $\ell=\mu$ and $\ell=e$.
%We assume that the $J/\psi$ is identified through its decay to $\mu^+\mu^-$ 
%and the angular differential rates which we give below include an 
%explicit factor of the branching fraction 
%$\mathcal{B}(J/\psi\rightarrow\mu^+\mu^-)$, since the $\mu^+\mu^-$ pair is used 
%to define the angle $\theta_{J/\psi}$. 
The partial rates are obtained from
the full differential decay rate assuming the $J/\psi$ decay is purely electromagnetic and summing over $\mu^+\mu^-$ helicities (assuming that the $\mu^+\mu^-$ are massless and hence pure helicity eigenstates)~\cite{Cohen:2018vhw}. 
This gives
\begin{align}
\label{dgammadq2}
&\frac{d\Gamma}{dq^2} =\frac{G_F^2}{(2\pi)^3}|\eta_\mathrm{EW}V_{cb}|^2\frac{(q^2-{m_\ell^2})^2|\vec{p'}|}{12M_{B_c}^2q^2}\nonumber\\
&\times\Big[\left({H_-}^2+{H_0}^2+{H_+}^2\right) \nonumber\\
+&\frac{{m_\ell^2}}{2q^2}{ \left({H_-}^2+{H_0}^2+{H_+}^2+3 {H_t}^2\right)}\Big],
\end{align}
%\begin{align}
%\label{dgammadq2}
%&\frac{d\Gamma(B_c \rightarrow J/\psi \ell \overline{\nu})}{dq^2} =\frac{G_\mathrm{F}^2}{(2\pi)^3}|\eta_{\mathrm{EW}}|^2|V_{cb}|^2\frac{q^2|\vec{p'}|}{12M_{B_c}^2}\Big[\nonumber\\
%&\left({H_-}^2+{H_0}^2+{H_+}^2\right)\Big],
%\end{align}
\begin{align}
\label{dgammadcosw}
&\frac{d\Gamma}{dq^2 d\cos(\theta_W)} =\nonumber\\ 
&\frac{G_F^2}{(2\pi)^3}|\eta_\mathrm{EW}V_{cb}|^2\frac{(q^2-{m_\ell^2})^2|\vec{p'}|}{16M_{B_c}^2q^2}\mathcal{B}(J/\psi\rightarrow \mu^+\mu^-)\Big[\nonumber\\
&H_-^2 \frac{1}{2}(1+\cos(\theta_W))^2+H_+^2 \frac{1}{2} (1-\cos(\theta_W))^2\nonumber\\
+&H_0^2 \sin ^2(\theta_W)\nonumber\\
+&\frac{{m_\ell}^2}{2 q^2} \Big(\left(H_-^2+H_+^2\right) \sin ^2(\theta_W)\nonumber\\
+&2 (H_t-H_0 \cos (\theta_W))^2\Big)\Big],
\end{align}
%\begin{align}
%\label{dgammadcosw}
%&\frac{d\Gamma(B_c \rightarrow J/\psi(\rightarrow\mu^+\mu^-) \ell \overline{\nu})}{dq^2 d\cos(\theta_W)} =\nonumber\\ 
%&\frac{G_\mathrm{F}^2}{(2\pi)^3}|\eta_{\mathrm{EW}}|^2|V_{cb}|^2\frac{q^2|\vec{p'}|}{16M_{B_c}^2}\mathcal{B}(J/\psi\rightarrow\mu^+\mu^-)\Big[\nonumber\\
%&H_-^2 \frac{1}{2}(1+\cos(\theta_W))^2+H_+^2 \frac{1}{2} (1-\cos(\theta_W))^2\nonumber\\
%+&H_0^2 \sin ^2(\theta_W)\Big],
%\end{align}
\begin{align}
\label{dgammadtheta}
&\frac{d\Gamma}{dq^2 d\cos(\theta_{J/\psi})} =\nonumber\\ 
&\frac{G_F^2}{(2\pi)^3}|\eta_\mathrm{EW}V_{cb}|^2\frac{(q^2-{m_\ell^2})^2|\vec{p'}|}{16M_{B_c}^2q^2}\mathcal{B}(J/\psi\rightarrow \mu^+\mu^-)\Big[\nonumber\\
&{(H_-^2+H_+^2 )\frac{1}{2}\left(1+  \cos^2 (\theta_{J/\psi})\right)}+{H_0^2 \left(1-  \cos^2 (\theta_{J/\psi})\right)}\nonumber\\
+&\frac{{m_\ell^2}}{2q^2} \Big({(H_-^2+H_+^2) \frac{1}{2}\left(1+  \cos^2 (\theta_{J/\psi})\right)}\nonumber\\
+&{(H_0^2+3H_t^2 )\left(1-  \cos^2 (\theta_{J/\psi})\right)}\Big)\Big],
\end{align}
%\begin{align}\label{dgammadtheta}
%&\frac{d\Gamma(B_c \rightarrow J/\psi(\rightarrow\mu^+\mu^-) \ell \overline{\nu})}{dq^2 d\cos(\theta_{J/\psi})} =\nonumber\\ 
%&\frac{G_\mathrm{F}^2}{(2\pi)^3}|\eta_{\mathrm{EW}}|^2|V_{cb}|^2\frac{q^2|\vec{p'}|}{16M_{B_c}^2}\mathcal{B}(J/\psi\rightarrow\mu^+\mu^-)\Big[\nonumber\\
%&{(H_-^2+H_+^2 )\frac{1}{2}\left(1+  \cos^2 (\theta_{J/\psi})\right)}+{H_0^2 \left(1-  \cos^2 (\theta_{J/\psi})\right)}\Big],
%\end{align}
and 
\begin{align}
\label{dgammadchi}
&\frac{d\Gamma}{dq^2 d\chi} =\nonumber\\ 
&\frac{G_F^2}{(2\pi)^3}|\eta_\mathrm{EW}V_{cb}|^2\frac{(q^2-{m_\ell^2})^2|\vec{p'}|}{16M_{B_c}^2q^2}\mathcal{B}(J/\psi\rightarrow \mu^+\mu^-)\frac{2}{3\pi}\Big[\nonumber\\
&{ H_-^2}+{ H_0^2}+{ H_+^2}+{\frac{1}{2}H_- H_+ \cos (2 \chi)}\nonumber\\
&+\frac{{m_\ell^2}}{ 2q^2} \Big({H_-^2}+{H_0^2}+{H_+^2}+{3H_t^2}-{H_- H_+ \cos (2 \chi)}\Big)\Big].
\end{align}
%\begin{align}
%\label{dgammadchi}
%&\frac{d\Gamma(B_c \rightarrow J/\psi(\rightarrow\mu^+\mu^-) \ell \overline{\nu})}{dq^2 d\chi} =\nonumber\\ 
%&\frac{G_\mathrm{F}^2}{(2\pi)^3}|\eta_{\mathrm{EW}}|^2|V_{cb}|^2\frac{q^2|\vec{p'}|}{16M_{B_c}^2}\frac{2}{3\pi}\mathcal{B}(J/\psi\rightarrow\mu^+\mu^-)\Big[\nonumber\\
%&{ H_-^2}+{ H_0^2}+{ H_+^2}+{\frac{1}{2}H_- H_+ \cos (2 \chi)}\Big] .
%\end{align}
Here $p'$ is the momentum of the $J/\psi$, $p$ is the momentum of the $B_c^-$, $q=p-p'$, $|\vec{p'}|$ is the magnitude of the $J/\psi$ spatial momentum in the $B_c^-$ rest frame and $m_\ell$ is the mass of the final state lepton $\ell$. $G$ is the Fermi constant defined from muon lifetime, $V_{cb}$ is the appropriate Cabibbo-Kobayashi-Maskawa matrix element and 
$\eta_{\mathrm{EW}}$ is a structure-independent electroweak correction. For 
the electrically neutral $J/\psi$ final state this factor is close to 1~\cite{Sirlin:1981ie}, 
$1+\alpha\log(M_Z/\mu)/\pi$. 
Taking $\mu=M_{B_c}$, the mass of the $B_c^-$ meson, 
but allowing for variation by a factor 
of 2~\cite{Bailey:2014tva}, 
gives $\eta_{\mathrm{EW}}$ = 1.0062(16). 
Note that we include in the expressions for the differential rate terms with factors of $m_\ell^2/q^2$, which are negligible for $\ell=e$ but have a very small visible effect near $q^2=0$ for $\ell =\mu$.

The helicity amplitudes are defined as
\begin{align}
H_\pm(q^2) =& (M_{B_c}+M_{J/\psi})A_1(q^2) \mp \frac{2M_{B_c}|\vec{p'}|}{M_{B_c}+M_{J/\psi}}V(q^2),\nonumber\\
H_0(q^2) =& \frac{1}{2M_{J/\psi} \sqrt{q^2}} \Big(-4\frac{M_{B_c}^2{\vec{p'}}^2}{M_{B_c}+M_{J/\psi}}A_2(q^2)\nonumber\\
& +   (M_{B_c}+M_{J/\psi})(M_{B_c}^2 - M_{J/\psi}^2 - q^2)A_1(q^2) \Big),\nonumber\\
H_t(q^2) =& \frac{2M_{B_c}|\vec{p'}|}{\sqrt{q^2}}A_0(q^2)\label{helicityamplitudes}
\end{align}
and correspond to the nonzero values of $\overline{\epsilon}^*_\mu(q,\lambda^\prime)\langle J/\psi(p^\prime,\lambda)|\bar{c}\gamma^\mu (1-\gamma^5)  b|B_c^-(p)\rangle$ for the different combinations of the $J/\psi$ and $W^-$ polarisations $\lambda$ and $\lambda^\prime$ respectively. The form factors in Eq.~(\ref{helicityamplitudes}) are the standard Lorentz invariant ones, their relations to the matrix elements are given by \cite{RevModPhys.67.893}
\begin{align}
& \langle  J/\psi(p',\lambda)|\bar{c}\gamma^\mu  b|B_c^-(p)\rangle=\nonumber\\
& \frac{2i V(q^2)}{M_{B_c} + M_{J/\psi}} \varepsilon^{\mu\nu\rho\sigma}\epsilon^*_\nu(p',\lambda) p'_\rho p_\sigma \nonumber \\
&\langle  J/\psi(p',\lambda)|\bar{c}\gamma^\mu \gamma^5 b|B_c^-(p)\rangle = \nonumber\\
& 2M_{J/\psi}A_0(q^2)\frac{\epsilon^*(p',\lambda)\cdot q}{q^2} q^\mu \nonumber\\
&  +(M_{B_c}+M_{J/\psi})A_1(q^2)\Big[ \epsilon^{*\mu}(p',\lambda) - \frac{\epsilon^*(p',\lambda)\cdot q}{q^2} q^\mu \Big]\nonumber \\
&-A_2(q^2)\frac{\epsilon^*(p',\lambda)\cdot q}{M_{B_c}+M_{J/\psi}}\Big[ p^\mu + p'^\mu - \frac{M_{B_c}^2-M_{J/\psi}^2}{q^2}q^\mu \Big]. \label{formfactors}
\end{align}
We also have
\begin{equation}
\langle 0 |\bar{c}\gamma^\nu c|J/\psi(p',\lambda)\rangle = N_{J/\psi}\epsilon^\nu(p',\lambda),
\end{equation}
\begin{equation}
\langle B_c^-(p) |\bar{b}\gamma^5 c|0)\rangle = N_{B_c},
\end{equation}
and
\begin{equation}\label{spinsumident}
\sum_{\lambda} \epsilon_\nu(p',\lambda)\epsilon_{\mu}^{*}(p',\lambda) = -g_{\nu\mu}+\frac{p'_\nu p'_\mu}{M^2}
\end{equation}
where $N_{J/\psi}$ and $N_{B_c}$ are amplitudes proportional to the decay constant of the corresponding meson and $\epsilon$ is the $J/\psi$ polarisation vector. We will make use of these when we come to extract the form factors in Eq.~(\ref{formfactors}) from our lattice correlation functions. 

\section{Computational Strategy}
\label{sec:calcdetails}
We follow the strategy of previous heavy-HISQ calculations of form-factors~\cite{EuanBsDsstar,EuanBsDs,Cooper:2020wnj}. 
We work with multiple heavy quarks with masses $m_h$ between the physical $c$ 
and $b$ masses on lattices with a range of fine lattice spacings. 
Most of our $m_h$ masses are below the $b$ mass but we are able 
to reach a mass very close to the physical $b$ quark mass on our finest 
lattices. An important point is that, at every lattice spacing, we 
are able to cover the full $q^2$ range of the decay for the heavy 
quark masses that we can reach at that lattice spacing, i.e. the 
$q^2$ range expands on finer lattices in step with the heavy quark 
mass range~\cite{EuanBsDs}. 
We compute form factors by extracting combinations of the 
relevant matrix elements defined in Eq.~(\ref{formfactors}) from 
correlation functions computed on the lattice. We then fit the form factors 
as a function of lattice spacing and heavy quark mass to determine 
their functional form in the continuum limit 
at the physical $b$ mass.  

The correlation functions that we calculate, for general choices $\nu$ and $\Gamma$ of $J/\psi$ polarisation and current operator respectively, are
\begin{align}
C_\text{2pt}^{J/\psi}(t,0) =& \langle  0|\bar{c}\gamma^\nu c(t) \left(\bar{c}\gamma^\nu c(0)\right)^\dagger| 0 \rangle,\nonumber\\
%=& \sum_{a,n}\frac{(-1)^{at}e^{-tE_{n^a}}}{2E_{n^a}}\langle  0|\bar{c}\gamma^\nu c(0)| n^a\rangle \langle n^a|\left(\bar{c}\gamma^\nu c(0)\right)^\dagger| 0 \rangle
C_\text{2pt}^{H_c}(t,0) =& \langle  0|\left(\bar{h}\gamma^5 c(t)\right)^\dagger\bar{h}\gamma^5 c(0) | 0 \rangle,\nonumber\\
%=& \sum_{b,m}\frac{(-1)^{at}e^{-tM_{n^a}}}{2M_{n^a}}\langle  0|\left(\bar{b}\gamma^5 c(0)\right)^\dagger| n^a\rangle \langle n^a|\bar{b}\gamma^5 c(0)| 0 \rangle
C_\text{3pt}(T,t,0) =& \langle  0|\bar{c}\gamma^\nu c(T) ~ \bar{c}\Gamma h(t) ~ \bar{h}\gamma^5 c(0)| 0 \rangle. \label{threepointcorr}
%= \sum_{a,b,n,m}\frac{(-1)^{(T+t)a + tb}}{2E_{n^a}2M_{m^b}}&\langle 0| \bar{c}\gamma^\nu c|n^a \rangle\nonumber\\
%&\times \langle  n ^a|\bar{c}\Gamma^\mu b |m^b\rangle\nonumber\\
%&\times \langle m^b| \bar{b}\gamma^5 c|0  \rangle e^{-(T-t)E_{n^a} - tM_{m^b}},
\end{align}
Note that in this section, for notational simplicity, we consider the matrix elements in terms of continuum current operators. The nonperturbative renormalisation of our lattice current operators is discussed in Section~\ref{sec:renorm}. 

By considering the insertion of complete sets of states we may express these correlation functions in terms of general fit forms (folding the two-point 
functions about the mid-point of the lattice so that $t$ extends from 
0 to $N_t/2$)
\begin{align}
{C}_\text{2pt}^{J/\psi}(t,0) =\sum_{n}\Big((A^n)^2{e^{-tE_{n}}}\nonumber+(-1)^{t}(A^n_o)^2{e^{-tE_n^o}}\Big)
\end{align}
\begin{align}
{C}_\text{2pt}^{H_c}(t,0) =\sum_{n}\Big((B^n)^2{e^{-tM_{n}}}\nonumber+(-1)^{t}(B^n_o)^2{e^{-tM_n^o}}\Big)
\end{align}
and
\begin{align}\label{threepointfit}
{C}_\text{3pt}(T,t,0) &=\sum_{n,m}\Big({  A^n B^m J^{nm} e^{-(T-t)E_{n} - tM_{m}} }\nonumber\\
+&{(-1)^{T+t}}  A^n_o B^m J^{nm}_{oe} e^{-(T-t)E_n^o - tM_{m}} \nonumber\\
+&{(-1)^{t}}  A^n B^m_o J^{nm}_{eo} e^{-(T-t)E_{n} - tM_m^o} \nonumber\\
+&{(-1)^{T}}  A^n_o B^m_o J^{nm}_{oo} e^{-(T-t)E_n^o - tM_m^o} \Big) .
\end{align}
Here $n$, $m$ correspond to on shell particle states with quantum numbers 
resulting in nonzero amplitudes and the $o$ labels indicate 
energies and amplitudes corresponding to the time-doubled states 
typical of staggered quarks. 
The lowest energy, non-oscillating states are those corresponding 
to the $J/\psi$ and $H_c^-$. We work with the $H_c^-$ at rest, choosing 
to access the full $q^2$ range by giving momentum only to 
the $J/\psi$, and extract the matrix elements of these states 
from our lattice correlators. This gives:
\begin{align}
A^0=&\frac{N_{J/\psi} }{\sqrt{2E_{J/\psi}}}\left(1+\frac{\vec{p'}_{\nu}^{2}}{M_{J/\psi}^2}\right)^{1/2},
\end{align}
\begin{align}
B^0 = \frac{N_{H_c}}{\sqrt{2M_{H_c}}}
\end{align}
and
\begin{equation}
J^{00}_{(\nu,\Gamma)} = \sum_{\lambda}\frac{\epsilon^\nu(p',\lambda) \langle  J/\psi(p',\lambda ) |\bar{c}\Gamma b |H_c^-\rangle}{\sqrt{2E_{J/\psi}2M_{H_c}\left(1+\vec{p'}_{\nu}^{2}/M_{J/\psi}^2\right)}}\label{relnorm}
\end{equation}
where $\vec{p'}_{\nu}$ is the $\nu$ component of the $J/\psi$ spatial momentum, with $\nu$ corresponding to the choice of polarisation in Eq.~(\ref{threepointcorr}), with current $\overline{c}\Gamma h$. In the subsequent subsections we discuss the combinations of $\nu$ and $\Gamma$ for which we must compute correlation functions in order to extract the full set of correlation functions defined in Eq.~(\ref{formfactors}).
\subsection{Extracting $V(q^2)$}
The choice of operators used to extract $V(q^2)$ is the most straightforward since the matrix element of the vector current $\overline{c}\gamma^\mu b$ involves only $V$. With $p=(M_{H_c},0,0,0)$ we have
\begin{align}
 \sum_{\lambda}&{\epsilon^\nu(p',\lambda) \langle  J/\psi(p',\lambda ) |\bar{c}\gamma^\mu h |H_c^-\rangle}\nonumber\\
=\sum_{\lambda}&\frac{2i V(q^2)}{M_{H_c} + M_{J/\psi}}\varepsilon^{\mu\kappa\rho\sigma}\epsilon^\nu(p',\lambda)\epsilon^*_\kappa(p',\lambda) p'_\rho p_\sigma \nonumber \\
=&\frac{-2i \epsilon^{\mu\nu\rho0} p'_\rho M_{H_c} }{M_{H_c} + M_{J/\psi}}V(q^2)
\end{align}
In this calculation we give the $J/\psi$ spatial momentum $\vec{p'} = (k,k,0)$.
In order to isolate all the form factors we need one component of 
$\vec{p'}$ to be zero. Keeping both of the others non-zero minmises the discretisation 
errors for a given magnitude of $p'$. Here we choose $\mu=3$ and $\nu=1$ and find 
\begin{align}
V(q^2)=&\Phi(k)\frac{M_{H_c} + M_{J/\psi}}{2i k M_{H_c} }J^{00}_{(1,\gamma^3)}
\end{align}
where we have defined the relativistic normalisation
\begin{equation}
\Phi(k) = \sqrt{2E_{J/\psi}2M_{H_c}\left(1+k^{2}/M_{J/\psi}^2\right)}
\end{equation}
with $k$ the $\nu$ component of $p^{\prime}$. 
\subsection{Extracting $A_0(q^2)$}
In order to isolate $A_0(q^2)$, following \cite{PhysRevD.90.074506}, we make use of the partially conserved axial-vector current (PCAC) relation $\langle \partial A \rangle = (m_c+m_h) \langle P \rangle$ where $A=\overline{c}\gamma^5 \gamma^\nu h$ and $P=\overline{c}\gamma^5 h$. From Eq.~(\ref{formfactors}) we have
\begin{align}
\sum_{\lambda}\epsilon^\nu(p',\lambda)& q_\mu\langle  J/\psi(p',\lambda)|\bar{c}\gamma^\mu \gamma^5 h|H_c^-\rangle \nonumber\\
=& \sum_{\lambda}\epsilon^\nu(p',\lambda)2M_{J/\psi}A_0(q^2){\epsilon^*(p',\lambda)\cdot q}\nonumber\\
=& \frac{2kE_{J/\psi}M_{H_c}}{M_{J/\psi}}A_0(q^2).
\end{align}
Taking $\Gamma^\mu = \gamma^5$ and $\nu = 1$ in Eq.~(\ref{relnorm}) and multiplying by $m_c+m_b$ we then have
\begin{equation}
A_0(q^2) = \Phi(k)\frac{(m_c+m_b)M_{J/\psi}}{2kE_{J/\psi}M_{H_c}} J^{00}_{(1,{\gamma^5})}
\end{equation}
\subsection{Extracting $A_1(q^2)$}
In order to isolate $A_1$ we use the axial-vector current $\Gamma = \bar{c}\gamma^\mu \gamma^5 h$ and $J/\psi$ operator $\bar{c}\gamma^\nu c$ and choose $\mu =\nu= 3$ along the spatial direction with zero $J/\psi$ momentum. Using Eq.~(\ref{formfactors}) this gives 
\begin{align}
\sum_{\lambda}\epsilon^3(p',\lambda)\langle  J/\psi(p',\lambda)|&\bar{c}\gamma^3 \gamma^5 h|H_c^-\rangle =\nonumber\\
&(M_{J/\psi}+M_{H_c})A_1(q^2)
\end{align}
which gives, in terms of $J^{00}$ 
\begin{equation}
A_1(q^2) = \Phi(0) \frac{J^{00}_{(3,{\gamma^3\gamma^5})}}{M_{J/\psi}+M_{H_c}}.
\end{equation}
\subsection{Extracting $A_2(q^2)$}
The extraction of $A_2$ is more complicated than the extraction of the other form factors since no trivial choice of directions in axial-vector and $J/\psi$ operators isolates the contribution of $A_2$ relative to $A_1$ or $A_0$. We use axial-vector current operator $\Gamma = \bar{c}\gamma^1 \gamma^5 h$ and $J/\psi$ operator $\bar{c}\gamma^1 c$. This yields contributions from each form factor, 
\begin{align}
&\Phi(k)J^{00}_{(1,\gamma^1 \gamma^5)}=\sum_{\lambda}\epsilon^1(p',\lambda)\langle  J/\psi(p',\lambda)|\bar{c}\gamma^1 \gamma^5 b|H_c^-\rangle= \nonumber\\
& -\frac{2k^2E_{J/\psi}M_{H_c}}{q^2M_{J/\psi}}A_0(q^2)\nonumber\\
&+(M_{H_c} + M_{J/\psi})\left(1+\frac{k^2}{M_{J/\psi}^2}+\frac{E_{J/\psi}M_{H_c}k^2}{M_{J/\psi}^2q^2}\right)A_1(q^2)\nonumber\\
&-A_2(q^2)\frac{k^2E_{J/\psi}M_{H_c}}{M_{J/\psi}^2(M_{H_c}+M_{J/\psi})}\left(1+\frac{M_{H_c}^2-M_{J/\psi}^2}{q^2}\right).\label{A2subbit}
\end{align}
We must then subtract the $A_0$ and $A_1$ contributions to obtain $A_2$. 

\section{Lattice Calculation}
\label{sec:lattdeets}
\begin{figure}
\centering
\includegraphics[scale=0.4]{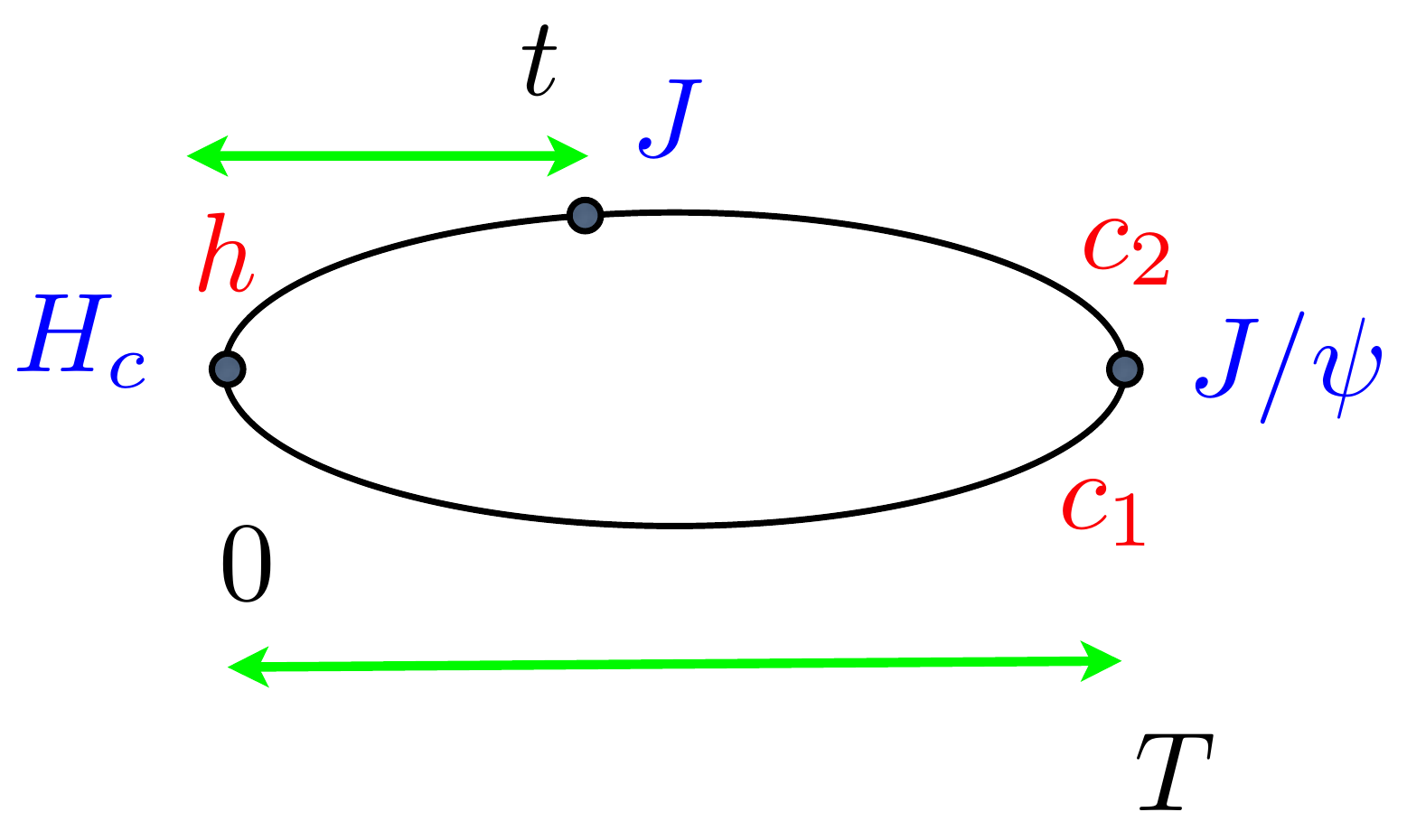}
\caption{\label{3pt} Arrangement of propagators in the three point function, we refer to $c_1$ and $c_2$ as the `spectator' and `active' charm quarks respectively and $h$ as the `extended' heavy quark. $J$ represents the insertion of a vector or 
axial-vector current. }
\end{figure}

In this section we give details of the gauge field configurations used in this calculation, as well as the values of masses, momenta and staggered spin-taste operators used in the calculation. We use the the second generation MILC gluon ensembles including light, strange and charm sea quarks \cite{PhysRevD.87.054505,PhysRevD.82.074501} and compute HISQ charm and heavy quark propagators. The details of these gauge configurations are given in table~\ref{lattdets}.
\begin{table}
\caption{Details of the gauge field configurations used in our calculation \cite{PhysRevD.87.054505,PhysRevD.82.074501}. Set 1 is referred to as `fine', set 2 as `superfine', set 3 as `ultrafine' and set 4 as `physical fine'. The physical value of $\omega_0$ was determined in \cite{PhysRevD.88.074504} to be 0.1715(9)fm and the values of $\omega_0/a$ were taken from \cite{PhysRevD.96.034516,PhysRevD.91.054508,EuanBsDs}.  \label{lattdets}}
\begin{tabular}{c c c c c c c}\hline
 Set &$\omega_0/a$ & $N_x\times N_t$ &$am_{l0}$&$am_{s0}$& $am_{c0}$  &$n_\text{configs}$ \\ \hline
1 & $1.9006(20)$  & $32\times96 $    &$0.0074$ &$0.037$  & $0.440$ & 980\\
2 & $2.896(6)$  & $48\times144  $    &$0.0048$ &$0.024$  & $0.286$ & 500\\
3 & $3.892(12)$  &$ 64\times192  $    &$0.00316$ &$0.0158$  & $0.188$ &374\\
4 & $1.9518(7)$  &$ 64\times96  $    &$0.0012$ &$0.0363$  & $0.432$ &300\\\hline
\end{tabular}
\end{table}
As stated earlier, we work with the $H_c$ at rest on the lattice. The arrangement of operators in the three point correlation functions is shown in figure~\ref{3pt}. We refer to $c_1$ as the `spectator' charm quark, $c_2$ as the `active' quark 
and $h$ as the `extended' heavy quark. 
We use a single random wall source at time $T$ for both the spectator and active quark, with a phase patterning the source for the active quark to achieve an appropriate 
staggered spin-taste for the meson. 
We use the spectator propagator at time $0$ as the source for the extended heavy quark propagator with a spin-taste operator corresponding to the $H_c$ quantum numbers and then tie the heavy quark propagator together with the active charm quark propagator at the current at time $t$. 

\begin{table}
\centering
\caption{Spin-taste operators used to isolate form factors. The first column is the operator used for the $H_c$, the second for the $J/\psi$ and the third column is the operator used at the current. \label{spintastetable}}
\begin{tabular}{c | c c c }\hline
 &$\mathcal{O}_{H_c}$ & $\mathcal{O}_{J/\psi}$ & $\mathcal{O}_J$   \\
\hline
$V$ & $\gamma_0\gamma_5\otimes \gamma_0\gamma_5$ & $\gamma_1\otimes \gamma_1\gamma_2$ & $\gamma_3\otimes \gamma_3$  \\
$A_0$ & $\gamma_5\otimes \gamma_5$ & $\gamma_1\otimes 1$ & $\gamma_5\otimes \gamma_5$  \\
$A_1$ & $\gamma_5\otimes \gamma_5$ & $\gamma_3\otimes \gamma_3$ & $\gamma_3\gamma_5\otimes \gamma_3\gamma_5$ \\
$A_2$ & $\gamma_5\otimes \gamma_5$ & $\gamma_1\otimes \gamma_1$ & $\gamma_1\gamma_5\otimes \gamma_1\gamma_5$ \\ \hline    
\end{tabular}
\end{table}

\begin{table}
\centering
\caption{Details of the charm and heavy valence masses. \label{masses}}
\begin{tabular}{c c c }\hline
 Set &$am_h^\text{val}$ & $am_c^\text{val} $ \\ \hline
1 & $0.6,0.65,0.8$  & $0.449$      \\
2 & $0.427,0.525,0.65,0.8$  & $0.274$     \\
3 & $0.5,0.65,0.8$  & $0.194$     \\
4 & $0.5,0.65,0.8$  & $0.433$    \\ \hline
\end{tabular}
\end{table}

Following the notation of \cite{PhysRevD.75.054502} for staggered
spin-taste operators, we choose lattice meson and current operators which have the same quantum numbers as those detailed in section~\ref{sec:calcdetails}. We always take the currents to be the local ones, $\gamma_i \otimes\gamma_i$ and $\gamma_i\gamma_5 \otimes\gamma_i\gamma_5$. This means that the current insertions only need to be renormalised once. In order to make the desired matrix element an overall taste singlet, this requires that we use different tastes of $J/\psi$ and $H_c$.
The specific combinations we use are given in table~\ref{spintastetable}. 
This means that we have three different tastes of $J/\psi$ and 
two different tastes of $H_c$. Different tastes of a meson differ in 
mass by discretisation effects that are small for HISQ 
quarks~\cite{PhysRevD.75.054502, PhysRevD.87.054505} 
and particularly small for heavy mesons.  
When fitting two- and three-point correlators the energies 
corresponding to different spin-taste operators use separate 
priors (i.e. we assume that they are different) except for $\gamma_3\otimes \gamma_3$ 
and $\gamma_1\otimes \gamma_1$ whose energies we fix to be equal.
We will show later how close the different taste meson masses are. 

We also calculate pseudoscalar heavyonium two-point correlation 
functions 
with spin-taste $\gamma_5 \otimes \gamma_5$. This allows 
us to determine the mass of the ground-state heavyonium meson that 
we denote $\eta_h$. 
The mass of the $\eta_h$ is a useful physical proxy 
for the heavy quark mass when we map out the dependence on heavy quark 
mass of our results. 

We choose values of $am_h$ for the heavy quark masses at each value of 
the lattice spacing following \cite{EuanBsDsstar}. 
They range from above the $c$ mass up to a value of 0.8. This corresponds 
to a growing range in $m_h$ as the lattice spacing becomes finer. 
The valence masses used to compute propagators are given in table~\ref{masses}.
We also choose the values of $T$ for our 3-point correlation functions 
following~\cite{EuanBsDsstar}.

Since we must compute heavy quark propagators at multiple masses, from multiple $T$, the choice to insert momentum via the active charm quark minimises 
the number of inversions. 
We put momentum on the $c$ quark propagator via a 
momentum twist~\cite{Sachrajda:2004mi, Guadagnoli:2005be}. 
The twists were chosen to span evenly the physical $q^2$ range for the largest value of $am_h$, approximated from $aM_{H_c}$ values given in \cite{EuanBsDsstar} and the physical $J/\psi$ mass. The values of twists and $T$ are given in table~\ref{twists}. Note that because we choose twists to evenly span the physical $q^2$ range for the maximum value of $am_h^\mathrm{val}$ on a given lattice, the greatest values of twist, $\theta$, will correspond to negative values of $q^2$ for the smaller values of $am_h^\mathrm{val}$ used. We include these negative $q^2$ points in our analysis, since the form factors are analytic for $-\infty<q^2<M_\mathrm{pole}^\mathrm{min}$ where $M_\mathrm{pole}^\mathrm{min}$ is the lowest mass subthreshold resonance.
\begin{table}
\centering
\caption{Details of the twists and three-point ranges, $T$, used in our 
calculation. The twists are given in units of $\pi/L$ and are applied along both the $x$ and $y$ direction. The momentum component of the $J/\psi$ in each
of these directions in lattice units is then $\pi\theta/N_x$. \label{twists}}
\begin{tabular}{ c c c }\hline
Set & $\theta $ &$T$  \\ \hline
1   & $0,0.361,0.723,1.084,1.446,1.807 $    &$14,17,20$   \\
2   & $0,0.826,1.651,2.477,3.302,4.128$   &$22,25,28$   \\
3   &$ 0,1.241,2.483,3.724,4.966,6.207 $    &$31,36,41$   \\
4   & $0,0.677,1.354,2.030,2.707,3.384 $    &$14,17,20$   \\\hline
\end{tabular}
\end{table}

\subsection{Non-Perturbative Current Renormalisation}
\label{sec:renorm}
The renormalisation factors relating 
our lattice current operators to the partially conserved lattice currents 
were computed in~\cite{EuanBsDsstar} for the axial-vector 
and in~\cite{EuanBsDs} for the vector current using the PCAC 
and PCVC relations respectively. 
These factors were not computed for $am_h=0.6$ on set 1 and $amh=0.65$ 
on set 4 and for these we interpolate between 
the other values given in~\cite{EuanBsDs, EuanBsDsstar}, noting that the 
differences between results on set 1 and set 4 are very small. 
In order to account for correlations between $Z$ factors, 
which are neglected in our interpolation, we take the conservative approach of setting 
the uncertainty on our interpolated values equal to the largest uncertainty 
of the computed values on the corresponding set.
We tabulate the renormalisation factors in table~\ref{Zfactors}, 
where 
we also include the $ma$-dependent discretisation correction terms, $Z^\text{disc}$, 
for the HISQ-quark tree level wavefunction renormalisation 
computed in~\cite{ronrenormdisc}. $Z^\text{disc}$ starts at $(am_h)^4$ because 
of the high level of improvement in the HISQ action, so is very close to 1. 
The total renormalisation factor is given by $Z^{V,A}Z^\text{disc}$.
Note that the pseudoscalar current used to determine $A_0$ is 
absolutely normalised and needs no $Z$ factor. 

\begin{table}
\centering
\caption{$Z$ factors from \cite{EuanBsDsstar} and \cite{EuanBsDs} for the axial-vector and vector operators used in this work, together with the discretisation corrections, $Z^\text{disc}$. $Z^A$ and $Z^V$ values for $am_h=0.6$ on set 1 and $am_h=0.65$ on set 4 were obtained by interpolation from the other values for those sets, with uncertainties set equal to the largest uncertainty of the other values. \label{Zfactors}}
\begin{tabular}{ c c c c c }
\hline
Set & $am_h$ & $Z^A$& $Z^V$& $Z^\text{disc}$ \\\hline
1 & 0.6 & 1.03526(58)& 1.0217(35) & 0.99711 \\
 & 0.65 & 1.03740(58)& 1.0254(35) & 0.99635 \\
 & 0.8 & 1.04367(56)& 1.0372(32) & 0.99306 \\
\hline
2 & 0.427 & 1.0141(12)& 1.0025(31) & 0.99931 \\
 & 0.525 & 1.0172(12)& 1.0059(33) & 0.99859 \\
 & 0.65 & 1.0214(12)& 1.0116(37) & 0.99697 \\
 & 0.8 & 1.0275(12)& 1.0204(46) & 0.99367 \\
\hline
3 & 0.5 & 1.00896(44)& 1.0029(38) & 0.99889 \\
 & 0.65 & 1.01363(49)& 1.0081(43) & 0.99704 \\
 & 0.8 & 1.01968(55)& 1.0150(49) & 0.99375 \\
\hline
4 & 0.5 & 1.03184(47)& 1.0134(24) & 0.99829 \\
 & 0.65 & 1.03717(47)& 1.0229(29) & 0.99645 \\
 & 0.8 & 1.04390(39)& 1.0348(29) & 0.99315 \\
\hline
\end{tabular}
\end{table}

The statistical uncertainty of each lattice matrix element is typically at 
least an order of magnitude greater than that of the corresponding 
current renormalisation factor. We therefore ignore correlations 
between different $Z$ factors and between $Z$ factors and our correlator data.

\section{Results}
\label{sec:results}
The correlator fits discussed in Section \ref{sec:calcdetails} were done 
simultaneously for all correlators on each set using 
the \textbf{corrfitter} python package \cite{corrfitter}. This allows us 
to determine the ground-state fit parameters that we need along with 
a complete covariance matrix for them that can be fed into our continuum 
extrapolation fits. 
The priors for our correlator fits are listed in table~\ref{priortable} 
and take the same form on all sets. 
Our correlator fits include an SVD cut  on the correlation matrix following 
the tools included in~\cite{corrfitter} (see Appendix D of~\cite{Dowdall:2019bea} 
for a discussion of this). 
We also truncate the time-length of the three point and two point correlators 
that we fit to remove excited-state dominated data points at early times 
and increase fit stability. 
The specifics of these fit choices are given in table~\ref{fitparams}, along 
with variations used to test the stability of our analysis in Section~\ref{stabsec}.

\begin{table}
\centering
\caption{Correlator fit priors. $\Delta E_i = E_{i+1}-E_{i}$, $\Omega_{J/\psi} =(3.0^2+p'^2)^\frac{1}{2}$ and $\Omega_{H_c}=M^\text{max}_{H_c}\left(\frac{am_h}{0.8}\right)^\frac{1}{2}$ where $M^\text{max}_{H_c}$ is the value of $M_{H_c}$ corresponding to the largest $am_h$, taken from \cite{EuanBsDsstar}. While $\Omega_{J/\psi}$ was chosen to follow the relativistic dispersion relation, $\Omega_{H_c}$ was chosen heuristically to give prior values approximately following the observed $H_c$ masses on each set while remaining suitably loose so as not to constrain the fit results. \label{priortable}}
\begin{tabular}{ c c c c c }\hline
 Prior & $\eta_b$ & $\eta_c$ & $J/\psi(p')$ & $H_c$\\\hline
$E_0/\text{GeV}$	&$m_h2.5(0.5)$	&$3.0(0.6)$  	&$\Omega_{J/\psi} 1.0(0.2)$ 	&$\Omega_{H_c}1.0(0.2)$ 	\\
$\Delta E_i/\text{GeV}$	&$m_h1.25(1.9)$	&$1.5(2.25)$	&$\Omega_{J/\psi} 0.5(0.75)$ 	&$\Omega_{H_c}0.5(0.75)$	\\
$E^o_0/\text{GeV}$	&$-$		&$-$		&$\Omega_{J/\psi} 1.2(0.5)$	&$\Omega_{H_c}1.2(0.5)$		\\
$\Delta E^o_i/\text{GeV}$&$-$		&$-$		&$\Omega_{J/\psi} 0.5(0.75)$	&$\Omega_{H_c}0.5(0.75)$	\\
$A(B)_{(o)}^n$		&$0.1(5.0)$	&$0.1(5.0)$	&$0.1(5.0)$			&$1.1(5.5)$\\\hline
\end{tabular}
\end{table}
\begin{table}
\centering
\caption{Details of fit parameters, together with variations used in section \ref{stabsec} to check stability. $\Delta T$ indicates the number of data points at the extremities of correlation functions not included in the fit.\label{fitparams} Bold values are those used to produce our final results. $\chi^2/\text{dof}$ is estimated by introducing svd and prior noise as in \cite{corrfitter}. We do not compute $\chi^2$ values including prior and svd noise for those fits with $n_\text{exp}=4$.}
\begin{tabular}{ c c c c c c c c }\hline
Set & $n_\text{exp}$ & $\Delta T_\text{3pt}$ & $\Delta T_\text{2pt}^{J/\psi}$& $\Delta T^{H_c}_\text{2pt}$ & SVD cut & $\chi^2/\text{dof}$ & $\delta$ \\ \hline
1  & \textbf{3} & \textbf{2} & \textbf{5} & \textbf{5} & \textbf{0.025} & \textbf{1.0} & 0\\
  & 3 & 2 & 5 & 5 & 0.05 & 0.99 & 1\\
  & 3 & 3 & 7 & 7 & 0.025 & 0.96& 2 \\
  & 4 & 2 & 5 & 5 & 0.025 & $-$ & 3\\\hline
2  & \textbf{3} & \textbf{3} & \textbf{6} & \textbf{6} & \textbf{0.05} & \textbf{1.0}& 0\\
  & 3 & 3 & 6 & 6 & 0.1 & 0.98& 1\\
  & 3 & 4 & 9 & 9 & 0.05 & 1.0& 2\\
  & 4 & 3 & 6 & 6 & 0.05 & $-$ & 3\\\hline
3  & \textbf{3} & \textbf{2} & \textbf{5} & \textbf{5} & \textbf{0.025} & \textbf{0.93}& 0\\
  & 3 &  2 & 5 & 5 & 0.1 & 0.94& 1\\
  & 3 &  3 & 7 & 7 & 0.025 & 0.97& 2\\
  & 4 &  3 & 6 & 6 & 0.025 & $-$ & 3\\\hline
4  & \textbf{3} &  \textbf{2} & \textbf{5} & \textbf{5} & \textbf{0.075} & \textbf{0.97}& 0\\
  & 3 &  2 & 5 & 5 & 0.1 & 0.97& 1\\
  & 3 &  3 & 7 & 7 & 0.075 & 0.97& 2\\
  & 4 &  2 & 5 & 5 & 0.075 & $-$ & 3\\\hline
\end{tabular}
\end{table}

\begin{table}
\centering
\caption{Lattice form factor results for set 1. $ak$ here is the value of the $x$ and $y$ components of the lattice momentum for the $J/\psi$. $ak$ is calculated from the corresponding twist in Table \ref{twists}.  \label{set1}}
\begin{tabular}{ c c c c c c }
\hline
$am_h$& $ak$ & $A_0$& $A_1$& $A_2$& $V$\\\hline
0.6&	0.0	&$-$	&0.9001(67)	&$-$	&$-$	\\
&	0.0354858	&0.86(12)	&0.8983(69)	&1.9(3.6)	&1.82(31)	\\
&	0.0709715	&0.858(63)	&0.8937(71)	&1.8(1.7)	&1.82(17)	\\
&	0.106457	&0.848(43)	&0.8864(71)	&1.8(1.2)	&1.80(12)	\\
&	0.141943	&0.833(34)	&0.8764(71)	&1.9(1.0)	&1.772(92)	\\
&	0.177429	&0.815(28)	&0.8639(71)	&1.97(97)	&1.734(76)	\\
\hline
0.65&	0.0	&$-$	&0.9028(69)	&$-$	&$-$	\\
&	0.0354858	&0.87(13)	&0.9010(71)	&1.5(3.6)	&1.81(31)	\\
&	0.0709715	&0.865(67)	&0.8965(73)	&1.5(1.3)	&1.81(17)	\\
&	0.106457	&0.855(46)	&0.8893(73)	&1.50(83)	&1.79(12)	\\
&	0.141943	&0.840(35)	&0.8793(73)	&1.52(67)	&1.760(91)	\\
&	0.177429	&0.822(29)	&0.8669(73)	&1.54(61)	&1.723(76)	\\
\hline
0.8&	0.0	&$-$	&0.9109(76)	&$-$	&$-$	\\
&	0.0354858	&0.89(16)	&0.9091(79)	&1.2(6.1)	&1.79(31)	\\
&	0.0709715	&0.891(81)	&0.9045(81)	&1.2(1.6)	&1.79(16)	\\
&	0.106457	&0.881(56)	&0.8974(81)	&1.17(74)	&1.77(12)	\\
&	0.141943	&0.867(43)	&0.8875(81)	&1.16(46)	&1.743(91)	\\
&	0.177429	&0.849(35)	&0.8752(82)	&1.15(34)	&1.708(75)	\\
\hline
\end{tabular}
\end{table}
\begin{table}
\centering
\caption{Lattice form factor results for set 2. $ak$ here is the value of the $x$ and $y$ components of the lattice momentum for the $J/\psi$. $ak$ is calculated from the corresponding twist in Table \ref{twists}.  \label{set2}}
\begin{tabular}{ c c c c c c }
\hline
$am_h$& $ak$ & $A_0$& $A_1$& $A_2$& $V$\\\hline
0.427&	0.0	&$-$	&0.8720(99)	&$-$	&$-$	\\
&	0.0540337	&0.861(94)	&0.864(10)	&1.0(1.4)	&1.73(22)	\\
&	0.108067	&0.825(49)	&0.8412(99)	&1.00(70)	&1.66(12)	\\
&	0.162101	&0.771(35)	&0.8056(99)	&1.01(60)	&1.562(88)	\\
&	0.216135	&0.705(28)	&0.760(11)	&1.02(66)	&1.436(72)	\\
&	0.270169	&0.633(25)	&0.709(12)	&1.10(93)	&1.297(66)	\\
\hline
0.525&	0.0	&$-$	&0.8725(98)	&$-$	&$-$	\\
&	0.0540337	&0.88(10)	&0.8647(98)	&0.9(1.7)	&1.69(22)	\\
&	0.108067	&0.848(52)	&0.8424(98)	&0.91(47)	&1.63(12)	\\
&	0.162101	&0.795(37)	&0.8076(98)	&0.88(29)	&1.534(87)	\\
&	0.216135	&0.729(30)	&0.763(10)	&0.85(27)	&1.414(71)	\\
&	0.270169	&0.657(27)	&0.713(12)	&0.81(31)	&1.281(66)	\\
\hline
0.65&	0.0	&$-$	&0.8724(98)	&$-$	&$-$	\\
&	0.0540337	&0.92(11)	&0.8648(99)	&0.8(2.3)	&1.67(22)	\\
&	0.108067	&0.882(57)	&0.8430(98)	&0.87(55)	&1.61(12)	\\
&	0.162101	&0.829(40)	&0.8090(99)	&0.84(25)	&1.519(87)	\\
&	0.216135	&0.763(32)	&0.766(11)	&0.80(17)	&1.404(72)	\\
&	0.270169	&0.690(29)	&0.716(12)	&0.75(17)	&1.275(67)	\\
\hline
0.8&	0.0	&$-$	&0.874(10)	&$-$	&$-$	\\
&	0.0540337	&0.96(12)	&0.866(10)	&0.8(3.0)	&1.67(22)	\\
&	0.108067	&0.928(65)	&0.845(10)	&0.86(72)	&1.61(12)	\\
&	0.162101	&0.874(45)	&0.811(10)	&0.83(31)	&1.522(89)	\\
&	0.216135	&0.807(36)	&0.769(11)	&0.79(18)	&1.410(74)	\\
&	0.270169	&0.733(33)	&0.720(13)	&0.73(14)	&1.284(69)	\\
\hline
\end{tabular}
\end{table}
\begin{table}
\centering
\caption{Lattice form factor results for set 3. $ak$ here is the value of the $x$ and $y$ components of the lattice momentum for the $J/\psi$. $ak$ is calculated from the corresponding twist in Table \ref{twists}.  \label{set3}}
\begin{tabular}{ c c c c c c }
\hline
$am_h$& $ak$ & $A_0$& $A_1$& $A_2$& $V$\\\hline
0.5&	0.0	&$-$	&0.8573(79)	&$-$	&$-$	\\
&	0.0609372	&0.915(25)	&0.8391(79)	&0.69(77)	&1.638(48)	\\
&	0.121874	&0.841(16)	&0.7913(81)	&0.75(17)	&1.499(31)	\\
&	0.182812	&0.735(16)	&0.7244(95)	&0.700(89)	&1.318(27)	\\
&	0.243749	&0.626(16)	&0.650(13)	&0.61(10)	&1.123(31)	\\
&	0.304686	&0.521(20)	&0.567(30)	&0.49(21)	&0.931(55)	\\
\hline
0.65&	0.0	&$-$	&0.8515(82)	&$-$	&$-$	\\
&	0.0609372	&0.976(29)	&0.8338(82)	&0.7(1.1)	&1.630(50)	\\
&	0.121874	&0.899(19)	&0.7874(84)	&0.75(25)	&1.495(32)	\\
&	0.182812	&0.789(19)	&0.7222(98)	&0.71(11)	&1.320(28)	\\
&	0.243749	&0.676(19)	&0.650(14)	&0.623(80)	&1.129(33)	\\
&	0.304686	&0.565(23)	&0.567(30)	&0.51(13)	&0.941(58)	\\
\hline
0.8&	0.0	&$-$	&0.8475(82)	&$-$	&$-$	\\
&	0.0609372	&1.040(33)	&0.8303(82)	&0.6(1.3)	&1.644(53)	\\
&	0.121874	&0.960(21)	&0.7848(84)	&0.76(31)	&1.511(34)	\\
&	0.182812	&0.846(21)	&0.7209(99)	&0.73(14)	&1.338(30)	\\
&	0.243749	&0.728(21)	&0.649(14)	&0.638(89)	&1.149(35)	\\
&	0.304686	&0.610(25)	&0.568(31)	&0.52(12)	&0.960(62)	\\
\hline
\end{tabular}
\end{table}
\begin{table}
\centering
\caption{Lattice form factor results for set 4. $ak$ here is the value of the $x$ and $y$ components of the lattice momentum for the $J/\psi$. $ak$ is calculated from the corresponding twist in Table \ref{twists}.  \label{set4}}
\begin{tabular}{ c c c c c c }
\hline
$am_h$& $ak$ & $A_0$& $A_1$& $A_2$& $V$\\\hline
0.5&	0.0	&$-$	&0.8928(86)	&$-$	&$-$	\\
&	0.0332236	&0.86(18)	&0.8911(90)	&5(24)	&1.88(46)	\\
&	0.0664472	&0.850(92)	&0.8865(94)	&6(14)	&1.86(24)	\\
&	0.0996708	&0.839(63)	&0.8796(95)	&7(12)	&1.84(17)	\\
&	0.132894	&0.825(48)	&0.8701(96)	&11(16)	&1.80(13)	\\
&	0.166118	&0.807(39)	&0.8582(98)	&52(66)	&1.77(11)	\\
\hline
0.65&	0.0	&$-$	&0.8995(91)	&$-$	&$-$	\\
&	0.0332236	&0.87(21)	&0.8977(95)	&1.2(5.6)	&1.82(44)	\\
&	0.0664472	&0.87(11)	&0.893(10)	&1.3(1.9)	&1.81(23)	\\
&	0.0996708	&0.858(72)	&0.887(10)	&1.4(1.1)	&1.78(16)	\\
&	0.132894	&0.844(55)	&0.877(10)	&1.37(88)	&1.76(12)	\\
&	0.166118	&0.827(45)	&0.866(10)	&1.38(77)	&1.72(10)	\\
\hline
0.8&	0.0	&$-$	&0.907(10)	&$-$	&$-$	\\
&	0.0332236	&0.90(26)	&0.905(11)	&0.9(9.4)	&1.80(44)	\\
&	0.0664472	&0.90(13)	&0.901(11)	&1.1(2.5)	&1.79(23)	\\
&	0.0996708	&0.886(87)	&0.894(11)	&1.1(1.2)	&1.77(16)	\\
&	0.132894	&0.872(67)	&0.885(11)	&1.10(70)	&1.74(12)	\\
&	0.166118	&0.855(54)	&0.874(12)	&1.10(51)	&1.71(10)	\\
\hline
\end{tabular}
\end{table}
\begin{table}
\centering
\caption{$J/\psi$ masses for the local spin-taste operator $\gamma_1\otimes \gamma_1$ and $1-$link operators $\gamma_1\otimes 1$ and $\gamma_1\otimes \gamma_1\gamma_2$ used in our calculation, see table \ref{spintastetable}. \label{charmMasses}}
\begin{tabular}{ c c c c c }
\hline
& & $aM_{J/\psi}$ & \\\hline
Set & $\gamma_1\otimes \gamma_1$&  $\gamma_1\otimes 1$ & $\gamma_1\otimes \gamma_1\gamma_2$\\\hline
1&1.41391(20)	&1.41432(27)	&1.41410(26)	\\
\hline
2&0.92990(24)	&0.92997(31)	&0.92990(32)	\\
\hline
3&0.69210(16)	&0.69207(24)	&0.69209(24)	\\
\hline
4&1.37833(24)	&1.37867(32)	&1.37844(32)	\\
\hline
\end{tabular}
\end{table}
\begin{table}
\centering
\caption{$\eta_h$ masses and $H_c$ masses for the local spin-taste operators $\gamma_5\otimes \gamma_5$ and $\gamma_0\gamma_5\otimes \gamma_0\gamma_5$ that we use in our calculation, see table \ref{spintastetable}. \label{Massesheavy}}
\begin{tabular}{ c c c c c }
\hline
Set & $am_h$ & $aM_{H_c}(\gamma_5\otimes \gamma_5)$& $aM_{H_c}(\gamma_0\gamma_5\otimes \gamma_0\gamma_5)$& $aM_{\eta_h}$\\\hline
1&	0.6	&1.52160(16)	&1.52326(14)	&1.67559(14)	\\
&	0.65	&1.57224(16)	&1.57396(14)	&1.77510(14)	\\
&	0.8	&1.72015(17)	&1.72208(14)	&2.06407(15)	\\
\hline
2&	0.427	&1.06720(16)	&1.06747(17)	&1.23354(15)	\\
&	0.525	&1.17255(15)	&1.17283(17)	&1.43949(13)	\\
&	0.65	&1.30314(15)	&1.30344(17)	&1.69388(12)	\\
&	0.8	&1.45421(15)	&1.45454(17)	&1.98757(11)	\\
\hline
3&	0.5	&1.01184(10)	&1.01197(11)	&1.342912(64)	\\
&	0.65	&1.16995(10)	&1.17008(11)	&1.650408(56)	\\
&	0.8	&1.32185(10)	&1.32199(11)	&1.945914(50)	\\
\hline
4&	0.5	&1.40004(20)	&1.40155(18)	&1.47014(18)	\\
&	0.65	&1.55423(19)	&1.55588(17)	&1.77380(17)	\\
&	0.8	&1.70243(21)	&1.70428(16)	&2.06296(17)	\\
\hline
\end{tabular}
\end{table}

As discussed in section~\ref{sec:calcdetails} we isolate form factors from the matrix elements extracted from correlation functions and combine these with current renormalisation factors to obtain $A_0$, $A_1$, $A_2$ and $V$ across a range 
of $q^2$ for several different heavy quark masses. 
The form factor results are given in 
Tables~\ref{set1},~\ref{set2},~\ref{set3} 
and~\ref{set4} and plotted in Figure~\ref{FFqsq}. 
We see that the least well-determined form factor 
is $A_2$, especially close to zero recoil. This can be 
traced back to the division by $k^2$ needed to determine it after the subtraction of the $\mathcal{O}(1)$ $A_1$ contribution in Eq.~(\ref{A2subbit}). In Figure~\ref{ksqA2} we plot ${A_2(q^2)|\vec{p^\prime}|^2}/{M_{J/\psi}^2}$, which is the combination of $A_2$ and $|\vec{p^\prime}|^2$ entering the helicity amplitudes in Eq.~(\ref{helicityamplitudes}). We see that this has much clearer behaviour as a function of $q^2$.

We also give values for the meson masses in tables~\ref{charmMasses} 
 for the $J/\psi$ and table~\ref{Massesheavy} for the $H_c$\footnote{Note that there is a systematic discrepancy between our values for $aM_{H_c}(\gamma_5\otimes \gamma_5)$ on set 1, in table~\ref{Massesheavy}, and those in~\cite{EuanBsDsstar}. We found that increasing the statistical uncertainties on our masses on set 1 by a factor of 10 to cover this discrepancy did not change our final results significantly.} and 
$\eta_h$. From table~\ref{spintastetable} we have 3 different $J/\psi$ tastes, 
using one local and two different 1-link point-split operators. The difference in 
mass of these different taste $J/\psi$ mesons is tiny, barely visible above statistical uncertainties (at a level of 1 MeV) even on our coarsest 
lattices. For the $H_c$ we use two spin-taste operators, both local. As expected 
for pseudoscalars the taste-differences are slightly larger than for the vector 
$J/\psi$ and they are visible above the smaller statistical errors in this case. 
The mass differences are still only a few MeV, however, for a meson of a mass 
of several GeV, and they fall on the finer lattices since they are a discretisation 
effect. 

We calculate the value of $q^2$ for each form-factor based on the meson masses 
of the corresponding meson tastes, along with the lattice momentum of 
the $J/\psi$ determined from the twists in Table~\ref{twists}. 

\subsection{Extrapolation to the Physical Point}
\label{sec:extrap}

\begin{table}
\centering
\caption{\label{physmasses}Values used in our fits for the physical masses 
of relevant mesons, in GeV. These are from the Particle Data 
Group~\cite{Tanabashi:2018oca} 
except for the unphysical $\eta_s$ which we take from lattice 
calculations of the mass of the pion and kaon~\cite{Dowdall:2013rya}. The 
$\eta_s$ mass is used to set mass mistuning terms in our 
fit and so include an uncertainty. The other masses are used as kinematic 
parameters in setting up our fit in $z$-space and used without uncertainties. 
The $J/\psi$ mass is also used to tune the $c$ quark mass, and there 
we do include its (negligible) uncertainty. }
\begin{tabular}{ c | c }\hline
meson & $M^\text{phys}$[GeV]\\\hline
${\eta_b}$&9.3889\\
${B_c}$   & 6.2749\\
${B}$     &5.27964 \\
${J/\psi}$& 3.0969\\
${D^*}$   &2.010 \\
${\eta_s}$   &0.6885(22)\\\hline
\end{tabular}
\end{table}

\begin{table}
\centering
\caption{\label{poletab} Expected $B_c$ pseudoscalar, 
vector and axial vector 
masses below $BD^*$ threshold that we use in our pole factor, Eq.~(\ref{poleformeq}). 
Pseudoscalar values for the ground-state and first radial excitation 
are taken from experiment~\cite{Aaij:2016qlz, Sirunyan:2019osb, Aaij:2019ldo, Tanabashi:2018oca}; the other values are taken from~\cite{Harrison:2017fmw} and are derived from 
lattice QCD calculations~\cite{Dowdall:2012ab} and model 
estimates~\cite{Eichten:1994gt,Godfrey:2004ya,Devlani:2014nda}.
}
\begin{tabular}{ c c c }\hline
$0^-/$GeV & ${1^-}/$GeV & ${1^+}/$GeV \\\hline
6.275 & 6.335 & 6.745\\
6.872 & 6.926 & 6.75\\
7.25 & 7.02 & 7.15\\
& 7.28 & 7.15\\\hline
\end{tabular}
\end{table}

\begin{figure*}
\centering
\includegraphics[scale=0.225]{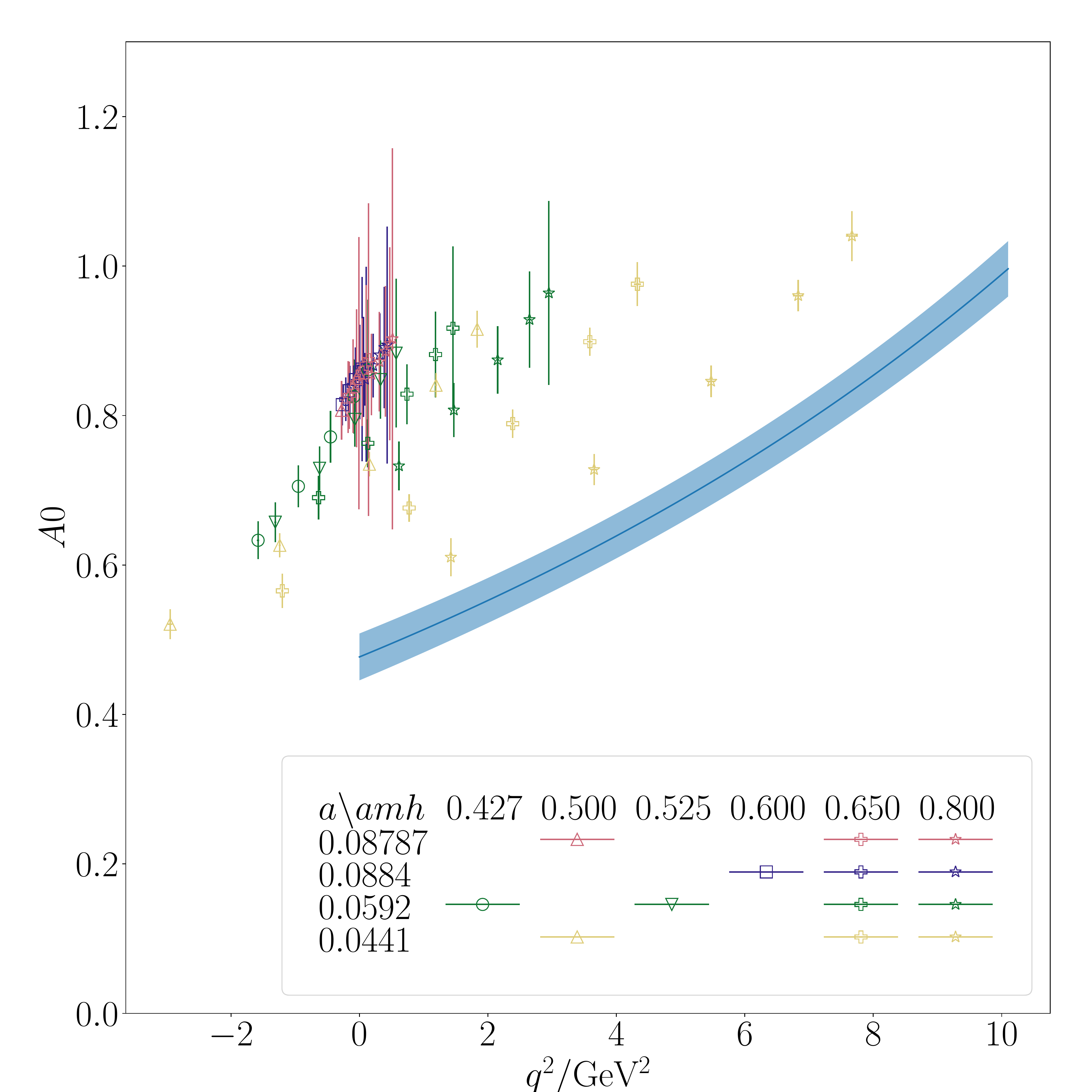}
\includegraphics[scale=0.225]{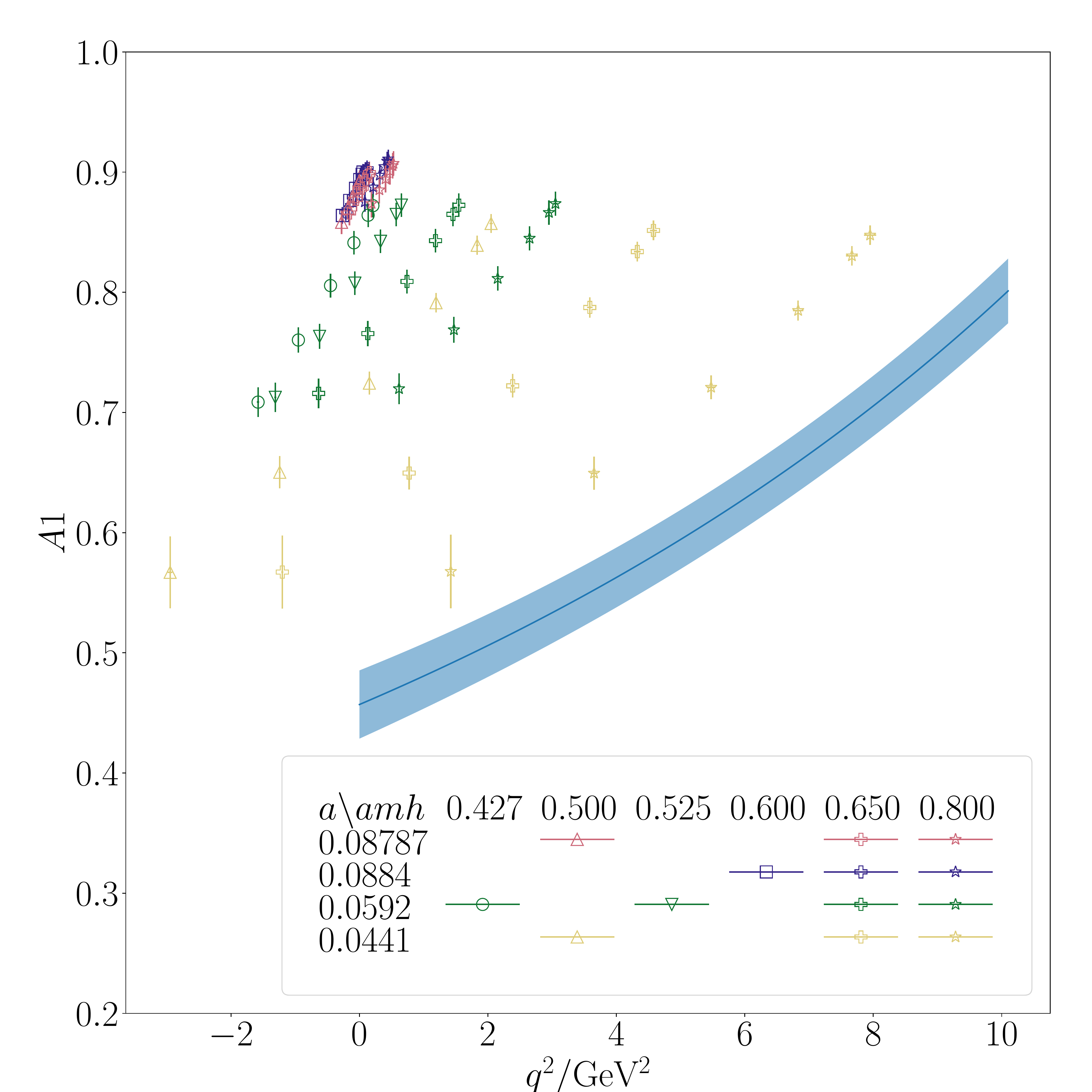}
\includegraphics[scale=0.225]{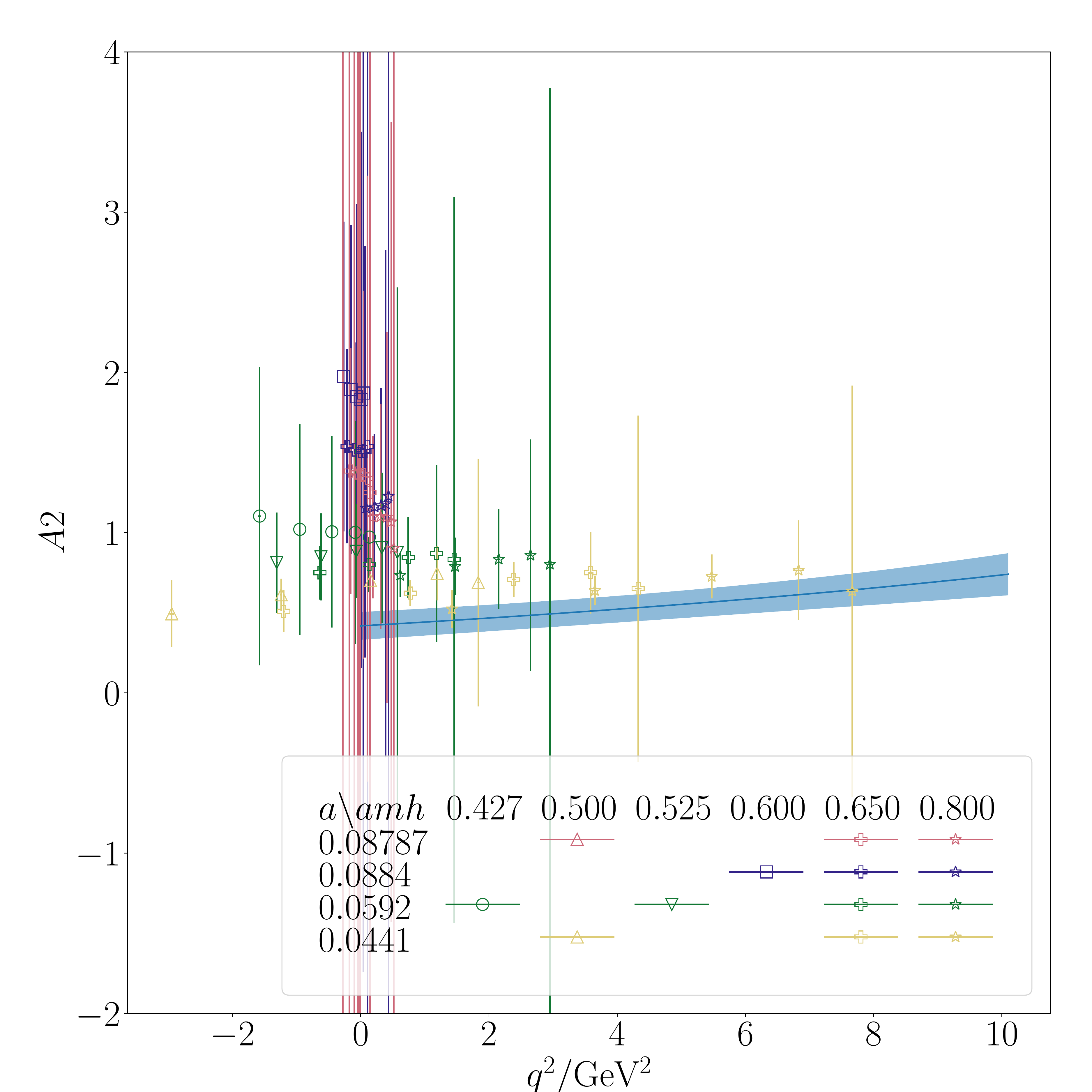}
\includegraphics[scale=0.225]{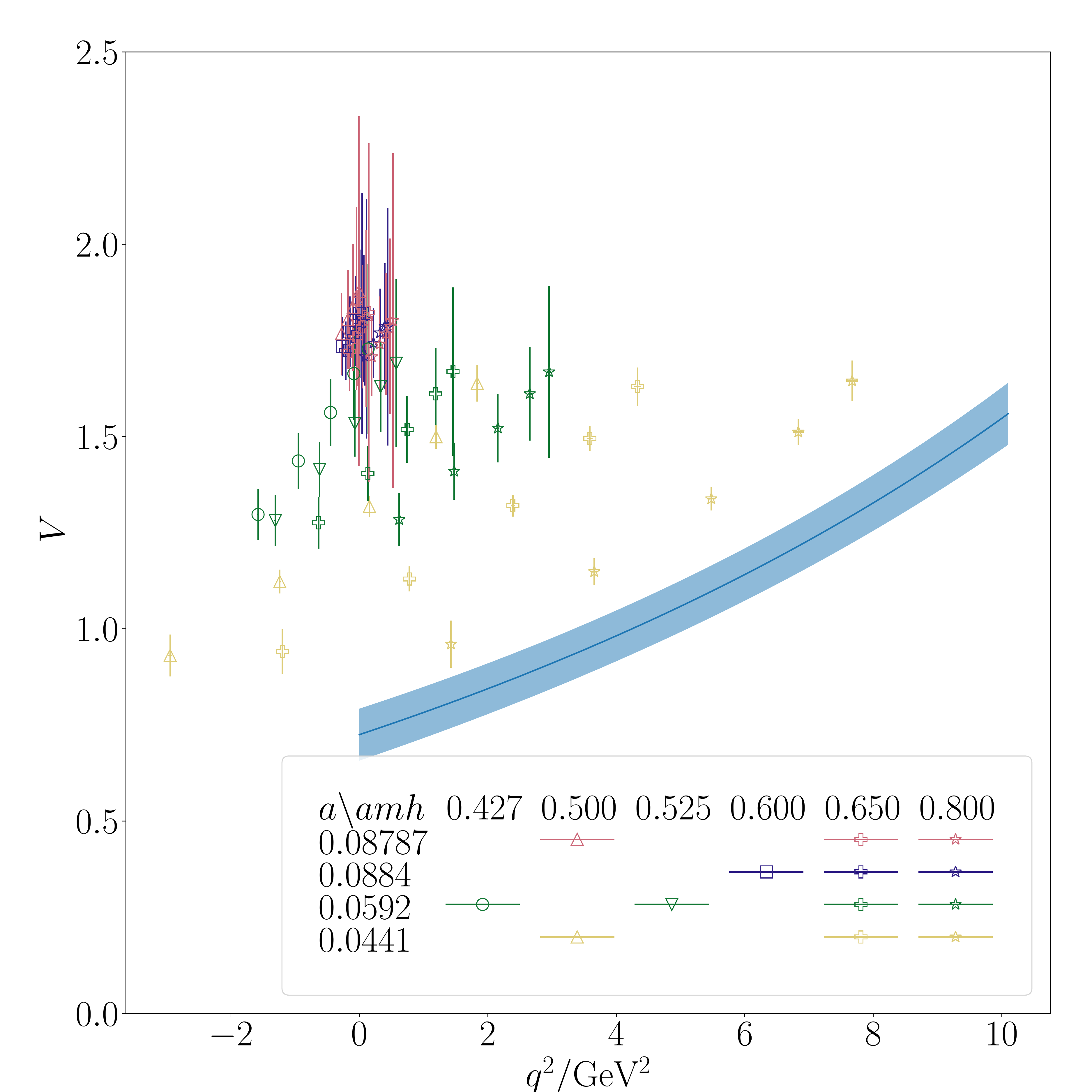}
\caption{\label{FFqsq} The points show our lattice QCD results for each 
form factor as given in Tables~\ref{set1},~\ref{set2},~\ref{set3} 
and~\ref{set4} as a function of squared 4-momentum transfer, $q^2$. 
The legend gives the mapping between symbol colour and shape and the 
set of gluon field configurations used, as given by the lattice 
spacing, and the heavy quark in lattice units. 
The blue curve with error band is the result of our fit in lattice 
spacing and heavy quark mass, evaluating the result in the continuum 
limit and for the $b$ quark mass, to give the physical form factor for $B_c \rightarrow J/\psi$. 
}
\end{figure*}

\begin{figure}
\centering
\includegraphics[scale=0.225]{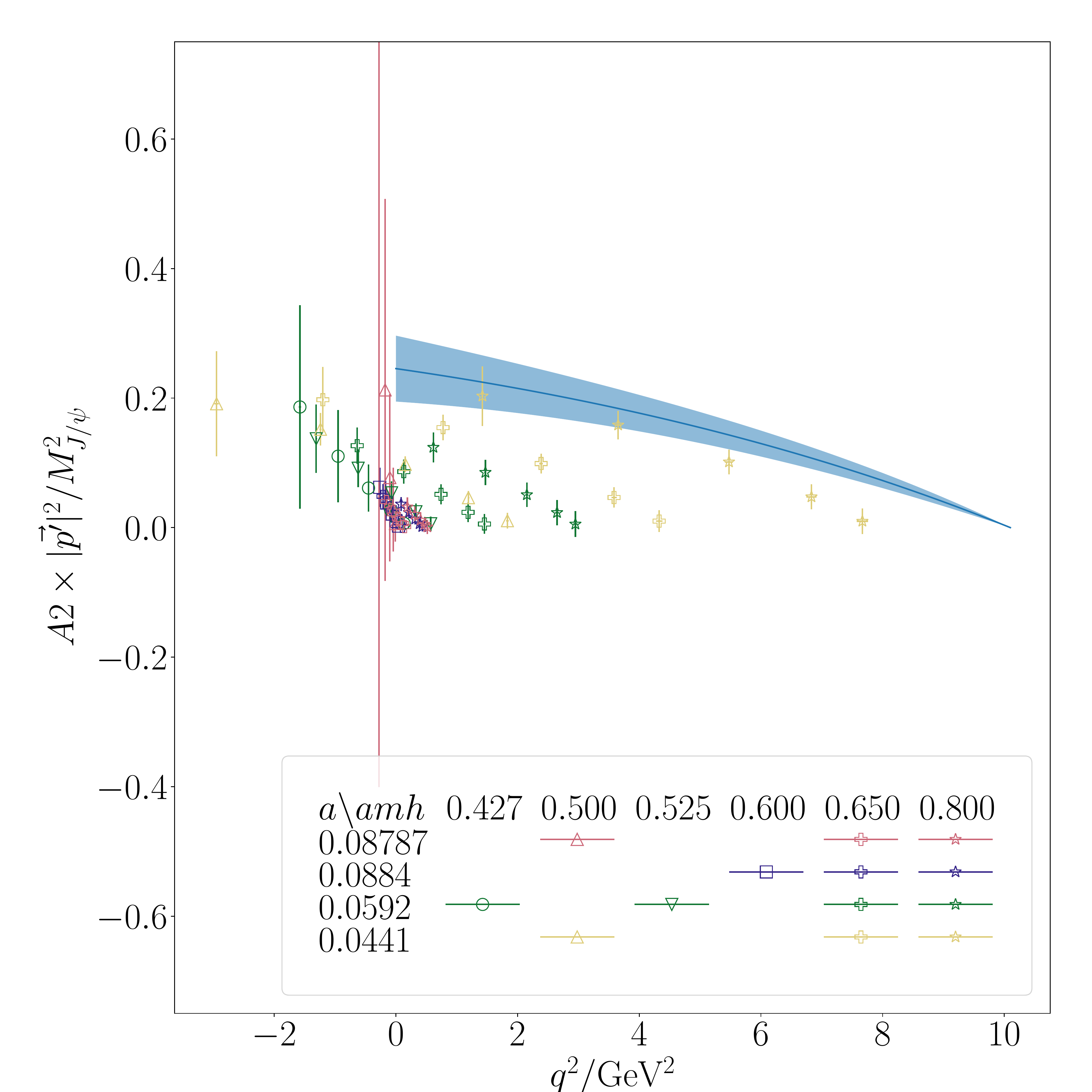}
\caption{\label{ksqA2}Plot showing ${A_2(q^2)|\vec{p^\prime}|^2}/{M_{J/\psi}^2}$ against $q^2$. The combination of $A_2$ and $|\vec{p^\prime}|^2$ entering the helicity amplitudes in Eq.~(\ref{helicityamplitudes}). This combination does not exhibit as much noise as $A_2$ alone, seen in Figure~\ref{FFqsq}.}
\end{figure}

In order to extract physical continuum form factors we must fit our lattice 
results to a fit function including discretisation effects 
arising from the use of large values of $am_h$ as well as the 
physical dependence upon $M_{H_c}$ and $q^2$. In order to fit 
the $q^2$ dependence we make use of the $z$-expansion 
(see, for example,~\cite{Boyd:1997kz, Caprini:1997mu, poleform}) which has 
become standard for the description of form factors. 
We map the physical $q^2$ range to within the unit circle via the change of variables
\begin{equation}\label{zdefinition}
z(q^2,t_0) = \frac{\sqrt{t_+-q^2} - \sqrt{t_+-t_0}}{\sqrt{t_+-q^2} + \sqrt{t_+-t_0}}.
\end{equation}
The physical $q^2$ range is from 0 to $t_-$ where
\begin{equation}
t_- = (M_{H_c}- M_{J/\psi})^2 .
\end{equation}
In Eq.~(\ref{zdefinition}) $t_+$ is the pair production threshold for a $b\overline{c}$ current, 
and the start of a cut in the complex $q^2$ plane. This point is mapped to 
$z=-1$.  
We take 
\begin{equation}
t_+ = (M_{H} + M_{D^*})^2.
\end{equation}
where $H$ is an $h\bar{u}$ pseudoscalar meson. Note there is a sizeable gap between $t_+^{1/2}$ and $t_-^{1/2}$ and it is 
independent of $m_h$ up to binding energy effects.
In this calculation we do not compute $M_H$ entering $t_+$. 
Instead we use an estimate of it derived from the $M_{H_c}$ values that 
we have: $M^\text{latt}_H = M_{H_c} - (M^\text{phys}_{B_c} - M^\text{phys}_B)$. 
This ensures that when we evaluate our fit function at the physical heavy 
quark mass then the correct physical continuum value of $t_+$ is 
recovered. We take the $D^*$ mass from experiment since our valence 
$c$ masses are tuned to the physical value throughout and the mistuning 
of the light quark masses in the sea is taken care of in our 
fit function. 

Table~\ref{physmasses} lists the physical masses we use for the 
$B_c$, $B$ and $D^*$. These are used as numerical constants with no 
uncertainty to simplify the necessary covariances required to reconstruct our 
fit function. Their uncertainties are negligibly small in any case.   

The choice of $t_0$  in Eq.~(\ref{zdefinition}) determines the 
value of $q^2$ at which $z=0$. Here we choose $t_0=t_-$. This 
choice means that $z$ is simply related to the variable $w$ 
which is the dot product of the 4-velocities of the $B_c$ 
and $J/\psi$ and known to be a good 
variable for a description of $b \rightarrow c$ form factors 
in the Heavy Quark Effective Theory approach~\cite{Boyd:1997kz, Caprini:1997mu} 
to $B \rightarrow D^*$ decays. We expect that still to be a useful 
consideration here because the $B_c$ meson has similar properties to 
that of a heavy-light meson, interpolating between heavy-light 
and heavyonium~\cite{McNeile:2012qf}. Heavyonium mesons also display 
similar mass-suppressed differences in amplitudes between ground-state 
vector and pseudoscalar mesons, so that we expect a $B_c \rightarrow J/\psi$ decay 
to behave rather similarly to $B \rightarrow D^*$.   

Physical particles with $\overline{b}c$ quark content, masses between 
$t_+^{1/2}$ and $t_-^{1/2}$ (i.e. above the physical 
$q^2$ range for the decay) and the appropriate quantum numbers to 
couple to the current operator result in the appearance of a sub-threshold 
pole in the corresponding form factor. Following~\cite{Boyd:1997kz, Caprini:1997mu} we 
include these poles in our fit using the form (Blaschke factor)
\begin{equation}\label{poleformeq}
P(q^2) = \prod_{M_\text{pole}} z(q^2,M_{\text{pole}}^2) .
\end{equation}
$P(q^2)$ is zero when $q^2=M_{\text{pole}}^2$ and has modulus 
1  along the cut in the $q^2$ plane for real $q^2 > t_+$ where 
multi-particle production can occur. 
For the pole masses corresponding to $B_c$ states we use the 
values listed in~\cite{Harrison:2017fmw}, which we reproduce 
here in table~\ref{poletab}. We use 4 masses in each of the vector and 
axial channels and 3 in the pseudoscalar channel. 
These values are derived from experiment and from lattice 
QCD calculations for the low-lying $B_c$ masses and from models for the 
higher-lying states. Since these masses are again simply numbers used 
in setting up our fit in $z$-space the values are used without uncertainties. 
As we will show in Section~\ref{stabsec} our results are not sensitive to the 
exact form of the pole expression above (Eq.~(\ref{poleformeq})).   
The pole masses are shifted from those of the $B_c$ 
 to account for our use of unphysical 
heavy-quark masses. We use 
$M_\text{pole} = M_{H_c}+M_\text{pole}^\text{phys}-M_{B_c}^\text{phys}$ 
again ensuring that when we extrapolate to the physical heavy quark 
mass the physical continuum values of Table~\ref{poletab} are recovered.

\begin{figure*}
\centering
\includegraphics[scale=0.225]{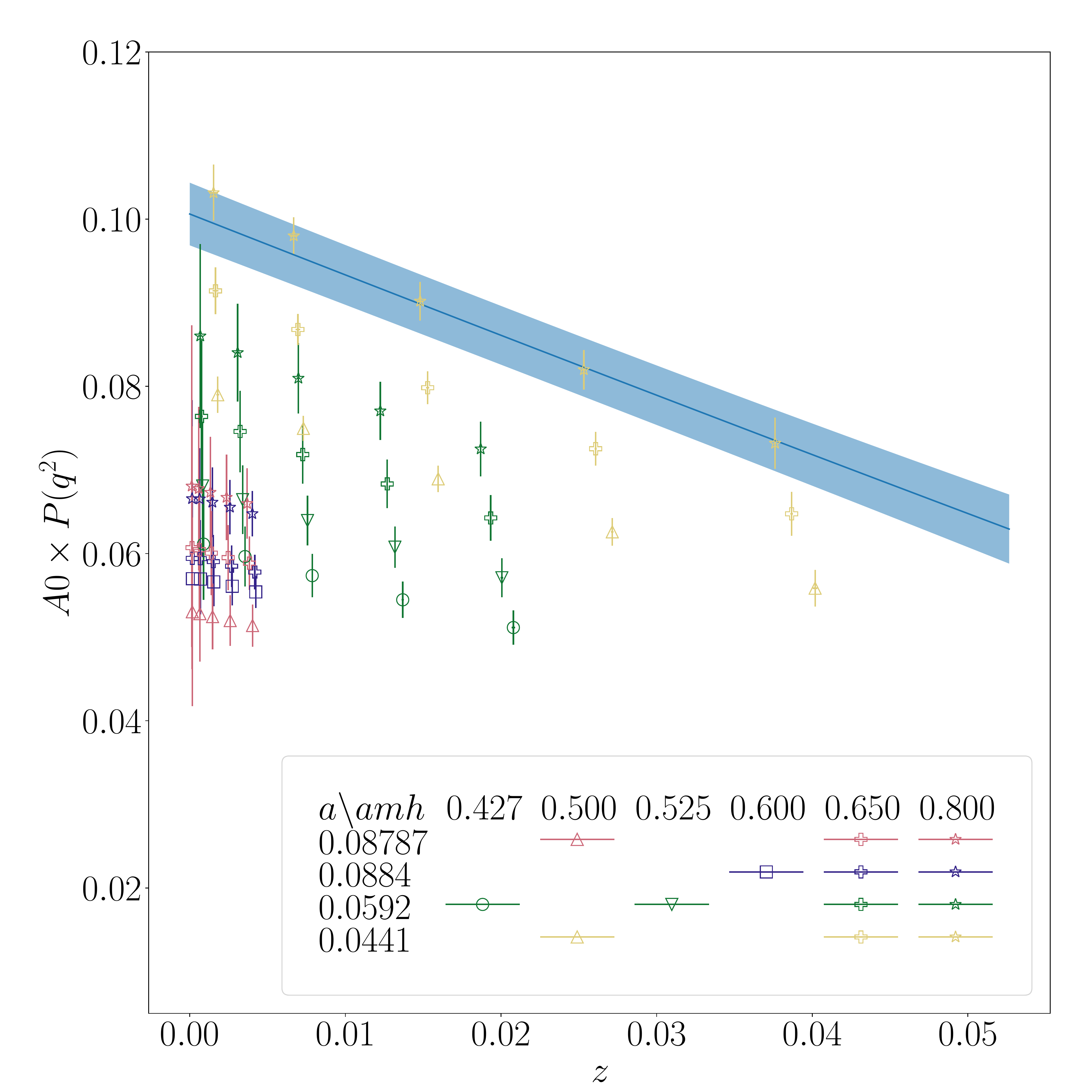}
\includegraphics[scale=0.225]{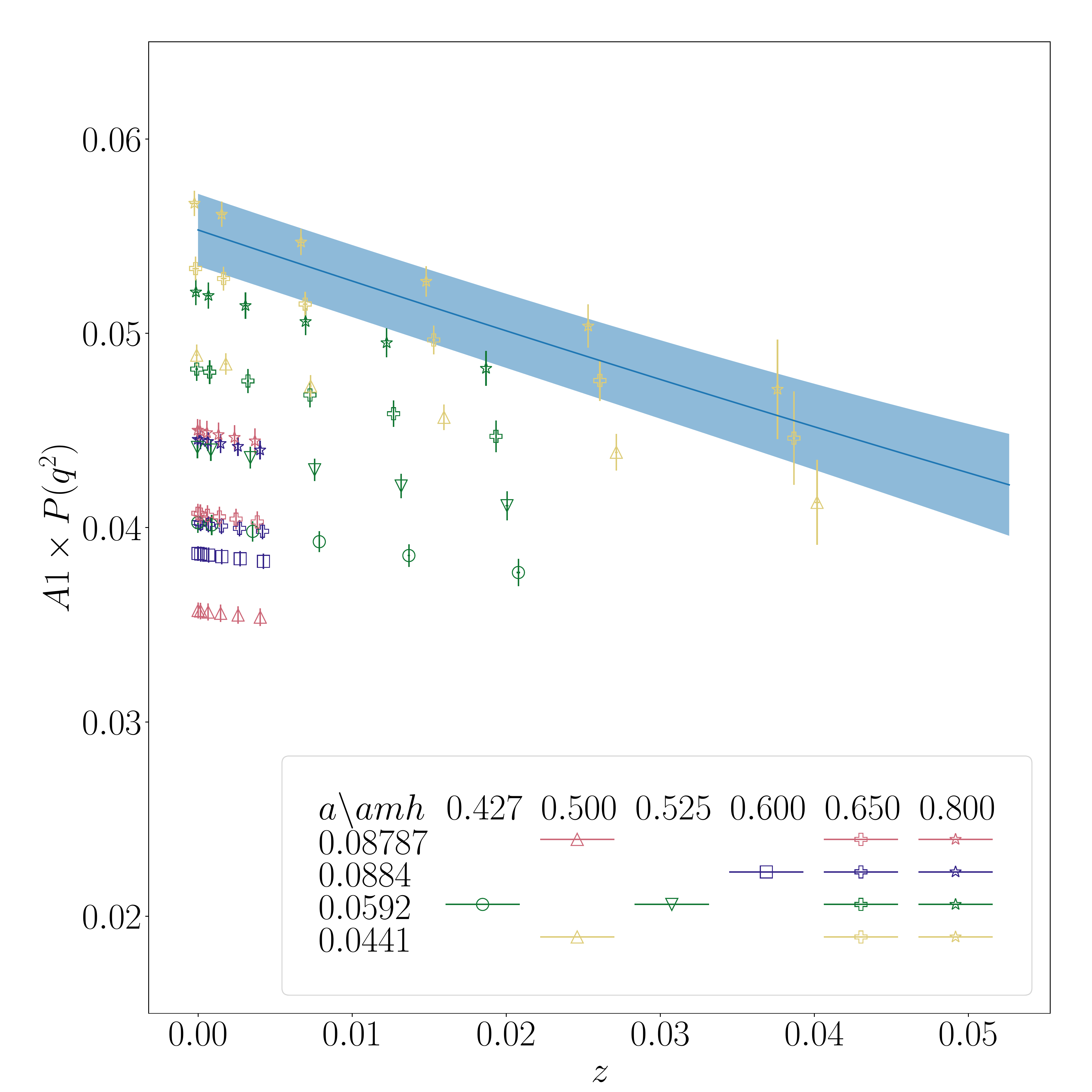}
\includegraphics[scale=0.225]{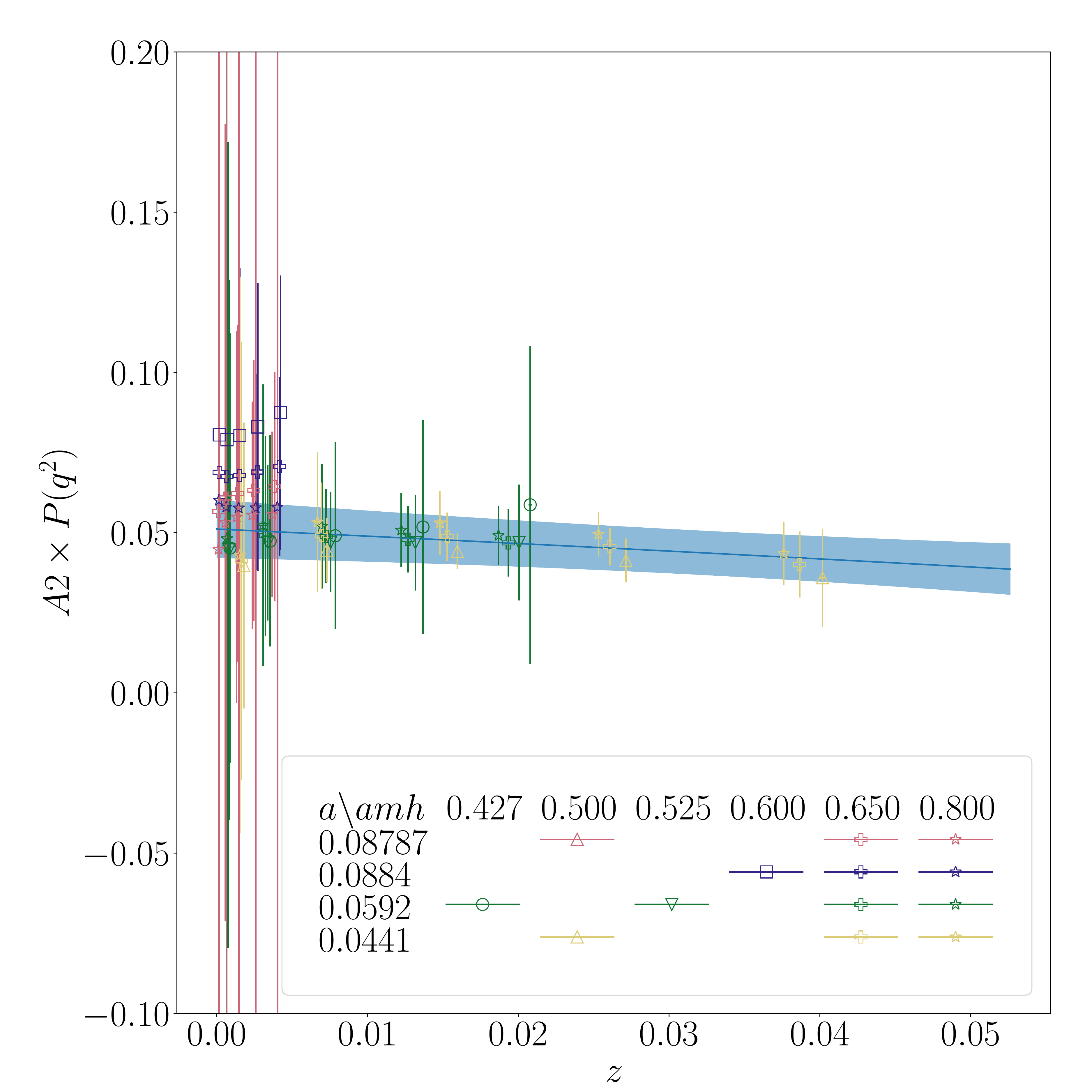}
\includegraphics[scale=0.225]{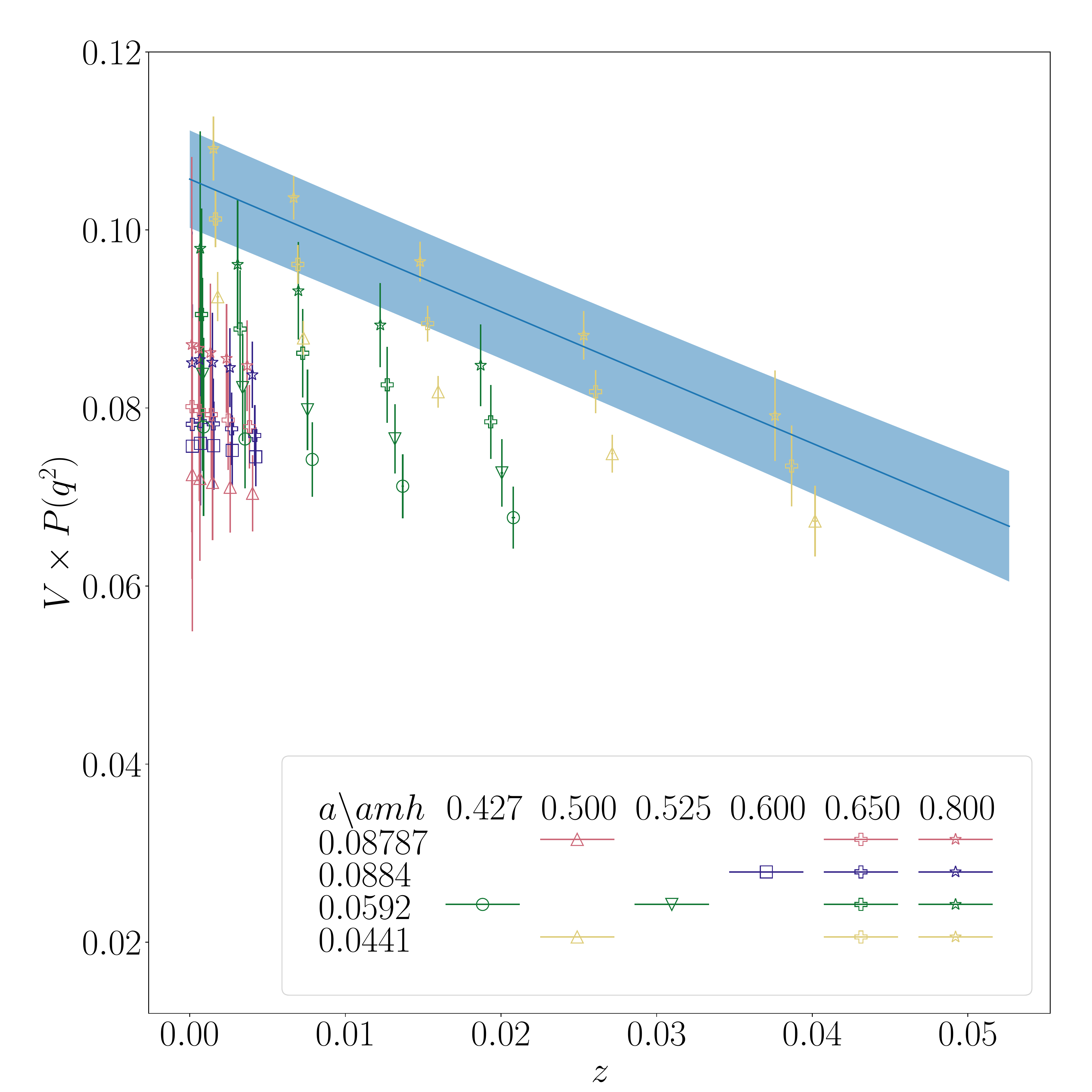}
\caption{\label{zspace} The points show our lattice QCD results for each 
form factor as given in Tables~\ref{set1},~\ref{set2},~\ref{set3} 
and~\ref{set4} multiplied by the pole function of Eq.~(\ref{poleformeq}) and plotted 
in $z$-space. 
The legend gives the mapping between symbol colour and shape and the 
set of gluon field configurations used, as given by the lattice 
spacing, and the heavy quark in lattice units. 
The blue curve with error band is the result of our polynomial fit in $z$
with lattice 
spacing and heavy quark mass dependence (Eq.~(\ref{fitfunctionequation})), evaluating the result in the continuum 
limit and for the $b$ quark mass, to give the physical form factor for $B_c \rightarrow J/\psi$. 
}
\end{figure*}

We fit each form factor, $F(q^2)$, to the fit function 
\begin{equation}
F(q^2) = \frac{1}{P(q^2)}\sum_{n=0}^3 a_n z^n \mathcal{N}_n \label{fitfunctionequation}
\end{equation}
where $P(q^2)$ is the appropriate pole form for that form 
factor (constructed using $1^-$ states for $V(q^2)$, $1^+$ states for $A_1(q^2)$ and $A_2(q^2)$ or $0^-$ states for $A_0(q^2)$) 
as in Eq.~(\ref{poleformeq}). 
The remainder of the fit function is a polynomial in $z$ with 
separate coefficients, $a_n$, for each form factor that take the form
\begin{equation}
\label{eq:anfitform}
a_n = \sum_{j,k,l=0}^3 b_n^{jkl}\Delta_{h}^{(j)} \left(\frac{am_c^\text{val}}{\pi}\right)^{2k} \left(\frac{am_h^\text{val}}{\pi}\right)^{2l}.
\end{equation}
The $\Delta_{h}^{(j)}$ allow for the dependence on the heavy quark 
mass using the $\eta_h$ mass as a physical proxy for this. 
We have $\Delta_{h}^{(0)}=1$ and
\begin{equation}
\Delta_{h}^{(j\neq 0)}=\left(\frac{2\Lambda}{M_{\eta_h}}\right)^j-\left(\frac{2\Lambda}{M_{\eta_b}^\text{phys}}\right)^j.
\end{equation}
We take $\Lambda$ = 0.5 GeV. 

The polynomials in $am_c/\pi$ and $am_h/\pi$ in Eq.~(\ref{eq:anfitform}) allow 
for discretisation effects that are either constant with the heavy quark 
mass (for example coming from the $c$-quark action) or that come from the 
heavy-quark action and so depend on $am_h$. 
Because of the form of the HISQ action only even powers of $a$ can appear 
here. 

The remainder of Eq.~(\ref{fitfunctionequation}), $\mathcal{N}_n$, takes into account 
the effect of mistuning the valence and sea quark masses for each form factor. 
\begin{equation}
\mathcal{N}_n = 1 + A_n \delta_{m_c}^\text{val}+ B_n \delta_{m_c}^\text{sea}+ C_n \delta_{m_s}^\text{sea} + D_n \delta_{m_l}^\text{sea}
\end{equation}
with
\begin{align}
\delta_{m_c}^\text{val} = (am_c^\text{val}-am_c^\text{tuned})/am_c^\text{tuned},\nonumber\\
\delta_{m_c}^\text{sea} = (am_{c}^\text{sea}-am_c^\text{tuned})/am_c^\text{tuned},\nonumber\\
\delta_{m_{s(l)}}^\text{sea} = (am_{s(l)}^\text{sea} - am_{s(l)}^\text{tuned})/(10am_{s}^\text{tuned}).\label{deltatermseq}
\end{align}

We use the $J/\psi$ meson mass to tune the $c$ quark 
mass for the reasons discussed in~\cite{Hatton:2020qhk}.  
We take
\begin{equation}
am_{c}^\text{tuned}=am_{c}^\text{val}\left(\frac{M_{J/\psi}^\text{phys}}{M_{J/\psi}}\right).
\end{equation}
We use the $\eta_s$ mass~\cite{Dowdall:2013rya} to determine 
the mistuning of the $s$ quark mass
in the sea, taking
\begin{equation}
am_{s}^\text{tuned}=am_{s}^\text{val}\left(\frac{M_{\eta_s}^\text{phys}}{M_{\eta_s}}\right)^2 .
\end{equation}
We use the $\eta_s$ masses in lattice units from~\cite{EuanBsDsstar}, together with the corresponding values of $am_{s}^\text{val}$, to do this. 
The values of the physical $J/\psi$ and $\eta_s$ masses that we use are given, 
with their uncertainties, in Table~\ref{physmasses}.
To determine the mistuning of the $u/d = l$ quark mass in the sea
we take
\begin{equation}
am_{l}^\text{tuned}=am_{s}^\text{tuned}/\Delta{[m_s/m_l]},
\end{equation}
with $\Delta{[m_s/m_l]} = 27.18(10)$ from~\cite{Bazavov:2017lyh}. 

We take priors of $0(1)$ for each $b_n$ for each form factor, 
multiplying terms of order $\mathcal{O}(a^2)$ by $0.5$ because $a^2$ errors are 
removed in the HISQ action at tree-level~\cite{PhysRevD.75.054502}.
We use priors of $0.0(0.1)$ for $B_n$ for each form-factor, following the results of the analysis 
of $m_c^\text{sea}$ effects on $w_0$ in \cite{PhysRevD.91.054508}, and priors of $0.0(0.5)$ 
for $C_n$ and $D_n$ for each form-factor since sensitivity to $u/d$ and $s$ sea 
quark masses enter only at 1-loop. All remaining priors are taken as $0(1)$. 
We have checked that the prior width is conservative using the empirical 
Bayes criterion~\cite{Lepage:2001ym}. 

In doing our fit to Eq.~(\ref{fitfunctionequation}) we impose the kinematical 
constraint 
\begin{equation}
\label{eq:contconstraint}
2M_{J/\psi}A_0(0) = (M_{J/\psi}+M_{H_c})A_1(0) - (M_{J/\psi}-M_{H_c})A_2(0).
\end{equation}
This condition holds in the continuum limit with physical sea quark 
masses for each heavy quark mass. 
We impose it using our lattice meson masses at each value of $am_h$ 
and allowing for discretisation and quark mass 
mistuning effects effects in this implementation. 
We do this by imposing 
\begin{eqnarray}
\label{eq:ourconstraint}
A_0(0) - (aM_{J/\psi}+aM_{H_c})/(2aM_{J/\psi})A_1(0) &+& \\
(aM_{J/\psi}-aM_{H_c})/(2aM_{J/\psi})A_2(0) &=& \Delta_\text{kin}. \nonumber 
\end{eqnarray}
$\Delta_\text{kin}$ here is a nuisance term made up of leading order 
discretisation and mistuning effects to account for the use of lattice 
masses rather than values in the physical continuum limit. 
We take 
\begin{eqnarray}
\label{eq:deltadef}
\Delta_\text{kin} &=& \sum_{i=1}^3\alpha_{c,i} (am^\text{val}_c/\pi)^{2i} + \alpha_{h,i}(am_h/\pi)^{2i} \nonumber \\
&+& \beta_c \delta_{m_c}^\text{val} + \beta'_c \delta_{m_c}^\text{sea} + \beta_l\delta_{m_{l}}^\text{sea} + \beta_s\delta_{m_{s}}^\text{sea}
\end{eqnarray} 
where $\alpha$ and $\beta$ are priors taken as $0(1)$. 
We find that the fit returns values for $\alpha$ and $\beta$ well 
within their prior widths. 

\begin{table}
\caption{\label{zexpcoefficients} Physical $z$-expansion coefficients for the pseudoscalar, axial-vector and vector form factors for $B_c \rightarrow J/\psi$ decay. The correlation matrices between the parameters are given in Appendix \ref{anfullcov}.}
\begin{tabular}{ c | c c c c }
\hline
& $a_0$	& $a_1$	& $a_2$	& $a_3$	\\\hline
${A0}$&	0.1006(37)&	-0.731(72)&	0.30(90)&	-0.02(1.00)\\
${A1}$&	0.0553(19)&	-0.266(40)&	0.31(70)&	0.11(99)\\
${A2}$&	0.0511(91)&	-0.22(19)&	-0.36(82)&	-0.05(1.00)\\
${V}$&	0.1057(55)&	-0.746(92)&	0.10(98)&	0.006(1.000)\\
\hline
\end{tabular}
\end{table}

The fit was done simultaneously across all form factors 
in order to preserve correlations important for constructing helicity amplitudes. We use the nonlinear least squares fitting routine from the python package \textbf{lsqfit}~\cite{lsqfit}.
It returns a $\chi^2/\mathrm{dof}$ of 0.18. 
The fit curves evaluated at the $b$ quark mass in the continuum limit are 
plotted along with the raw lattice results 
in Figure~\ref{FFqsq}. 

In Figure~\ref{zspace} we show the lattice results and the fit 
function in $z$-space in a way that makes clearer how the fit 
is operating. The figure shows that the form factor multiplied 
by the pole function takes a very simple, linear, form in $z$-space. 
The constant term in the $z$-space polynomial varies with heavy 
quark mass but the slope changes very little. 

The extrapolated continuum values of $a_n = b_n^{000}$ are 
given in Table~\ref{zexpcoefficients}. In the continuum 
limit at the physical $b$ mass Eq.~(\ref{fitfunctionequation})
becomes, for each form factor, 
\begin{equation}
F(q^2) = \frac{1}{P(q^2)}\sum_{n=0}^3 a_n z^n .\label{contfitfunction}
\end{equation}
$P(q^2)$ is the appropriate pole factor for that case (Eq.~(\ref{poleformeq})) 
using pole masses from Table~\ref{poletab}. 
To reconstruct the form factors in $q^2$ space both the values of the $z$-expansion 
coefficients from Table~\ref{zexpcoefficients} and their 
correlations are needed; 
the correlation matrices 
are given in Appendix~\ref{anfullcov}.
Table~\ref{zexpcoefficients} shows that our fit is able to pin 
down the constant and linear pieces of the $z$-expansion but 
is not able to pin down higher order terms in $z$. This is 
consistent with what we see in Figure~\ref{zspace}. Notice that the 
values obtained for $a_0$ and $a_1$ in each case are well within the 
prior width of 1.0 allowed for them in the fit.  

We choose the parameterisation of Eq.(\ref{fitfunctionequation}) and hence Eq.(\ref{contfitfunction}) for its simplicity. 
The alternative BGL scheme~\cite{Boyd:1997kz} includes additional outer functions along with the 
pole factor and allows unitarity constraints, which we do not make use of in our fit, to be applied. In Appendix~\ref{BGL_tests} 
we convert our physical continuum results to the BGL scheme and check these 
unitarity constraints. We see that the bounds are far from saturation and so 
implementing them would have no effect on our results.

\subsection{Heavy Mass Dependence}
\label{metabdependence}

\begin{figure*}
\centering
\includegraphics[scale=0.225]{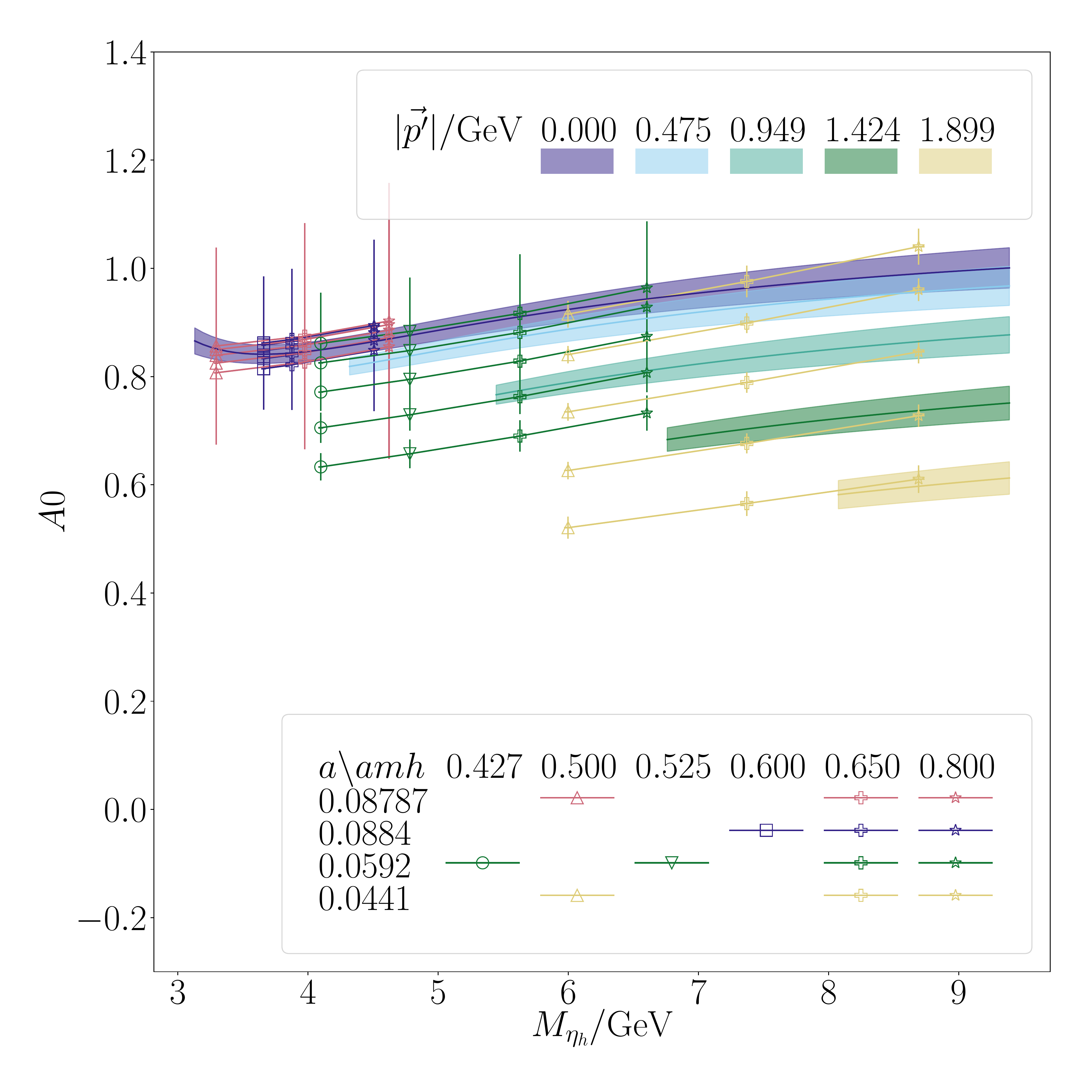}
\includegraphics[scale=0.225]{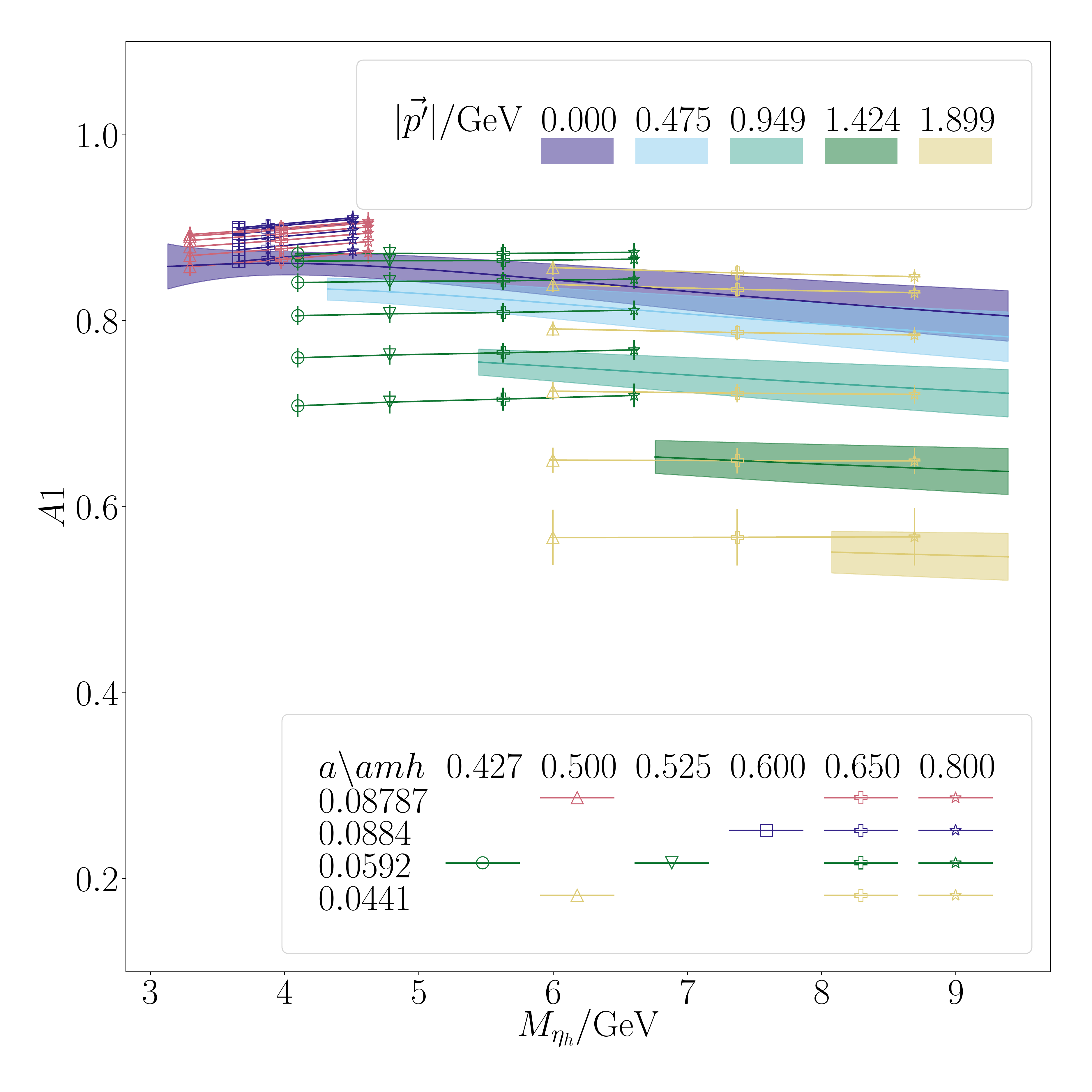}
\includegraphics[scale=0.225]{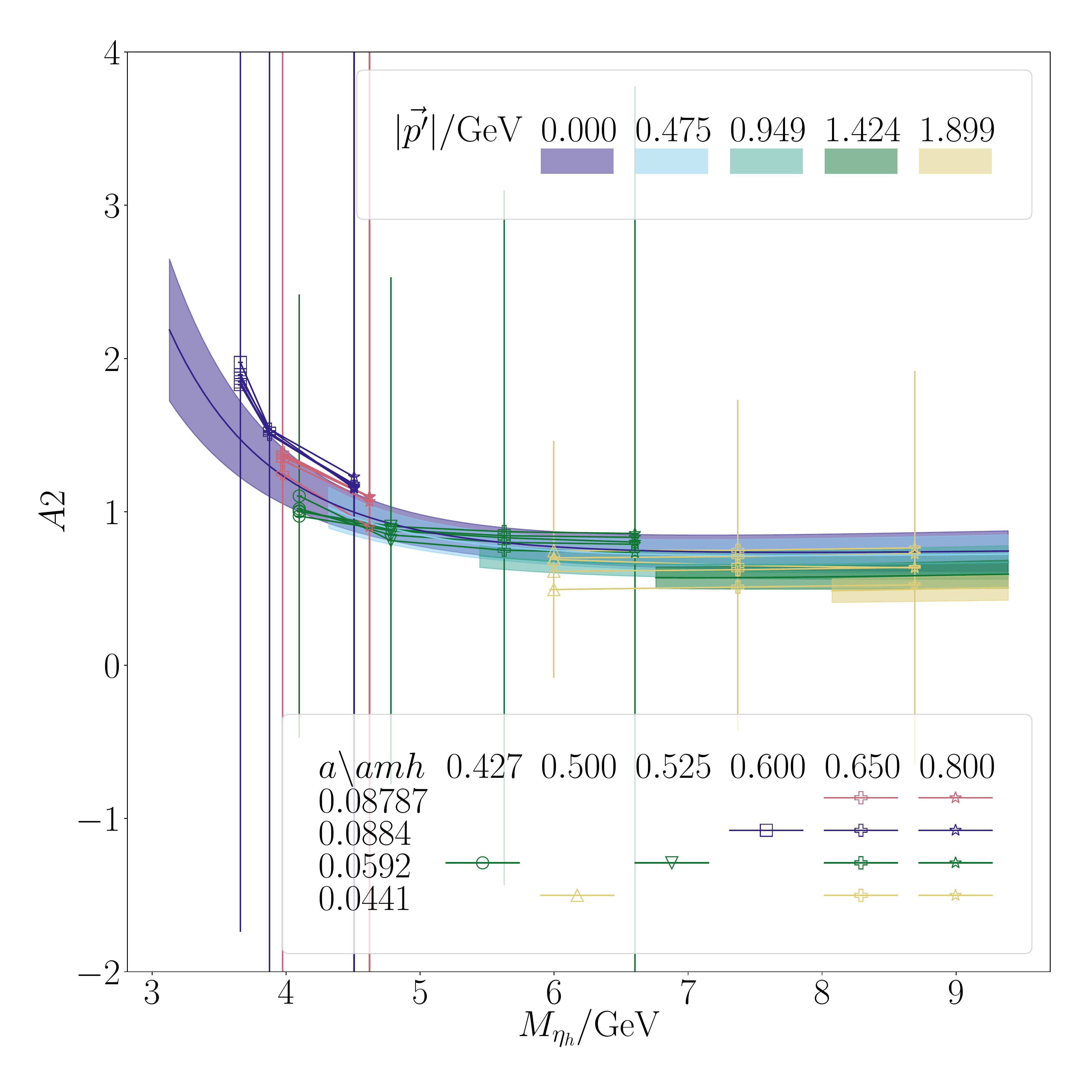}
\includegraphics[scale=0.225]{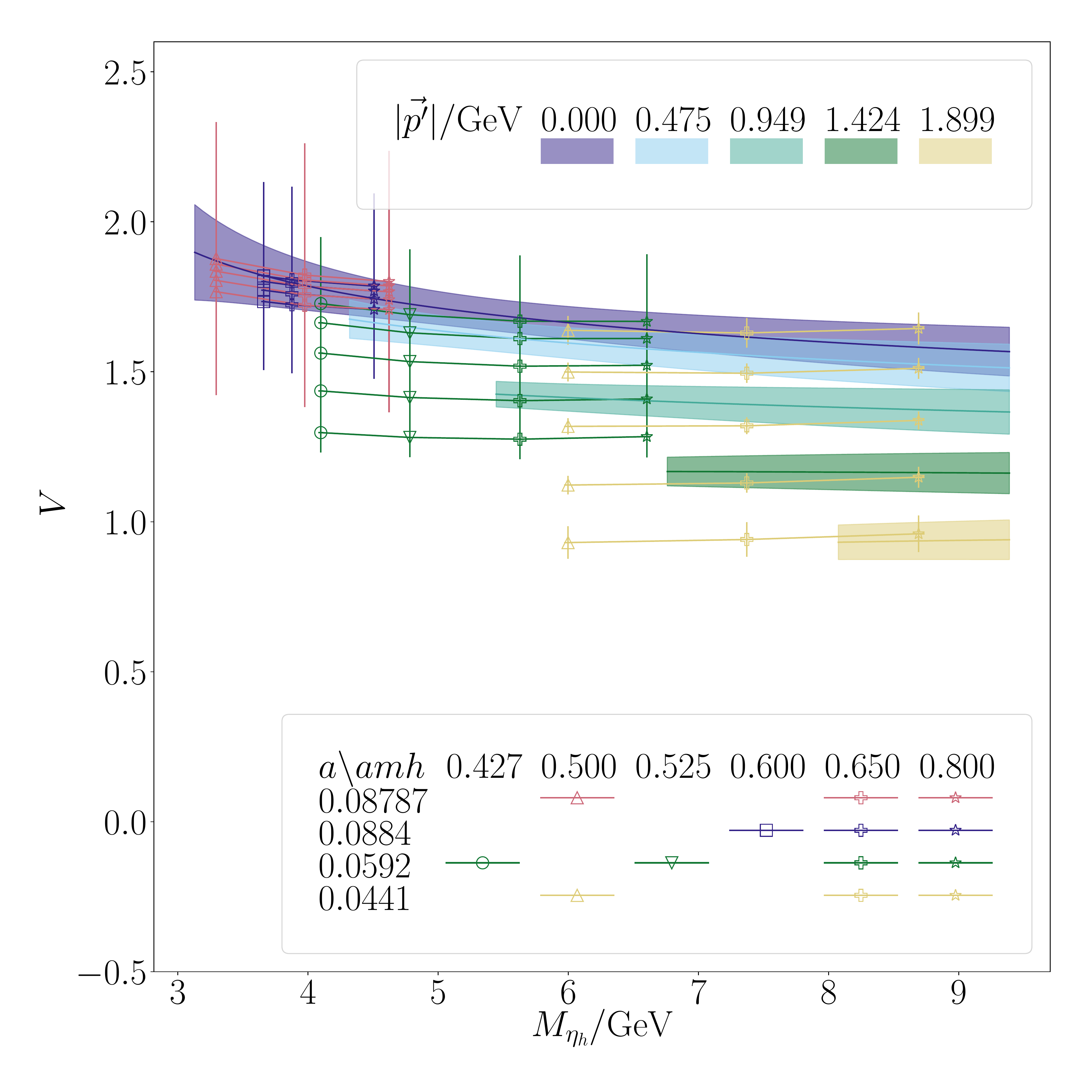}
\caption{\label{etahdepplot} The points show our lattice QCD results for each 
form factor as given in Tables~\ref{set1},~\ref{set2},~\ref{set3} 
and~\ref{set4} as a function of the $\eta_h$ mass $M_{\eta_h}$, with data points corresponding 
to the same $J/\psi$  spatial momentum (given in \cref{set1,set2,set3,set4}) connected. We also use Eq.~(\ref{mhcparam}) 
to plot our continuum result (solid coloured curves) at multiple, evenly spaced, fixed values of $J/\psi$ momentum within the semileptonic region $0\le q^2 \le q^2_\mathrm{max}$.
The legend gives the mapping between symbol colour and shape and the 
set of gluon field configurations used, as given by the lattice 
spacing, and the heavy quark in lattice units. Note that for the form factor $A_2$ we 
exclude from the plot the inaccurate lattice data for $am_h=0.5$ on set 4.
}
\end{figure*}

Our fits allow us to reconstruct the physical dependence of the form factors on the heavy quark mass. 
This allows us to check that the dependence is relatively benign and that we have adequate coverage 
of the heavy quark mass range. It could also be used to test theoretical understanding from model calculations. 

This test of the physical heavy mass dependence is complicated slightly by the fact that our continuum fit results depend on $m_h$ both through the $M_{\eta_h}$ dependence appearing in $\Delta^{(i)}_{h}$, and through 
$M_{H_c}$ appearing in $z$, $P(q^2)$ and our expression for the resonance masses. To evaluate our 
continuum form factors away from the physical $b$ mass we must first determine the 
$M_{\eta_h}$ dependence of $M_{H_c}$. To do this we fit our lattice data for the $\gamma_5 \otimes \gamma_5 $ $H_c$ 
and $\eta_h$ masses, together with the physical values of $M_{\eta_b}$ and $M_{B_c}$~\cite{pdg20}, using the simple form
\begin{align}
M_{H_c}=&(M^\mathrm{phys}_{\eta_c}+M_{\eta_h})/2  +\sum_{i=1}^4X_i\left(\frac{am_h}{\pi}\right)^{2i}\nonumber\\
&+ \sum_{i=1}^4Y_i\left(\frac{am_c}{\pi}\right)^{2i}+\sum_{i=1}^4Z_i\Delta^{(i)}_{hc}+\mathcal{N'}
\end{align}
where 
\begin{equation}
\Delta^{(i)}_{hc}=\left(\frac{2\Lambda_\mathrm{QCD}}{M^\mathrm{phys}_{\eta_c}}\right)^i-\left(\frac{2\Lambda_\mathrm{QCD}}{M_{\eta_h}}\right)^i
\end{equation}
and 
\begin{equation}
\mathcal{N'} = A' \delta_{m_c}^\text{val}+ B' \delta_{m_c}^\text{sea}+ C' \delta_{m_s}^\text{sea} + D' \delta_{m_l}^\text{sea}.
\end{equation}
This form ensures the correct value of $M_{H_c}$ as $m_h\rightarrow m_c$. We take $M^\mathrm{phys}_{\eta_c}=2.9839/\mathrm{GeV}$ from \cite{pdg20} and neglect its very small uncertainty. 
We take prior widths of $0(1)$ for $A'$, $B'$, $C'$, $D'$, $X_i$, $Y_i$ and $Z_i$
giving a good fit with $\chi^2/\mathrm{dof}=0.99$ and $Q=0.46$.
Our fitted parameters $Z_i$ may then be used to generate the continuum value of $M_{H_c}$ at a given $M_{\eta_h}$ by
\begin{equation}\label{mhcparam}
M_{H_c}=(M^\mathrm{phys}_{\eta_c}+M_{\eta_h})/2+\sum_{i=1}^4Z_i\Delta^{(i)}_{hc}.
\end{equation}
Note that this parameterisation of the $H_c$ mass is only used here to demonstrate the heavy mass 
dependence of the form factors and will not have any impact on subsequent results.

In Figure~\ref{etahdepplot} the form factors at fixed values of the $J/\psi$ momentum are 
plotted against $M_{\eta_h}$. Here we choose values of the $J/\psi$ momentum which evenly span the 
semileptonic range at the physical $b$ quark mass and only plot the mass region for which the 
resulting value of $q^2$ is between 0 and $q^2_\mathrm{max}$. We also plot our lattice 
data where we have connected points corresponding to a fixed $J/\psi$ spatial momentum.

Figure~\ref{etahdepplot} illustrates that the continuum form factors have only mild 
heavy mass dependence across the range of masses used 
in this study. This demonstrates that we have good coverage of the heavy quark mass 
range and that our extrapolation to the $b$ mass using these points is reliable.
This is consistent with what is seen for other $b\to c$ form factors, e.g.~\cite{EuanBsDsstar,EuanBsDs}.

\subsection{Tests of the Stability of the Analysis}
\label{stabsec}

\begin{figure}
\centering
\includegraphics[scale=0.4]{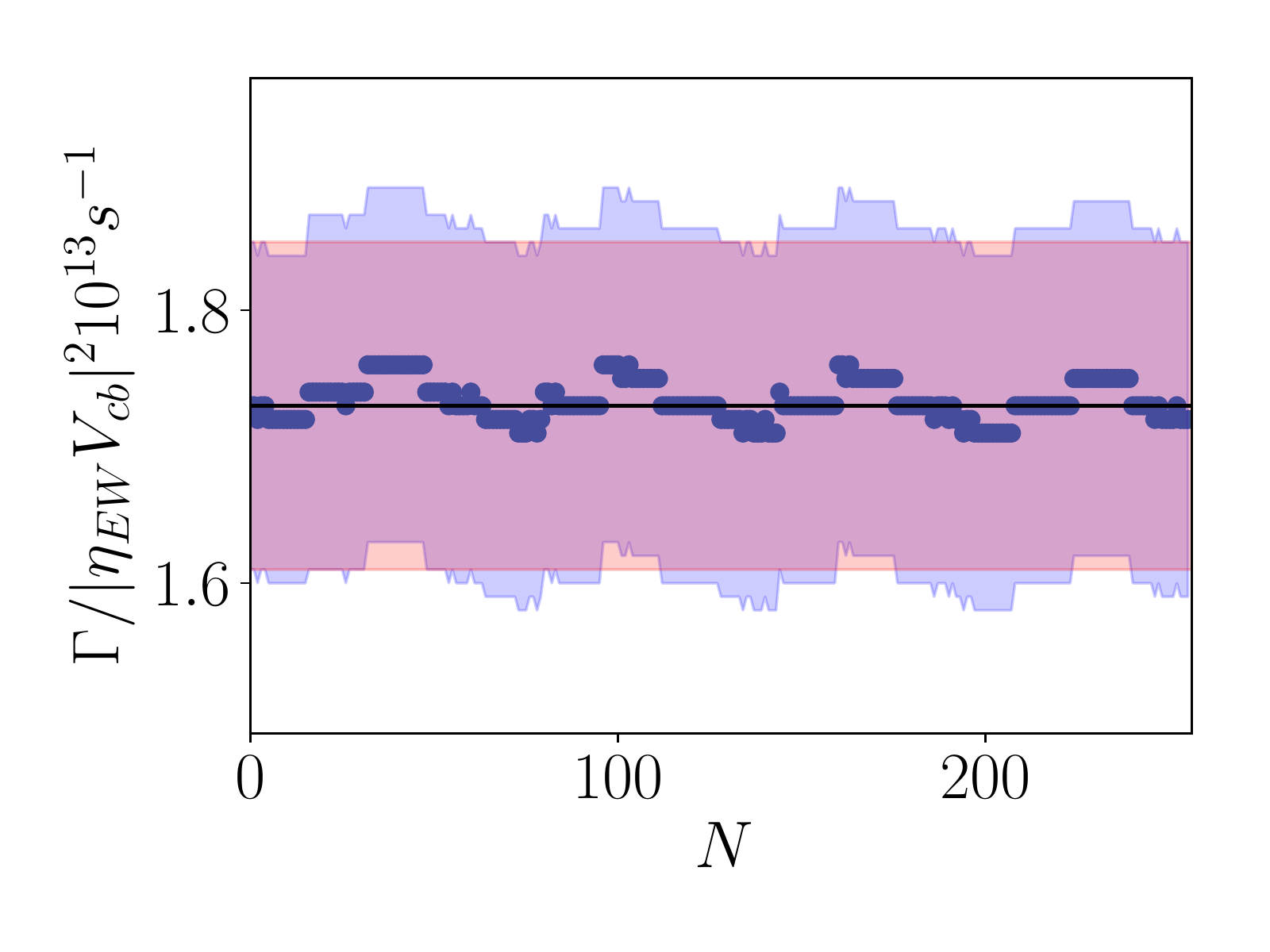}
\caption{\label{gammastabplot} Plot showing the stability of the total rate 
for $B_c^- \rightarrow J/\psi \mu^- \overline{\nu}_\mu$ under variations of the 
correlator fits. The $x$ axis value corresponds to 
$N = \delta_3 + 4\delta_2+16\delta_1+64\delta_4$ where $\delta_n$ is the 
value of $\delta$ corresponding to the fit given in Table~\ref{fitparams} 
for set $n$. The black horizontal line and red error band correspond to our 
final result and the blue points and blue error band correspond to the 
combination of fit variations associated to $N$. Our result for the total rate is 
very stable to these variations. }
\end{figure}

\begin{figure}
\centering
\includegraphics[scale=0.4]{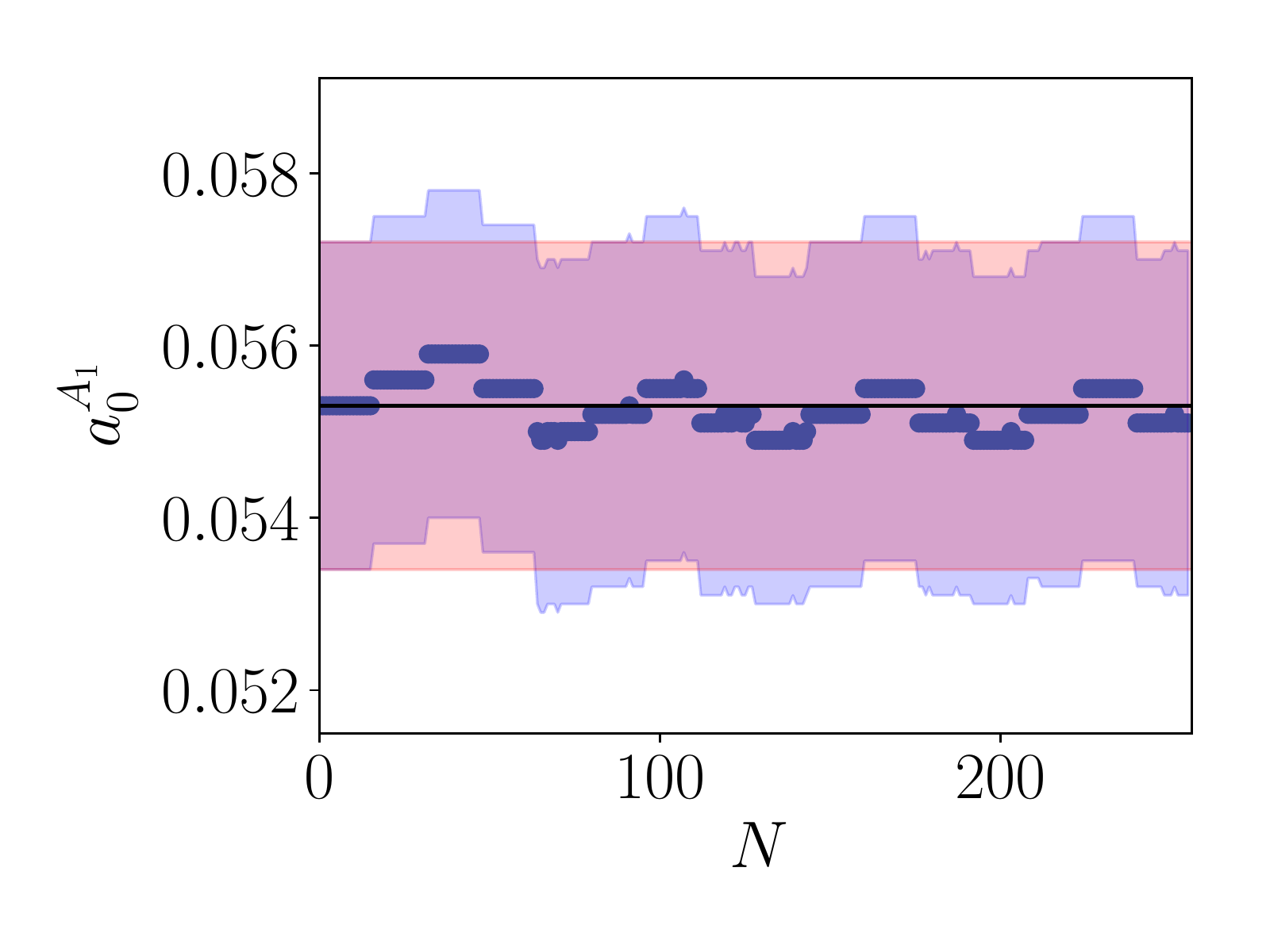}
\includegraphics[scale=0.4]{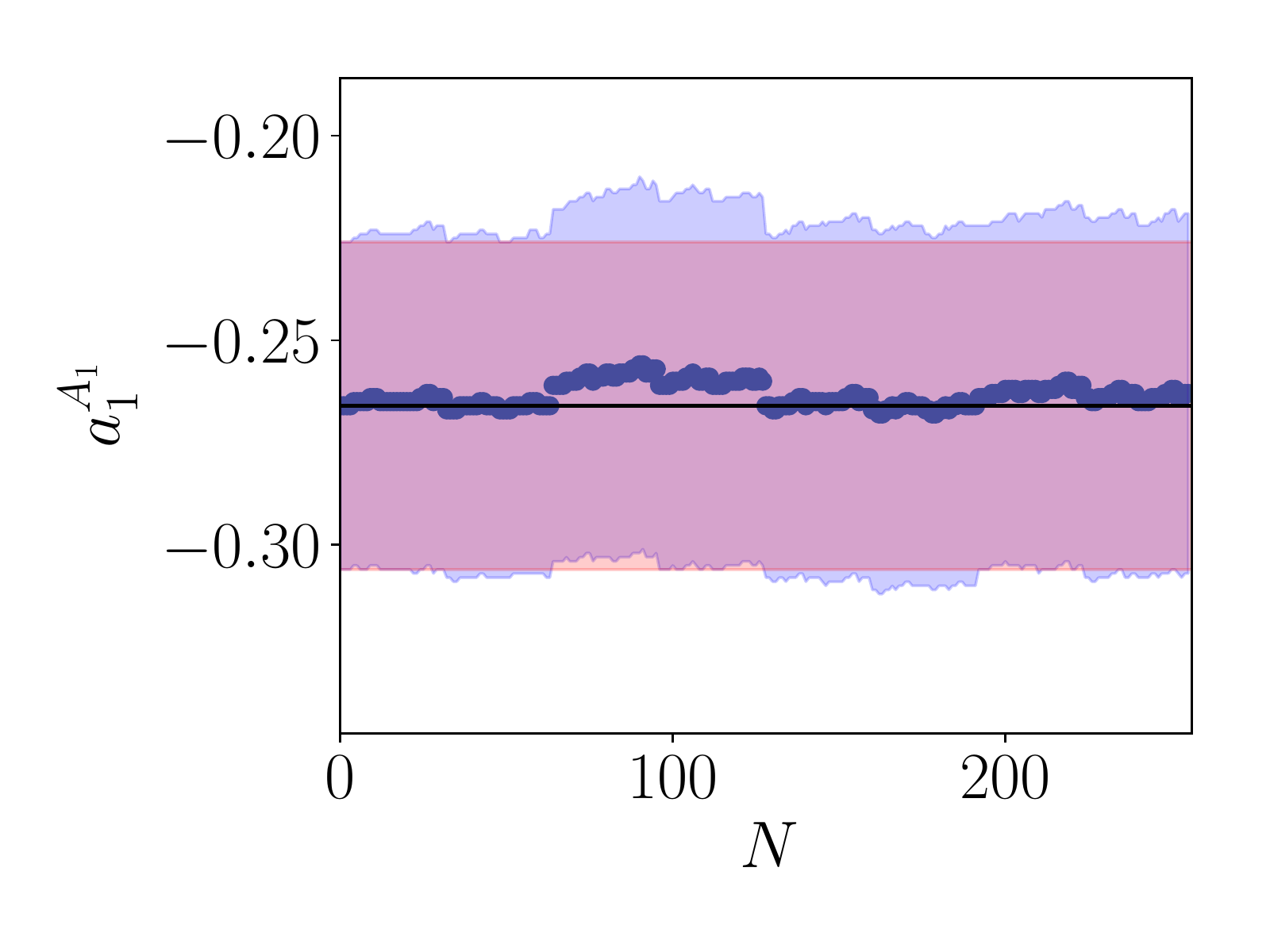}
\includegraphics[scale=0.4]{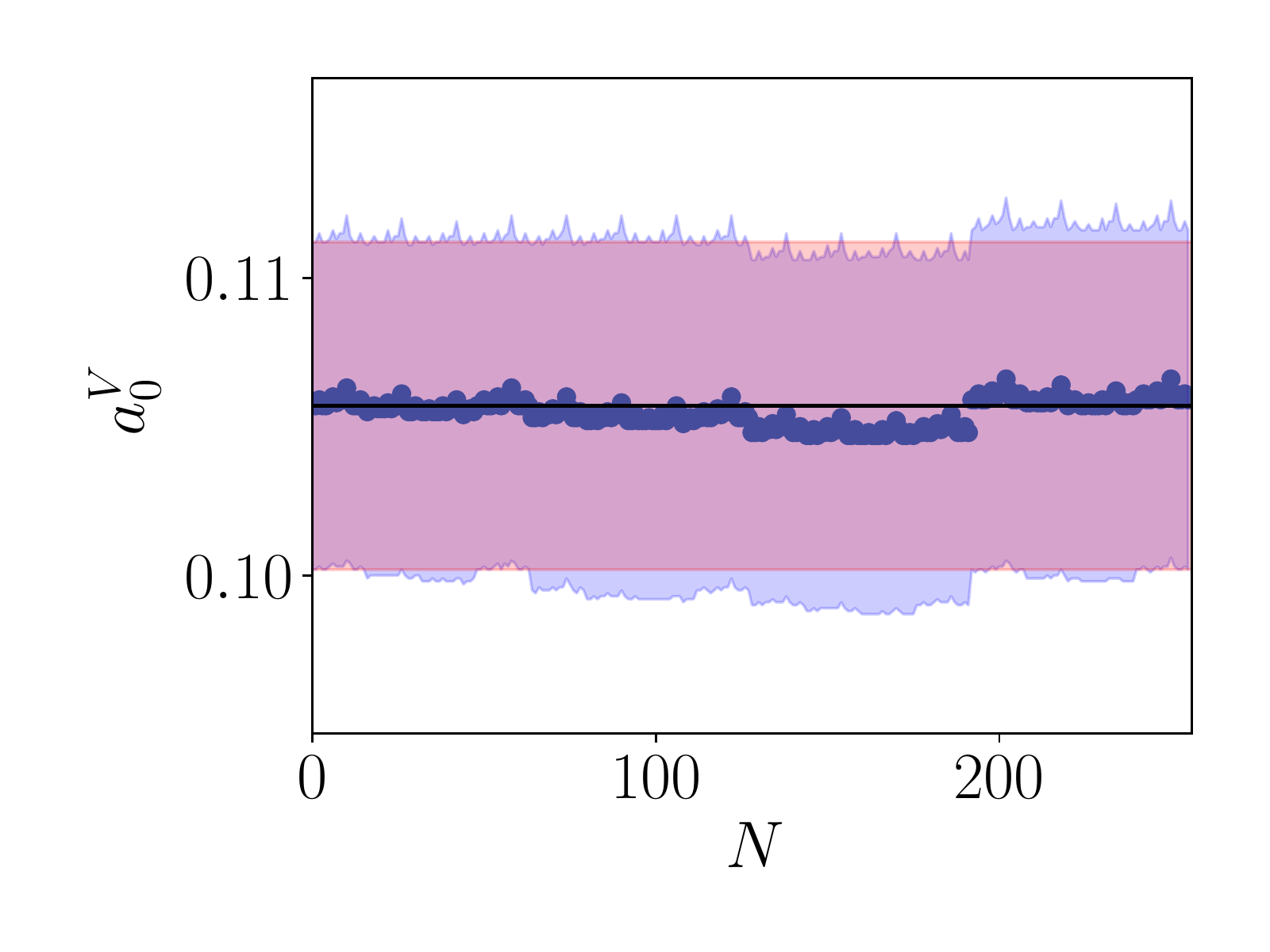}
\caption{\label{coeffstabplot} 
As for Figure~\ref{gammastabplot} showing the stability of the 
coefficients of the $z$-expansion for the form factors 
under variations of the correlator 
fits.  We include a subset of coefficients here; other plots look 
very similar. 
}
\end{figure}

Our results for the form factors are dependent to some extent 
upon choices made in the correlator fits as well as choices made in the 
continuum/heavy-quark mass fit. Here we test the impact of those choices 
on both the continuum fit coefficients and the final outcome of our calculation to make sure that it is robust.
It is convenenient in that test to have a single quantity which we can 
compare and we take that to be the total rate of 
$B_c^- \rightarrow J/\psi \ell^- \overline{\nu}_\ell$ 
decay, i.e. \
$\Gamma(B_c^-\rightarrow J/\psi \mu^-\overline{\nu}_\mu)/|\eta_{\mathrm{EW}}V_{cb}|^2$.
This is obtained by determining the helicity amplitudes from our form 
factors and then integrating in $q^2$ over the differential rate 
they give (see Eqs~(\ref{helicityamplitudes}) and~(\ref{dgammadq2})). 
The results for the differential rates and total rate will be discussed 
in more detail in Sec.~\ref{sec:discussion}; 
here we focus on the stability of the final result under variations of fit 
choice. 

We first look at the choices of the correlator fit parameters:  
$\Delta T_\text{3pt}$, $\Delta T_\text{2pt}^{J/\psi}$, 
$\Delta T^{H_c}_\text{2pt}$, the value of SVD cut and the number of 
exponentials used in the fit. In order to verify that our results 
are independent of such choices we repeat the full analysis using 
all combinations of the variations listed in Table~\ref{fitparams}. 
The total rate computed using each of these fit variations is 
plotted in Figure~\ref{gammastabplot}, where we see that our 
final result is insensitive to such variations. 

\begin{figure}
\centering
\includegraphics[scale=0.3]{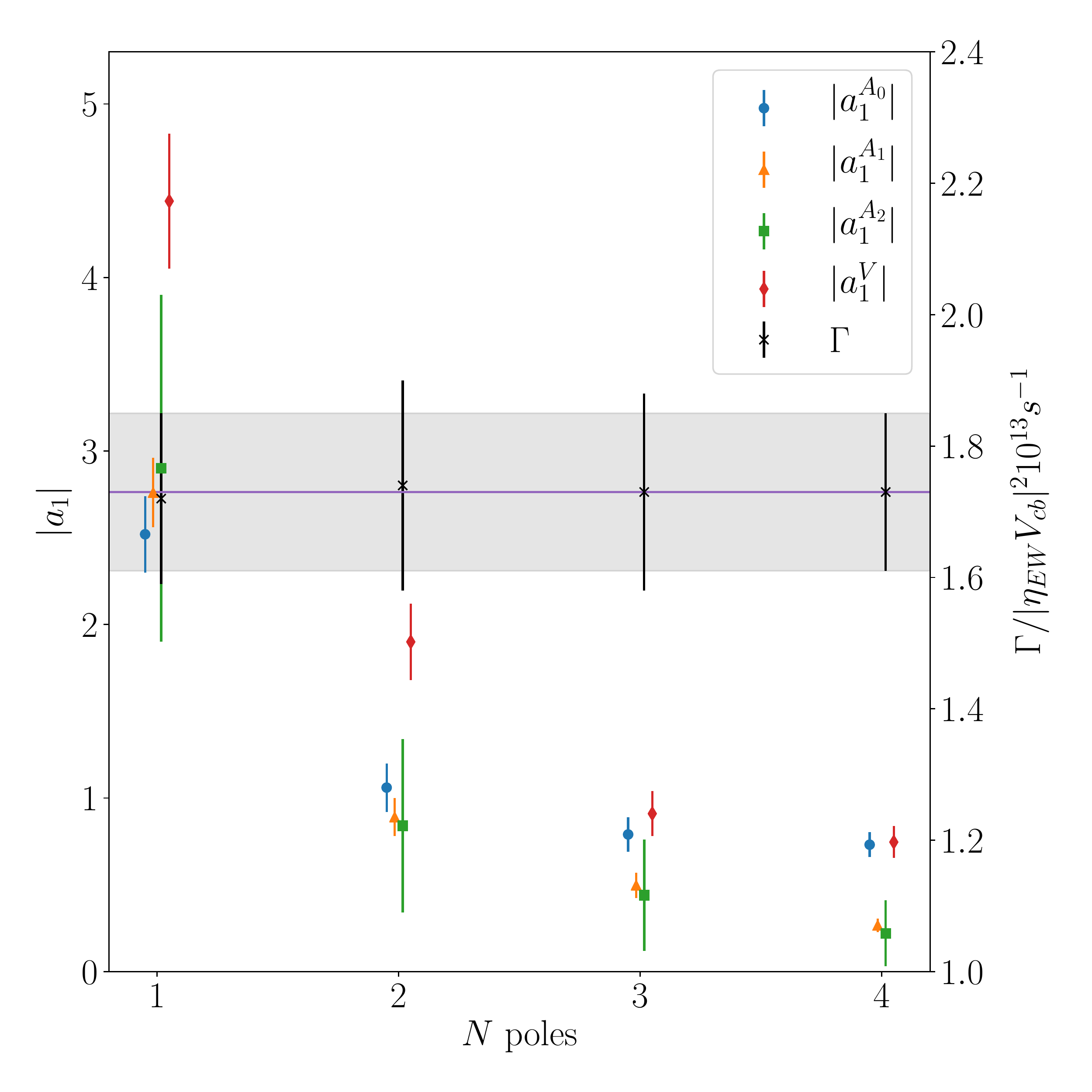}
\caption{\label{NPOLESPLOT} Magnitude of the $\mathcal{O}(z)$ coefficient, $a_1$, for each 
form factor plotted against the number of poles included in Eq.~(\ref{poleformeq}). 
The prior widths on the $b_n^{ijk}$ are scaled according to the number of 
poles, see text.  Note that the maximum number of poles included for $A_0$ is 3. 
The black crosses and error bars give the total width for the $\ell=\mu$ case, $\Gamma/|\eta_{\mathrm{EW}}V_{cb}|^2$, 
determined from that fit, using the right-hand $y$ axis. 
The grey band corresponds to our final result for the total width 
using $N_\text{poles}=4$, and prior values for $b_n^{ijk}$ of $0(1)$. This 
shows how the different coefficients as a function of $N_\text{poles}$ give 
a very stable result for the total width. }
\end{figure}

Figure~\ref{coeffstabplot} shows 
a similar stability plot for a subset of the $z$-expansion coefficients 
for separate form factors (Eq.~\ref{contfitfunction}). We again see 
that the $z$-expansion coefficients are stable, within their uncertainty 
band, under changes in the correlator fits. Plots for other form factors 
not shown here
are very similar. 

We now turn to the effect of choices in the extrapolation to the physical point. 
The pole form of Eq.~(\ref{poleformeq}) removes $q^2$ dependence from poles 
in the $q^2$ plane above the physical region for the decay. 
For $b \rightarrow c$ decays these poles are substantially above the physical 
region. For example, here $q^2_{\text{max}}$ is $(3.18 \,\mathrm{GeV})^2$ whereas 
the lowest pole, for the pseudoscalar form factor is at the $B_c$ mass of 6.275 GeV. 
Hence we do not expect the exact positions or number of poles to have a big effect 
on the fits. The magnitude of $P(q^2)$ does depend on how many poles are 
included~\cite{Harrison:2017fmw}, however, and this affects the normalisation of 
the quantity form factor times $P(q^2)$ that is expanded in $z$ 
(Eq.~(\ref{fitfunctionequation})). This in turn affects the prior width that 
must be allowed on the $z$-expansion coefficients to achieve a good fit. 
 
We have investigated the effect of including fewer poles in Eq.~(\ref{poleformeq}) 
by repeating our analysis including only the first $N_\text{poles}$ resonances 
listed in Table~\ref{poletab}. We then take a prior width on the $z$-expansion 
coefficients of $5.0 - N_\text{poles}$. 
We are able to obtain a good fit, with $\chi^2/\mathrm{dof}\approx 0.2$ in all cases.  
Since there are only 3 poles for $A_0$ expected below $t_+$, we include only 
3 poles for that form factor even in the $N_\text{poles}$ case. 

Figure~\ref{NPOLESPLOT} shows these results, plotting against the left-hand $y$-axis 
the magnitude of the coefficient corresponding 
to the order $z$ term, $a_1$, coming from the fits as a function of the number of poles 
included.  Results are given for each form factor.
We see that as we include fewer poles, increasingly large $z$-expansion 
coefficients are needed partly in order to account for the 
removal of physical $q^2$ dependence from missing poles but also because of 
the normalisation change. 

Figure~\ref{NPOLESPLOT} also gives, marked as black crosses and using the 
right-hand $y$ axis, the total width for $B_c \rightarrow J/\psi$ obtained 
from that fit. We see that the total width is very stable as a function of 
$N_\text{poles}$ as the change in $P(q^2)$ is compensated by the 
different coefficients obtained for the $z$-expansion.

\begin{figure}
\centering
\includegraphics[scale=0.275]{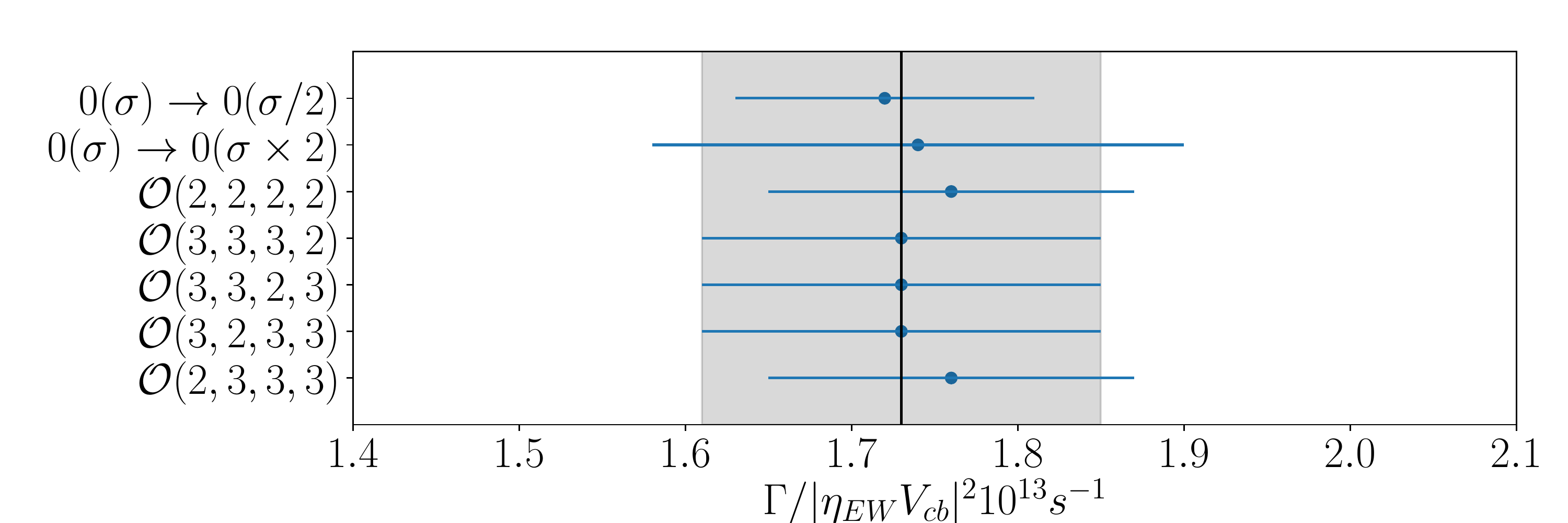}
\caption{\label{ntermstabplot} Plot showing the stability of the total rate 
for $B_c^- \rightarrow J/\psi \mu^-\overline{\nu}_\mu$ considering lower order truncations of $z$-expansion, discretisation and heavy mass dependent terms in Eq.~(\ref{fitfunctionequation}) and Eq.~(\ref{eq:anfitform}). $\mathcal{O}({n_1},{n_2},{n_3},{n_4})$ corresponds to the result including terms of highest order $\mathcal{O}( (2\Lambda/M_{\eta_h})^{n_1}, (am_c)^{2n_2}, (am_h)^{2n_3}, z^{n_4})$. The vertical black line is our final result, corresponding to $\mathcal{O}({3},{3},{3},{3})$, and the grey band is its uncertainty. We also include variations in which we multiply our prior widths either by a factor of 2 or 0.5, labelled as $0(\sigma)\rightarrow 0(\sigma\times 2)$ and $0(\sigma)\rightarrow 0(\sigma/ 2)$ respectively. Our result for the total rate is very stable to these variations.}
\end{figure}

We have also investigated the effects of including fewer $z$ terms in Eq.~(\ref{fitfunctionequation}) as well as fewer $am_c$, $am_h$ and $2\Lambda_{QCD}/M_{\eta_h}$ terms in Eq.~(\ref{eq:anfitform}). Figure~\ref{ntermstabplot} gives the total width obtained using these variations, where we see that our result is insensitive to the removal of the highest order terms. We also investigate the effect of increasing or decreasing the prior widths of the parameters $b_n^{ijk}$ in Eq.~(\ref{eq:anfitform}) by a factor of 2. These results are also shown in Figure~\ref{ntermstabplot} where we see only a very small effect on the central value of the total width.

\section{Discussion}
\label{sec:discussion}

\begin{figure}
\centering
\includegraphics[scale=0.225]{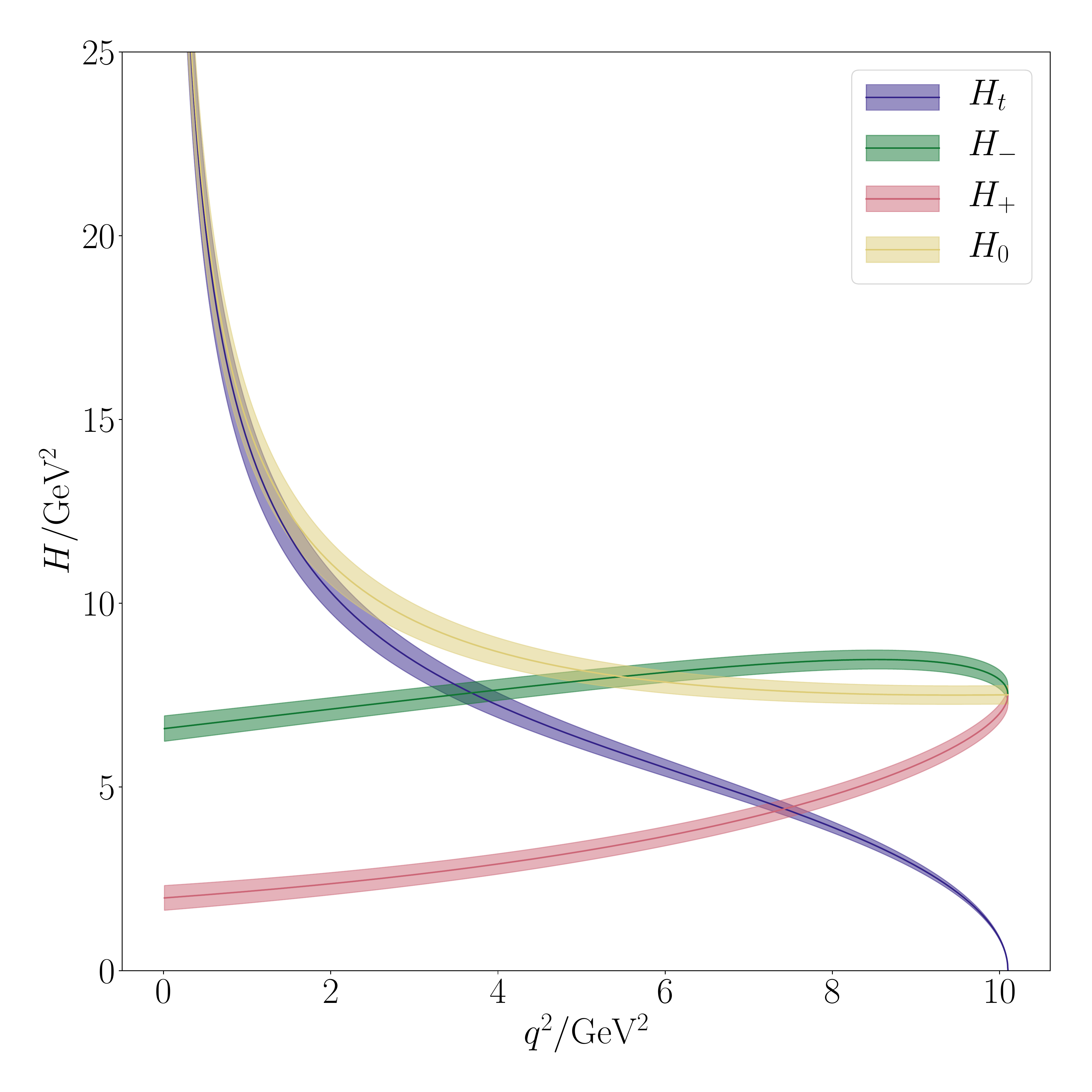}
\caption{\label{helicityamplitudesplot} Helicity amplitudes for $B_c^- \rightarrow J/\psi\ell^-\overline{\nu}_\ell$ plotted as a function of $q^2$.}
\end{figure}

\begin{figure}
\centering
\includegraphics[scale=0.18]{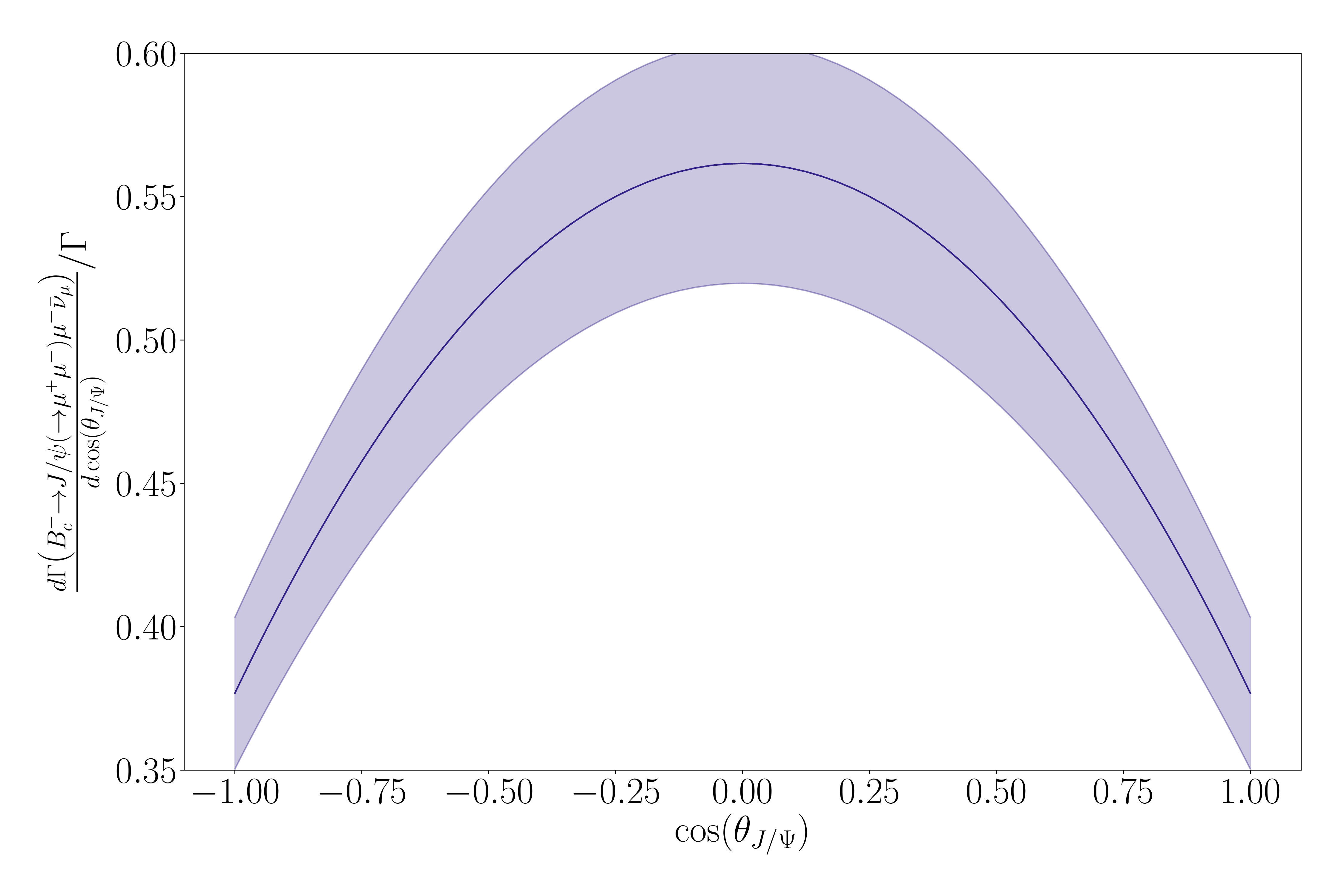}
\includegraphics[scale=0.18]{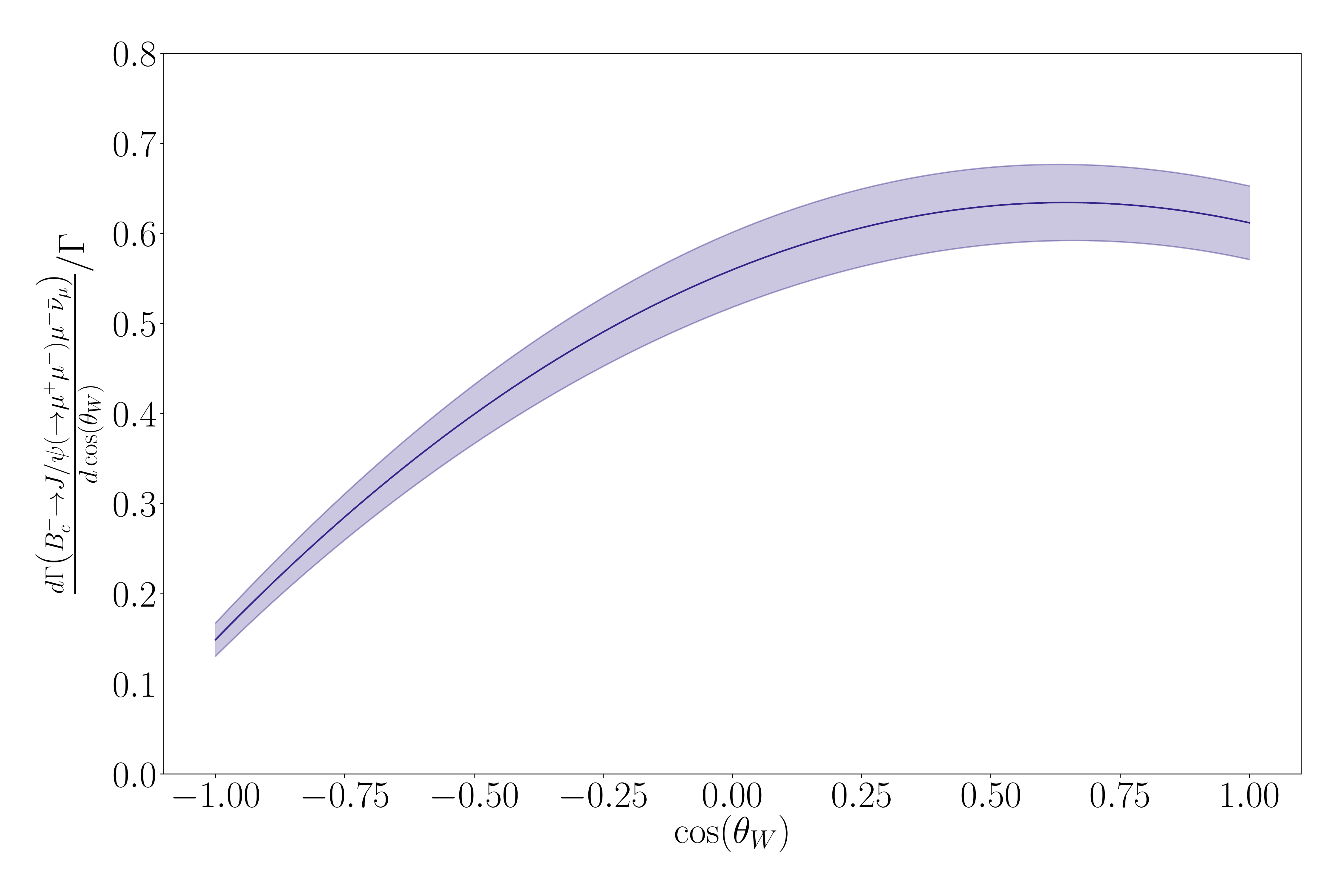}
\includegraphics[scale=0.18]{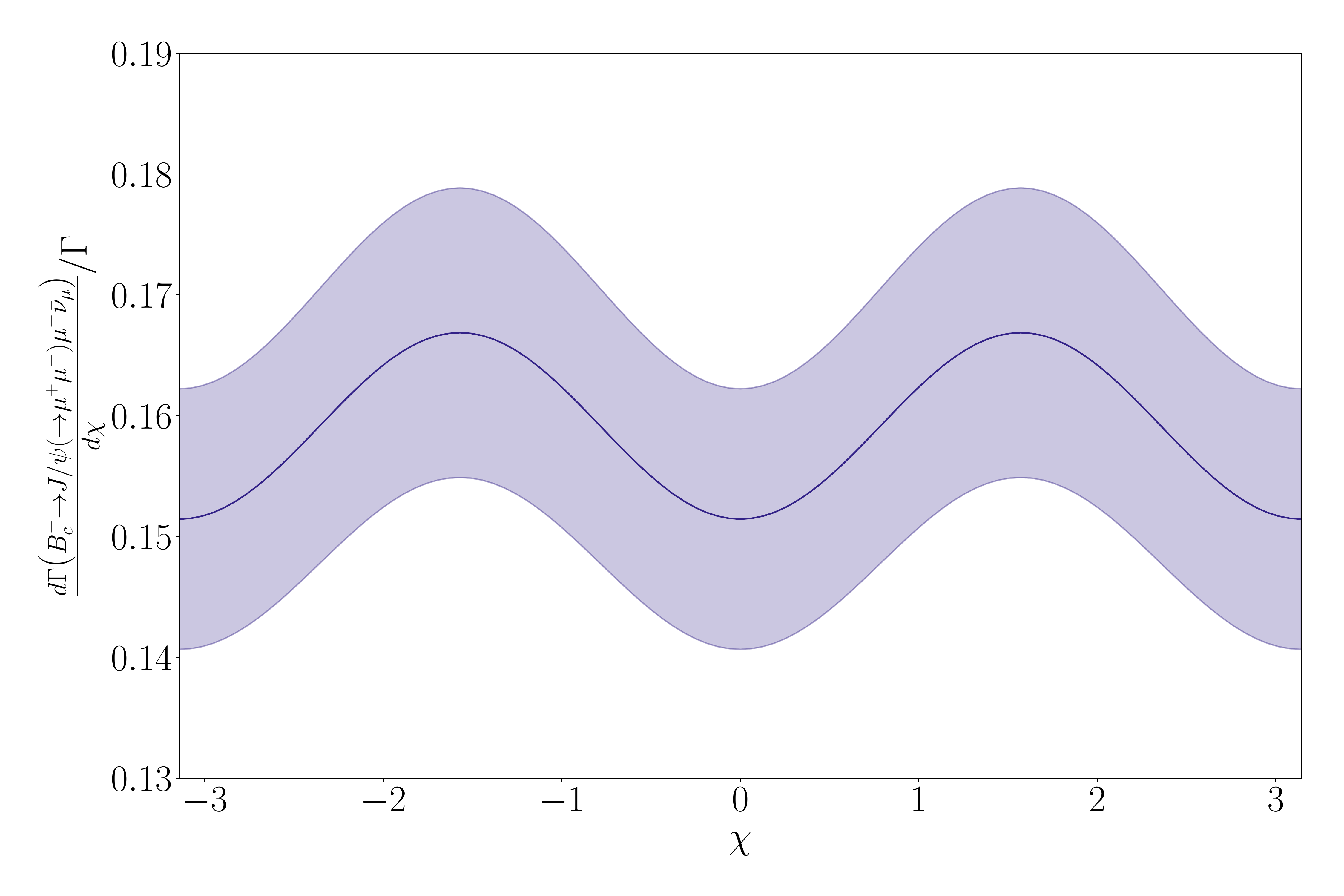}
\caption{\label{dgammadangles} Angular differential decay rates for $B_c^- \rightarrow J/\psi \ell^-\overline{\nu}_\ell$ for the $\ell=\mu$ case. 
From the top down, $d\Gamma/dq^2d\cos(\theta_{J/\psi})$, $d\Gamma/dq^2d\cos(\theta_W)$ and $d\Gamma/dq^2d\chi$. Each rate is normalised by the total decay rate $\Gamma(B_c^- \rightarrow J/\psi(\rightarrow\mu^+\mu^-) \mu^-\overline{\nu}_\mu$).}
\end{figure}

\begin{figure}
\centering
\includegraphics[scale=0.2]{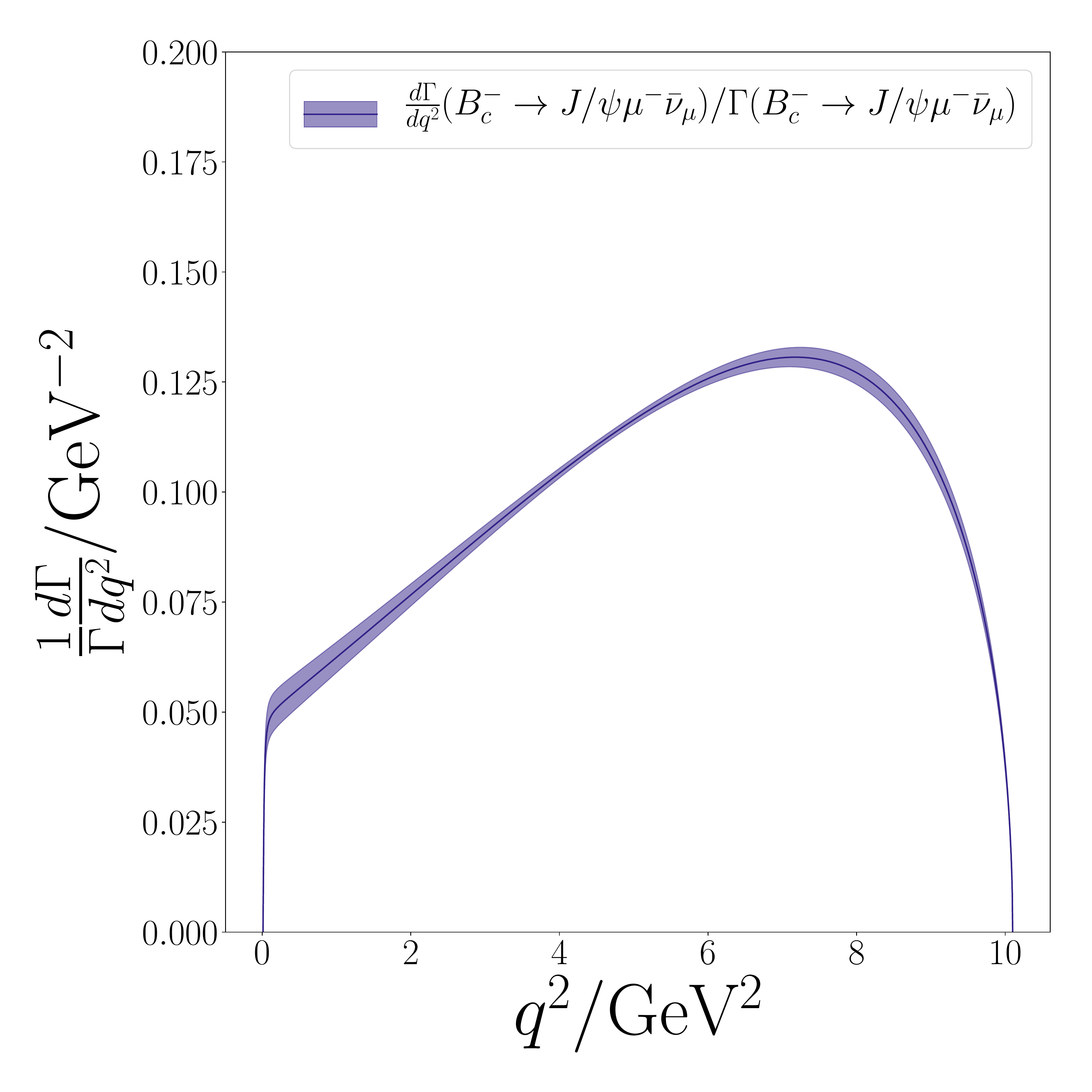}
\caption{\label{dgammadqsq} The differential rate $d\Gamma/dq^2$ for $B_c^- \rightarrow J/\psi \ell^- \overline{\nu}_\ell$ for the $\ell=\mu$ case, normalised by the total decay rate $\Gamma$.}
\end{figure}

Using the form factors from Section~\ref{sec:extrap} we construct the helicity 
amplitudes using Eq.~(\ref{helicityamplitudes}). 
The helicity amplitudes are plotted in Figure~\ref{helicityamplitudesplot}, 
where we see that $H_0$ and $H_t$ are singular at $q^2=0$ as we would 
expect from the $1/\sqrt{q^2}$ factors appearing in their definitions. 
This singular behaviour is cancelled by the $(q^2-m_\ell^2)^2$ factor appearing 
in the differential decay rate (Eq.~(\ref{dgammadq2})). 
Together with the factor of $|\vec{p^{\prime}}|$ this results in physical rates 
which vanish at the extremes of the physical range, $q^2=m_\ell^2$
and $q^2=q^2_\text{max}$

Using these helicity amplitudes we construct the differential decay rates 
using Eqs.~(\ref{dgammadq2}),~(\ref{dgammadcosw}),~(\ref{dgammadtheta}) 
and~(\ref{dgammadchi}). We take $m_\mu=105.66\mathrm{MeV}$ and $m_e=0.51100\mathrm{MeV}$~\cite{pdg20} where we do not include their negligible uncertainties.
The differential rates with respect to the angular variables defined 
in Figure~\ref{angles} are plotted in Figure~\ref{dgammadangles} and the 
differential rate $d\Gamma/dq^2$ is plotted in Figure~\ref{dgammadqsq}. 
In each case  we plot the rate for $l=\mu$ normalising each 
differential rate by the total integrated decay rate. This means that any 
constant factors, such as $|V_{cb}|^2$ or ${\mathcal{B}}(J/\psi \rightarrow \mu^+\mu^-)$ 
cancel out. 
Where an integration over $q^2$ is necessary we use a simple trapezoidal 
interpolation in order to ensure covariaces are carried through correctly, 
taking sufficiently many points that the results are insensitive to using 
additional points. 

The angular differential rates show the expected qualitative picture for a 
pseudoscalar to vector meson decay~\cite{RevModPhys.67.893}. 
The dependence on $\theta_{J/\psi}$ is that for the case where the vector 
meson is seen in final states that carry spin 
(i.e. here in $J/\psi \rightarrow \mu^+\mu^-)$. 
This differs from, for example, the case of $B \rightarrow D^*$ if 
the $D^*$ is seen in $D\pi$. 
The $\theta_W$ dependence is that of a $b$ quark decay via the $V-A$ 
weak interaction. This produces a virtual $W^-$ and charged lepton with, 
preferentially, a $(1+\cos \theta_W)^2$ distribution in the $W^-$ rest frame. 
The picture switches for a $c$ decay producing a virtual $W^+$~\cite{RevModPhys.67.893}. 

\begin{figure}
\centering
\includegraphics[scale=0.18]{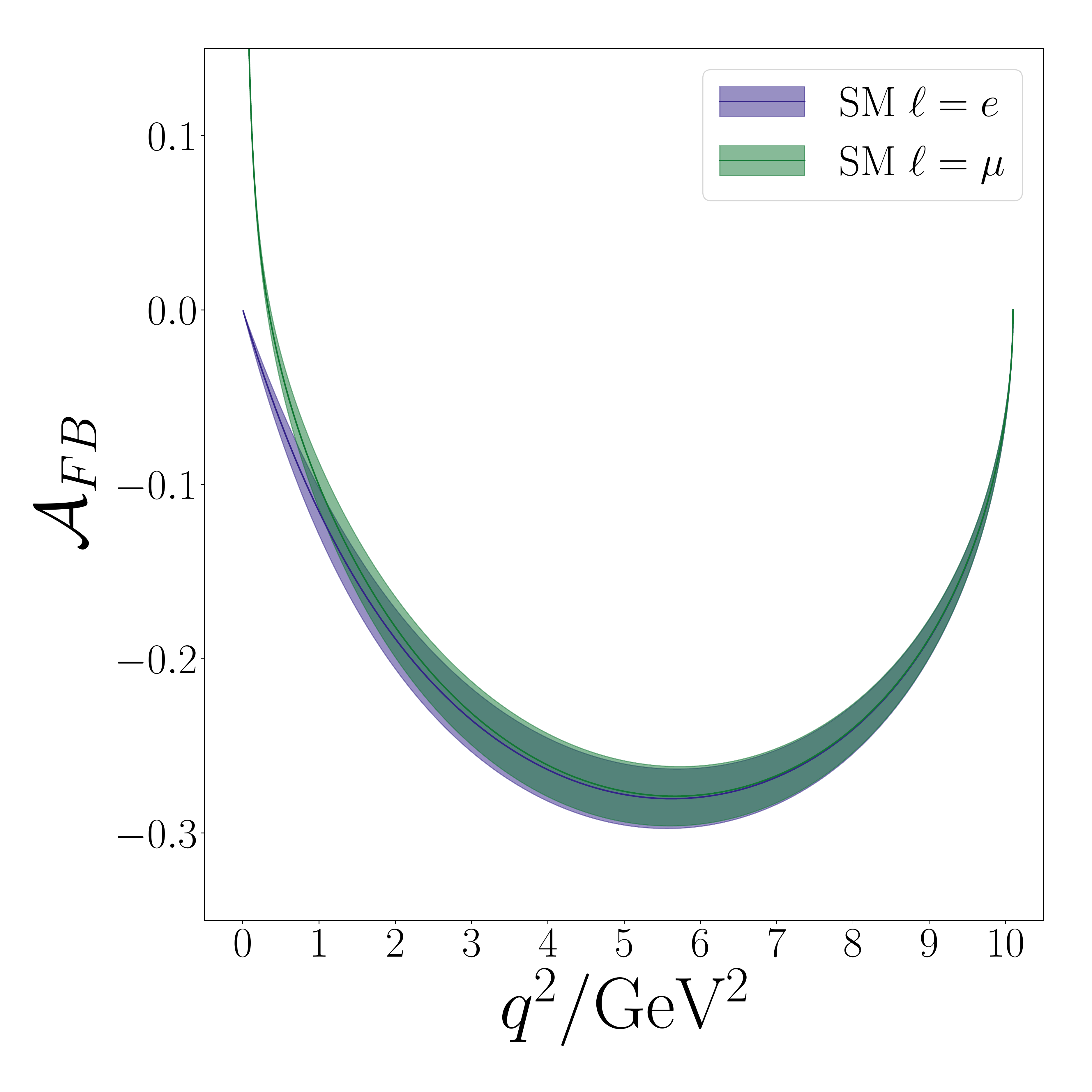}
\includegraphics[scale=0.18]{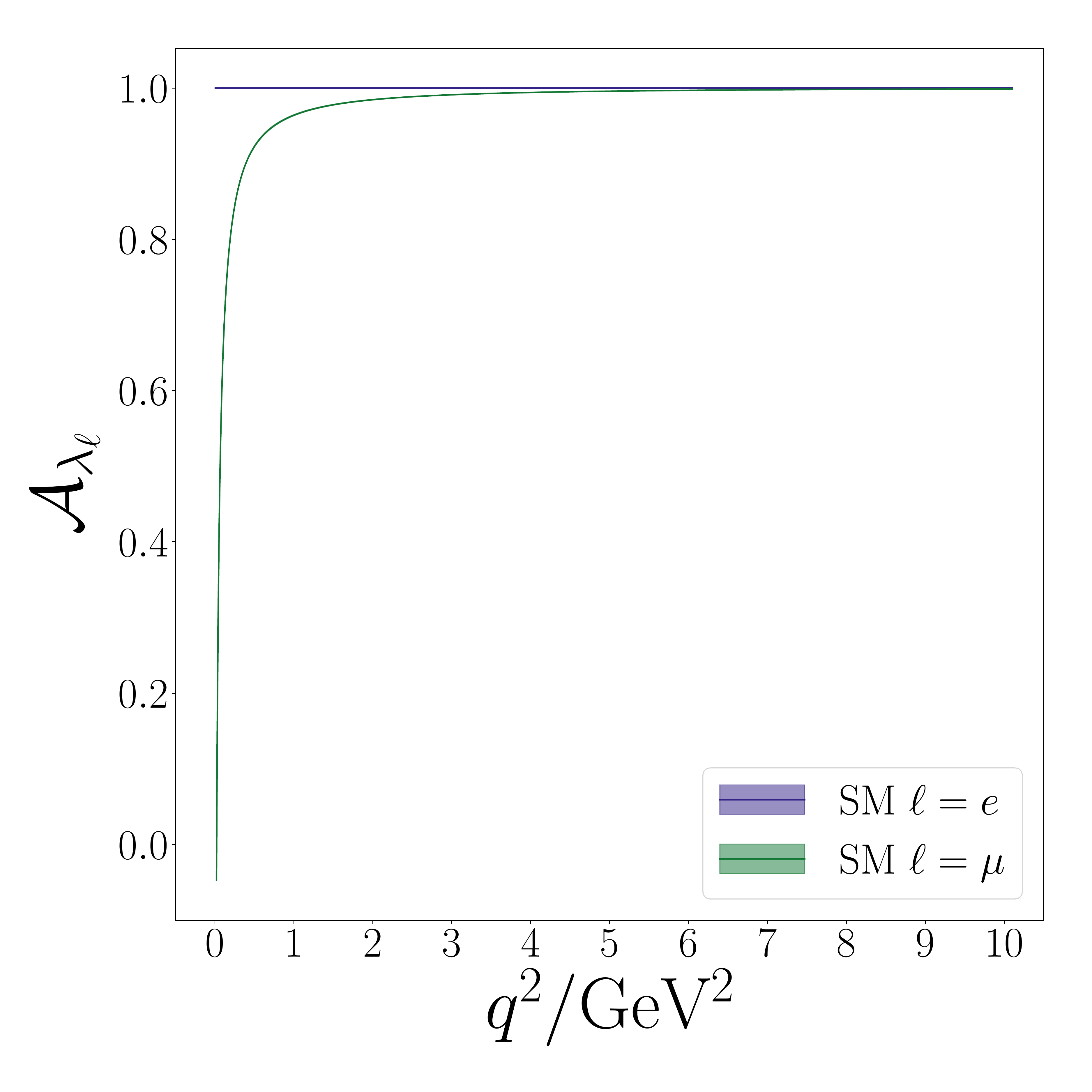}
\includegraphics[scale=0.18]{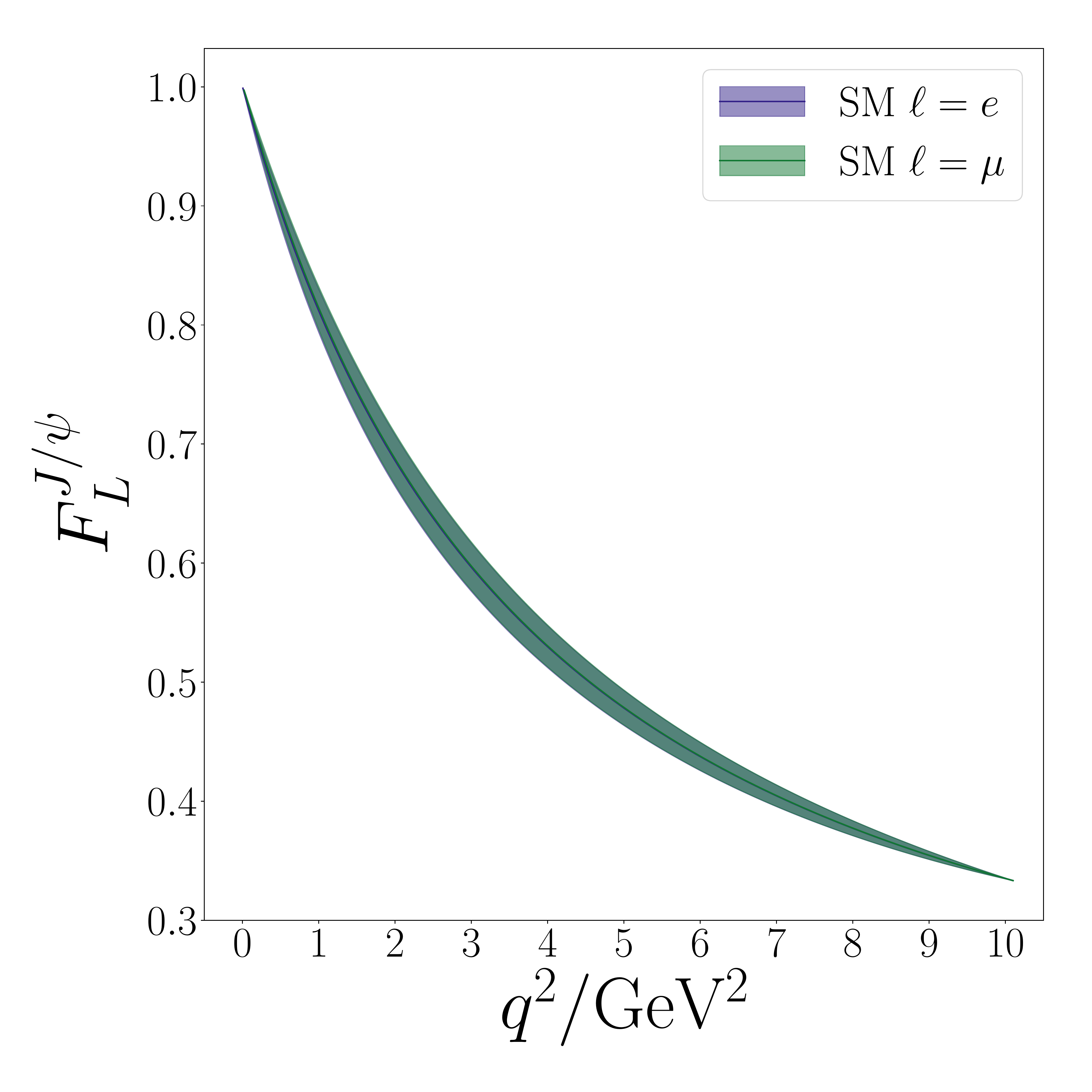}
\caption{\label{angasym} Angular asymmetry variables defined in Eq.~(\ref{angasymeq}) for the cases $\ell=e$ and $\ell=\mu$.}
\end{figure}

We also compute the forward-backward asymmetry, $\mathcal{A}_{FB}$, the lepton polarisation asymmetry (for the lepton $\ell$ produced from the virtual $W$), $\mathcal{A}_{\lambda_\ell}$, and the fraction of $J/\psi$ produced with longitudinal polarisation, $F_L^{J/\psi}$. These are defined as
\begin{align}
\label{angasymeq}
\mathcal{A}_{FB}(q^2) =& \frac{1}{d\Gamma/dq^2} \frac{2}{\pi}\int_0^\pi\frac{d\Gamma}{dq^2d\cos(\theta_W)}\cos(\pi-\theta_W)d\theta_W,\nonumber\\
\mathcal{A}_{\lambda_\ell}(q^2) =& \frac{d\Gamma^{\lambda_\ell=-1/2}/dq^2-d\Gamma^{\lambda_\ell=+1/2}/dq^2}{d\Gamma/dq^2},\nonumber\\
F_{L}^{J/\psi}(q^2) =& \frac{d\Gamma^{\lambda_{J/\psi}=0}/dq^2}{d\Gamma/dq^2}
\end{align}
where we have chosen the forward direction for the purpose of $\mathcal{A}_{FB}$ as being in the direction of the $J/\psi$ momentum in the $B_c$ rest frame. These quantities are plotted in Figure~\ref{angasym} for the cases $\ell=e$ and $\ell=\mu$. We see in the top plot that $\mathcal{A}_{FB}$ is negative except near $q^2=0$ for $\ell=\mu$. This may be be understood from Figure~\ref{helicityamplitudesplot} where we see that $H_+^2-H_-^2$, which is the dominant contribution to $\mathcal{A}_{FB}$ for $m_\ell^2<<q^2$, is less than or equal to zero across the full physical $q^2$ range. The behaviour of $\mathcal{A}_{FB}$ near $q^2=0$ in the $\ell=\mu$ case originates from the $-2m_\ell^2/q^2 H_tH_0 \cos(\theta_W)$ term in Eq.~\ref{dgammadcosw}. When $m_\ell^2/q^2\approx\mathcal{O}(1)$ it is apparent from Figure~\ref{helicityamplitudesplot} that this term will dominate over the $H_+^2-H_-^2$ contribution.
In the middle plot of Figure~\ref{angasym} we see that $\mathcal{A}_{\lambda_e}$ is equal to unity across the full $q^2$ range, in line with the expectation that in the massless limit the lepton is produced in a purely left handed helicity eigenstate.
In the bottom plot of Figure~\ref{angasym} we see that the longitudinal polarisation fraction, $F_L^{J/\psi}$, approaches unity near $q^2=0$ where $H_0$ and $H_t$ dominate the total rate, and goes to $1/3$ at $q^2_\mathrm{max}$ where $H_0=H_+=H_-$ and $H_t=0$.

We also compute the total decay rate for the cases $\ell=e$ and $\ell=\mu$. We find
\begin{eqnarray}
\label{gammatot}
\frac{\Gamma(B^-_c\rightarrow J/\psi \mu^-\overline{\nu}_\mu)}{|\eta_{\mathrm{EW}}V_{cb}|^2} &=& \GAMMATOTAL \\
&=& \GAMMAGEV \, .\nonumber
\end{eqnarray}
and
\begin{eqnarray}
\label{gammatote}
\frac{\Gamma(B^-_c\rightarrow J/\psi e^-\overline{\nu}_e)}{|\eta_{\mathrm{EW}}V_{cb}|^2} &=& \GAMMAeTOTAL \\
&=& \GAMMAeGEV \, \nonumber
\end{eqnarray}
with the ratio $\Gamma_e/\Gamma_\mu=\RATIOEMU$. We see that effects from including $m_\mu$ amount to $0.4\%$ of the rate. 

We can compare our results for the total rate to those from earlier, non-lattice QCD 
approaches such as potential models and QCD sum rules. 
In these other approaches it is much harder to quantify sources of uncertainty 
and derive an error budget. One way to obtain a picture of possible systematic 
errors is simply to compare results from different model calculations. 
Ref.~\cite{Ebert:2003cn}
uses a relativistic quark model to obtain a value $10.5\times 10^{-12}$ GeV 
for $\Gamma/|V_{cb}|^2$ and gives a table of comparison to other model 
approaches. Results vary over a range up to double this value, which implies 
a worrying lack of control of systematic effects in at least some of 
these calculations. In future it may 
be possible for such calculations to be tuned against our lattice QCD 
result for $B_c \rightarrow J/\psi$ and then yield useful information on 
$B_c$ decay to excited charmonium which is much harder to determine accurately 
in lattice QCD. 

From our value for $\Gamma/|\eta_{\mathrm{EW}}V_{cb}|^2$ in Eq.~(\ref{gammatot}) 
we can derive a result for the total width for 
$B_c^- \rightarrow J/\psi \mu^- \overline{\nu}_{\mu}$ using a value for $\eta_{\mathrm{EW}}$ 
(from Section~\ref{sec:theory}) and for 
$|V_{cb}|$. We take $|V_{cb}|= 41.0(1.4) \times 10^{-3}$ from an average of 
inclusive and exclusive determinations and with the error scaled by 2.4 
to allow for their inconsistency~\cite{pdg20}. 
This gives 
\begin{equation}
\label{eq:fullgamma}
\Gamma(B^-_c\rightarrow J/\psi \mu^-\overline{\nu}_\mu) = \GAMMATOTALincVCBETAEW . \\
\end{equation}
Here the first uncertainty is from our lattice QCD calculation and the second from 
the uncertainty in $|\eta_\mathrm{EW}V_{cb}|$. These uncertainties are approximately equal. 
  
Using the experimental average value for the $B_c$ lifetime 
of $0.510(9) \times 10^{-12}\,\mathrm{s}$~\cite{pdg20,Aaij:2014gka,Sirunyan:2017nbv}
then gives the branching fraction 
\begin{equation}
\label{eq:br}
\mathrm{Br}(B_c^-\rightarrow J/\psi \mu^-\overline{\nu}_\mu) = \BRANCHINGFRAC.
\end{equation}
The uncertainties are from lattice QCD, $|\eta_\mathrm{EW}V_{cb}|$ and the lifetime respectively. 

The accuracy of our result means that it can be used to normalise other 
$B_c$ decay modes. In the Particle Data Tables~\cite{pdg20} $B_c$ branching 
fractions are given in the modified form $B(\overline{b} \rightarrow B^+_c)\Gamma_i/\Gamma$.
Here the first factor is the probability for a produced $b$ quark to hadronise 
as a $B_c$ or one of its excitations. Using the value of the modified branching 
fraction for $B_c^- \rightarrow J/\psi \mu^- \overline{\nu}_{\mu}$ obtained by 
CDF~\cite{Aaltonen:2016dra} of $8.7(1.0)\times 10^{-5}$~\cite{pdg20} we infer
\begin{equation}
\label{bbc}
B(\overline{b} \rightarrow B_c^+) = \BbBc.  
\end{equation} 
The first uncertainty is the combined uncertainty from Eq.~(\ref{eq:br})
and the second uncertainty is from the experimental value of the modified branching 
fraction. 
This value can be compared to the value 
$B(\overline{b} \rightarrow B^+)=0.407(8)$~\cite{pdg20} to show how much 
less likely it is that a produced $b$ quark will form a $B_c$. 

We have focussed here on the rates for production of a $\mu$ or $e$
from the virtual $W$ in the final state. We will discuss results 
for $B_c^- \rightarrow J/\psi \tau^-\overline{\nu}_{\tau}$
in a follow-up paper~\cite{inprep}.

\subsection{Error Budget}

In order to estimate the contribution of the different sources of systematic uncertainty we use the 
inbuilt error budget routine in \textbf{lsqfit}~\cite{lsqfit}, which computes the partial variance of 
our result with respect to the priors and data entering the fit (see e.g.~\cite{Bouchard:2014ypa} Appendix A). The error budget for the total rate 
for the $\ell=\mu$ case is given in Table~\ref{errbudget}. The size of the systematic uncertainties reflect 
the extent to which the corresponding fit parameters are determined by the prior values together 
with how much their values impact the value of $\Gamma_\mu$.
Not surprisingly we see that large contributions come from 
the discretisation effects from the heavy quark mass and the statistical 
uncertainties on the finest (ultrafine, set 3) lattices. 
The relatively large uncertainty coming from sea charm quark mass mistuining terms reflects the fact that we do not have access to configurations with multiple sea charm masses at a given lattice spacing and hence this term is not well constrained. Excluding the sea charm mistuning term from our fits has only a small effect at of order $0.1\sigma$.

Note that several of the largest uncertainties, those from statistics on set 3 and $am_h$ discretisation effects and $M_{\eta_h}$ dependence, may be 
straightforwardly and systematically reduced in the future by increasing statistics on ultrafine 
lattices (set 3) and obtaining results on exafine ($a=$0.03 fm) lattices respectively. Sub-leading errors may also be improved,
with more precise determinations of $w_0$ or the inclusion of additional lattices with physical $u/d$ quarks.

\begin{table}
\centering
\caption{\label{errbudget} Error budget for the total 
rate for $B_c \rightarrow J/\psi \mu \overline{\nu}_\mu$. 
Errors are given as a percentage of the final answer. The top half gives sources of systematic errors: Extrapolation of data in $M_{\eta_h}$ to the physical value of $M_{\eta_b}$, errors proportional to powers of $am_c$, $am_h$, mass mistuning effects $\delta_m$, and uncertainties in determination of the lattice spacing. The second section gives a breakdown of the statistical uncertainties corresponding to each of our data sets. Finally `Other priors' includes all of the remaining sources of uncertainty such as $\Delta_\text{kin}$ and the current renormalisation factors $Z$.}
\begin{tabular}{ c | c }
\hline
 Source & $\% error$ \\\hline
$M_{\eta_h}\rightarrow M_{\eta_b}$ & 2.38\\ 
$am_c\rightarrow 0$ & 1.5\\ 
$am_h\rightarrow 0$ & 3.54\\ 
$\delta_{m_c}^\text{val}$ & 0.143\\ 
$\delta_{m_c}^\text{sea}$ & 3.2\\ 
$\delta_{m_s}^\text{sea}$ & 1.53\\ 
$\delta_{m_l}^\text{sea}$ & 0.671\\ 
$w_0/a$, $w_0$ & 1.37\\ 
 \hline
Statistics & \\\hline
Set 1 & 0.737\\ 
Set 2 & 1.8\\ 
Set 3 & 3.02\\ 
Set 4 & 0.435\\ 
 \hline
Other Priors & 1.61\\ 
 \hline
Total & 7.15\\ 
 \hline

\end{tabular}
\end{table} 

\section{Conclusions}
\label{sec:conclusion}
We have carried out the first lattice QCD calculation of the 
form factors within the Standard Model for 
$B_c^-\rightarrow J/\psi\ell^-\overline{\nu}_\ell$ decay. This is a process under 
active study by LHCb~\cite{PhysRevLett.120.121801, Bediaga:2018lhg} and 
more accurate form factors are needed (and crucially with quantified 
errors) to reduce uncertainties in 
the experimental determination of the lepton-universality violating 
ratio $R(J/\psi)$. 

We give all four form 
factors across the full $q^2$ range of the decay. The 
calculation is done using the HISQ action for all quarks, enabling us 
to normalise the lattice current operators that couple to the $W$ 
fully nonperturbatively. We have used the heavy-HISQ approach, obtaining results 
at multiple values of the heavy quark mass on a range of fine lattice spacings,
fitting the heavy-quark mass dependence to obtain a physical result 
for $B_c \rightarrow J/\psi$ in the continuum limit.  

We give each form factor in a parameterisation that allows it to 
be fully reconstructed, including correlations between the form 
factors (see Appendix~\ref{anfullcov}). This yields a total rate 
integrated over $q^2$ for a light lepton in the final state, 
up to a CKM factor, of (repeating Eq.~(\ref{gammatot})) 
\begin{equation}
\label{gammatot2}
\frac{1}{|\eta_{\mathrm{EW}}V_{cb}|^2}\Gamma(B^-_c\rightarrow J/\psi \mu^-\overline{\nu}_\mu) = \GAMMATOTAL .
\end{equation}
The 7\% uncertainty is broken down in an error budget in Table~\ref{errbudget}. 
Our result for the total rate gives a branching fraction (repeating Eq.~(\ref{eq:br}))
\begin{equation}
\label{eq:br2}
\mathrm{Br}(B_c^-\rightarrow J/\psi \mu^-\overline{\nu}_\mu) = \BRANCHINGFRAC.
\end{equation}

The first two uncertainties, from our lattice QCD calculation and from the current uncertainty in $V_{cb}$ are equal
here.

We will discuss the implications of our results for the case 
of a heavy ($\tau$) lepton in the final-state elsewhere~\cite{inprep}. 

We have demonstrated that the heavy-HISQ methodology is a viable procedure 
for these calculations and that the statistical challenge of 
analysing many lattice QCD correlation functions, while preserving 
covariances important for constructing and fitting $q^2$ and heavy mass dependence, 
can be handled. 
This work forms an important precursor for the calculation of 
$B_s \rightarrow D_s^*$ form factors across the full $q^2$ range using 
the same machinery, which is underway. This will allow the dominant 
uncertainty from external inputs to be reduced in the 
determination of $V_{cb}$ from experimental results~\cite{Aaij:2020hsi}. 

\subsection*{\bf{Acknowledgements}}

We are grateful to the MILC collaboration for the use of
their configurations and code. 
We thank C. Bouchard, B. Colquhoun, J. Koponen, P. Lepage, 
E. McLean and C. McNeile for useful discussions.
Computing was done on the Cambridge service for Data 
Driven Discovery (CSD3), part of which is operated by the 
University of Cambridge Research Computing on behalf of 
the DIRAC 
HPC Facility of the Science and Technology Facilities 
Council (STFC). The DIRAC component of CSD3 was funded by 
BEIS capital funding via STFC capital grants ST/P002307/1 and 
ST/R002452/1 and STFC operations grant ST/R00689X/1. 
DiRAC is part of the national e-infrastructure.  
We are grateful to the CSD3 support staff for assistance.
Funding for this work came from the UK
Science and Technology Facilities Council grants 
ST/L000466/1 and ST/P000746/1.

\begin{appendix}

\section{Reconstructing the Fit}
\label{anfullcov}

\begin{table}
\caption{\label{corr_A0} Correlation matrix for $z$-expansion coefficients of $A0$.}
\begin{tabular}{ c | c c c c }\hline
$\sigma^2 $& $a_0^{A0}$	& $a_1^{A0}$	& $a_2^{A0}$	& $a_3^{A0}$	\\\hline
$a_0^{A0}$&	1.0&	-0.3226&	0.03144&	-0.005681\\
$a_1^{A0}$&	-0.3226&	1.0&	-0.4749&	-0.001147\\
$a_2^{A0}$&	0.03144&	-0.4749&	1.0&	-0.0141\\
$a_3^{A0}$&	-0.005681&	-0.001147&	-0.0141&	1.0\\
\hline
\end{tabular}
\end{table}

\begin{table}
\caption{\label{corr_A0_A1} Correlation matrix for $z$-expansion coefficients of $A0$ and $A1$.}
\begin{tabular}{ c | c c c c }\hline
$\sigma^2 $& $a_0^{A1}$	& $a_1^{A1}$	& $a_2^{A1}$	& $a_3^{A1}$	\\\hline
$a_0^{A0}$&	0.5303&	-0.113&	0.06274&	0.01696\\
$a_1^{A0}$&	0.01653&	0.1351&	-0.2015&	-0.03253\\
$a_2^{A0}$&	-0.04758&	-0.0165&	0.1561&	0.01121\\
$a_3^{A0}$&	0.006368&	-0.01748&	0.03616&	0.008669\\
\hline
\end{tabular}
\end{table}

\begin{table}
\caption{\label{corr_A0_A2} Correlation matrix for $z$-expansion coefficients of $A0$ and $A2$.}
\begin{tabular}{ c | c c c c }\hline
$\sigma^2 $& $a_0^{A2}$	& $a_1^{A2}$	& $a_2^{A2}$	& $a_3^{A2}$	\\\hline
$a_0^{A0}$&	-0.3364&	0.2287&	-0.08114&	-0.008593\\
$a_1^{A0}$&	0.1069&	-0.3612&	0.1906&	0.01508\\
$a_2^{A0}$&	0.1271&	-0.1542&	-0.08251&	-0.003635\\
$a_3^{A0}$&	0.008374&	0.00399&	-0.04077&	-0.0044\\
\hline
\end{tabular}
\end{table}

\begin{table}
\caption{\label{corr_A0_V} Correlation matrix for $z$-expansion coefficients of $A0$ and $V$.}
\begin{tabular}{ c | c c c c }\hline
$\sigma^2 $& $a_0^{V}$	& $a_1^{V}$	& $a_2^{V}$	& $a_3^{V}$	\\\hline
$a_0^{A0}$&	0.02917&	-0.001606&	0.0001568&	6.563e-06\\
$a_1^{A0}$&	0.0007577&	0.01086&	0.0008211&	4.483e-05\\
$a_2^{A0}$&	-0.00359&	0.003377&	0.0004732&	2.63e-05\\
$a_3^{A0}$&	0.001619&	4.373e-05&	8.313e-05&	4.366e-06\\
\hline
\end{tabular}
\end{table}

\begin{table}
\caption{\label{corr_A1} Correlation matrix for $z$-expansion coefficients of $A1$.}
\begin{tabular}{ c | c c c c }\hline
$\sigma^2 $& $a_0^{A1}$	& $a_1^{A1}$	& $a_2^{A1}$	& $a_3^{A1}$	\\\hline
$a_0^{A1}$&	1.0&	-0.1026&	-0.01626&	-0.02278\\
$a_1^{A1}$&	-0.1026&	1.0&	-0.4741&	0.0354\\
$a_2^{A1}$&	-0.01626&	-0.4741&	1.0&	-0.09828\\
$a_3^{A1}$&	-0.02278&	0.0354&	-0.09828&	1.0\\
\hline
\end{tabular}
\end{table}

\begin{table}
\caption{\label{corr_A1_A2} Correlation matrix for $z$-expansion coefficients of $A1$ and $A2$.}
\begin{tabular}{ c | c c c c }\hline
$\sigma^2 $& $a_0^{A2}$	& $a_1^{A2}$	& $a_2^{A2}$	& $a_3^{A2}$	\\\hline
$a_0^{A1}$&	0.3107&	-0.09213&	0.08347&	0.01317\\
$a_1^{A1}$&	0.4329&	-0.054&	-0.2096&	-0.02255\\
$a_2^{A1}$&	-0.3841&	0.5468&	0.408&	0.03945\\
$a_3^{A1}$&	-0.01917&	-0.005971&	0.1006&	0.0114\\
\hline
\end{tabular}
\end{table}

\begin{table}
\caption{\label{corr_A1_V} Correlation matrix for $z$-expansion coefficients of $A1$ and $V$.}
\begin{tabular}{ c | c c c c }\hline
$\sigma^2 $& $a_0^{V}$	& $a_1^{V}$	& $a_2^{V}$	& $a_3^{V}$	\\\hline
$a_0^{A1}$&	0.0365&	-0.00378&	0.0003772&	1.863e-05\\
$a_1^{A1}$&	0.005846&	0.008809&	0.0002924&	1.149e-05\\
$a_2^{A1}$&	-0.007246&	0.005488&	4.225e-05&	2.69e-06\\
$a_3^{A1}$&	-0.004914&	0.0005791&	-0.0001413&	-7.189e-06\\
\hline
\end{tabular}
\end{table}

\begin{table}
\caption{\label{corr_A2} Correlation matrix for $z$-expansion coefficients of $A2$.}
\begin{tabular}{ c | c c c c }\hline
$\sigma^2 $& $a_0^{A2}$	& $a_1^{A2}$	& $a_2^{A2}$	& $a_3^{A2}$	\\\hline
$a_0^{A2}$&	1.0&	-0.6478&	0.1081&	0.006069\\
$a_1^{A2}$&	-0.6478&	1.0&	-0.2028&	0.005956\\
$a_2^{A2}$&	0.1081&	-0.2028&	1.0&	-0.05132\\
$a_3^{A2}$&	0.006069&	0.005956&	-0.05132&	1.0\\
\hline
\end{tabular}
\end{table}

\begin{table}
\caption{\label{corr_A2_V} Correlation matrix for $z$-expansion coefficients of $A2$ and $V$.}
\begin{tabular}{ c | c c c c }\hline
$\sigma^2 $& $a_0^{V}$	& $a_1^{V}$	& $a_2^{V}$	& $a_3^{V}$	\\\hline
$a_0^{A2}$&	0.00774&	-0.004908&	0.00013&	4.567e-06\\
$a_1^{A2}$&	-0.006189&	0.005253&	-0.0005569&	-3.031e-05\\
$a_2^{A2}$&	0.01841&	-0.0002758&	0.0007437&	3.818e-05\\
$a_3^{A2}$&	0.002944&	-0.0002532&	8.898e-05&	4.545e-06\\
\hline
\end{tabular}
\end{table}

\begin{table}
\caption{\label{corr_V} Correlation matrix for $z$-expansion coefficients of $V$.}
\begin{tabular}{ c | c c c c }\hline
$\sigma^2 $& $a_0^{V}$	& $a_1^{V}$	& $a_2^{V}$	& $a_3^{V}$	\\\hline
$a_0^{V}$&	1.0&	-0.2565&	0.01803&	0.0005269\\
$a_1^{V}$&	-0.2565&	1.0&	-0.3492&	-0.0105\\
$a_2^{V}$&	0.01803&	-0.3492&	1.0&	-0.001756\\
$a_3^{V}$&	0.0005269&	-0.0105&	-0.001756&	1.0\\
\hline
\end{tabular}
\end{table}

Our parameterisation of the form factors for $B_c \rightarrow J/\psi$ in the continuum limit 
is given in Eq.~(\ref{contfitfunction}). It consists of a pole factor with no uncertainty 
and a polynomial in $z$ for which the coefficients with their uncertainties are given 
in Table~\ref{zexpcoefficients}. 
In this section we give the correlations between the $z$-expansion coefficients which are necessary for reconstructing our results explicitly, as well as instructions for using the included ancillary files to load the $z$-expansion parameters together with their correlations automatically into \textbf{python}~\cite{python3}. 

The correlation between two coefficients is defined in the usual way as
\begin{equation}
\text{Corr}(X,Y) = \frac{\langle (\bar{X}-X)(\bar{Y}-Y) \rangle }{\sqrt{\sigma^2(X)\sigma^2(Y)}}
\end{equation}
where $\sigma^2(X)$ is the variance of $X$ and $\bar{X}$ is the mean of $X$. 
The values are tabulated in \cref{corr_A0,corr_A0_A1,corr_A0_A2,corr_A0_V,corr_A1,corr_A1_A2,corr_A1_V,corr_A2,corr_A2_V,corr_V}. 

In this calculation and in the ancillary files we use the \textbf{gvar} python package to track and propagate correlations. Included in the ancillary files are two text files; \textbf{CORRELATIONS.txt} contains a dictionary including the means and variances of the $z$-expansion parameters on the first line and a dictionary detailing the correlations between these parameters on the second line, \textbf{CHECKS.txt} contains arrays of $q^2$ values and form factor mean and standard deviation values at the corresponding values of $q^2$. This file is used by the python script \textbf{load\_fit.py} as a simple check that the fit has been loaded correctly. Running \textbf{python load\_fit.py} will load the parameters from \textbf{CORRELATIONS.txt} and compare values computed at hard coded intervals in $q^2$ to those in \textbf{CHECKS.txt} which were computed as part of this work. Running \textbf{python load\_fit.py} will also produce some simple plots of the form factors across the full $q^2$ range.  We have tested \textbf{load\_fit.py} using \textbf{Python 3.7.5}~\cite{python3}, \textbf{gvar 9.2.1}~\cite{gvar}, \textbf{numpy 1.18.2}~\cite{numpy} and \textbf{matplotlib 3.1.2}~\cite{matplotlib}.

\section{Converting to the BGL Parameterisation and Tests of Unitarity Bounds}
\label{BGL_tests}

\begin{table}
\caption{$\chi$ values entering the outer functions, $\phi(t,t_0)$, in the BGL scheme (Eq.(\ref{bglfit})) computed using the expressions in~\cite{Boyd:1997kz}.\label{chitab}}
\begin{tabular}{ c | c}\hline
 $\chi$ & $u=0.33$, $\alpha_s=0.22$\\\hline
$m_b^2\chi_T(u)/ \mathrm{GeV}^2$ & 0.0126835\\
$m_b^2\chi_T(-u)/ \mathrm{GeV}^2$ & 0.00737058\\
$\chi_L(u)$ & 0.00450115\\
$\chi_L(-u)$ & 0.0249307\\
\hline
\end{tabular}
\end{table}
\begin{table}
\caption{$a^\mathrm{BGL}_{n}$ values for our form factors converted to the helicity basis and then fitted using the BGL scheme (Eq.(\ref{bglfit})). The final column gives $\sum_{n=0}^3 (a^\mathrm{BGL}_n)^2$.\label{anbgl}}
\begin{tabular}{ c | c c c c | c}\hline
& $a^\mathrm{BGL}_0$	& $a^\mathrm{BGL}_1$	& $a^\mathrm{BGL}_2$	& $a^\mathrm{BGL}_3$	&$\sum (a^\mathrm{BGL}_n)^2$ \\\hline
$F_1$&0.002932(98)&-0.0149(29)&-0.028(49)&0.21(17)&0.044(72)\\
$F_2$&0.0374(14)&-0.192(24)&-0.34(21)&-0.16(26)&0.18(23)\\
$f$&0.01781(60)&-0.104(12)&0.30(16)&-0.23(18)&0.16(18)\\
$g$&0.0257(13)&-0.129(20)&-0.21(15)&-0.09(17)&0.070(95)\\
\hline
\end{tabular}
\end{table}

The now standard BGL parameterisation~\cite{Boyd:1997kz} makes use of the form factors in the helicity basis. These are given, in terms of the form factors used in this work, by~\cite{Cohen:2018dgz}:
\begin{align}
g=&\frac{2}{M_{B_c}+M_{J/\psi}}V\\
f=&(M_{B_c}+M_{J/\psi})A_1\\
F_1=&\frac{M_{B_c}+M_{J/\psi}}{M_{J/\psi}}\Big[ - \frac{2M_{B_c}^2 |\vec{p'}|^2}{(M_{B_c}+M_{J/\psi})^2}A_2 \nonumber\\
&- \frac{1}{2}(t-M_{B_c}^2+M_{J/\psi}^2)A_1\Big]\\
F_2=&2A_0
\end{align}
where $\vec{p'}$ is the $J/\psi$ spatial momentum in the $B_c$ rest frame. The BGL scheme then parameterises these form factors using the expansion in $z$ space
\begin{equation}
F_i(t)=\frac{1}{P_i(t)\phi(t,t_0)}\sum_{n=0}^\infty a^\mathrm{BGL}_{in}z(t,t_0)^n\label{bglfit}
\end{equation}
where the pole function $P_i$ is the same as we have defined in Eq.~(\ref{poleformeq}) and the outer functions, $\phi$, are defined in~\cite{Cohen:2018dgz}. Note that here we use the subthreshold resonance masses given in Table 2 of ~\cite{Cohen:2018dgz} in the pole functions $P_i$. In order to compute the outer functions we evaluate $\chi_{L(T)}(\pm u)$ using the expressions in~\cite{Boyd:1997kz}, using the pole mass $m_b=4.78\mathrm{GeV}$ from~\cite{pdg20} and $\alpha_s=0.22$ as in~\cite{Boyd:1997kz}. Their numerical values are given in Table~\ref{chitab}.

In order to convert our results into this parameterisation scheme we use our fitted continuum parameterisation to output form factor values at a large number of $q^2$ values in the semileptonic region $q^2<t_-$ and then fit these using Eq.~(\ref{bglfit}) truncated at $n=3$. We tabulate these values, together with $\sum_{n=0}^3 (a^\mathrm{BGL}_n)^2$,  in Table~\ref{anbgl} where we see that the unitarity bound $\sum_{n=0}^3 (a^\mathrm{BGL}_n)^2 \leq 1$ is satisfied comfortably for each form factor.
\end{appendix}

%===========================================================================8
\bibliographystyle{apsrev4-1}
\bibliography{BcJpsi}
%===========================================================================8

\end{document}